\documentclass[fleqn,usenatbib]{mnras}

\usepackage{newtxtext,newtxmath}

\usepackage[T1]{fontenc}
\usepackage{ae,aecompl}

\usepackage{graphicx}	% Including Figure files
\usepackage{amsmath}	% Advanced maths commands
\usepackage{epstopdf}
\usepackage{float}
\usepackage{longtable}
\usepackage{caption}
\usepackage{subcaption}
\usepackage{mathtools}
\usepackage{url}
\usepackage{tablefootnote}
\usepackage{booktabs}
\usepackage{makecell}
\usepackage{geometry}
\usepackage{pdflscape}

\title[Blackbody components in the X-ray spectra of GRBs]{A comprehensive view of blackbody components in the X-ray spectra of GRBs}

\author[V. Valan and J. Larsson]{
Vlasta Valan,$^{1}$\thanks{E-mail: vlasta2@kth.se}
Josefin Larsson,$^{1}$
\\
$^{1}$ KTH, Department of Physics and Oskar Klein Centre, Albanova, SE 106-91, Stockholm, Sweden\\
}

\date{Accepted XXX. Received YYY; in original form ZZZ}

\pubyear{2021}

\begin{document}
\label{firstpage}
\pagerange{\pageref{firstpage}--\pageref{lastpage}}
\maketitle

\begin{abstract}
A small fraction of gamma-ray bursts (GRBs) exhibit blackbody emission in the X-ray spectra, the origin of which is debated. In order to gain a more complete understanding of this phenomenon, we present a search for blackbody components in 116 GRBs with known redshifts observed by {\it Swift}~XRT. A time-resolved spectral analysis is carried out and the significance of the blackbody is assessed with respect to an absorbed power-law model. We report nine new detections and confirm the previously reported blackbody in GRB~171205A. Together with our previous results, there are a total of 19 GRBs with significant blackbody emission in a sample of 199 GRBs observed by {\it Swift} over 13 years. The detections include one short GRB and two low-luminosity GRBs.  We estimate fireball parameters from the blackbody components and note that the blackbody luminosity is correlated with the temperature and inferred Lorentz factor. There is a large spread in the properties of the blackbody components and the light curves, which points to different origins for the emission. In about a third of the GRBs, the blackbody  is clearly associated with late prompt emission from the jet. The rest of the sample includes cases that are fully consistent with the expectations from a cocoon, as well cases that may be explained by high-latitude emission or more energetic cocoons. These results indicate that thermal emission is associated with all parts of the jet. \end{abstract}


%%%%%%%%%%%%%%%%%%%%%%%%%%%%%%%%%%%%%%%%%%%%%%%%%%

%%%%%%%%%%%%%%%%% BODY OF PAPER %%%%%%%%%%%%%%%%%%

\section{Introduction}

Some supernovae and mergers of compact objects are accompanied by gamma-ray bursts (GRBs), which produce intense emission spanning the entire electromagnetic spectrum (see \citealt{kumar2015} for a review). This emission starts with gamma rays and X-rays during the so-called prompt phase, which lasts seconds to minutes, and is followed by afterglow emission at gradually lower energies, typically lasting weeks to months.  The prompt emission is thought to originate from a relativistic jet pointing toward the observer, while the afterglow is due to the subsequent interaction between the jet and the surrounding medium. The prompt emission is mainly observed by gamma-ray facilities, such as the  \textit{Fermi} Gamma-ray Burst Monitor (GBM, \citealt{Meegan2009}) and the \textit{Neil Gehrels Swift Observatory} Burst Alert Telescope (BAT, \citealt{Barthelmy2005}). The earliest soft X-ray observations are obtained by the \textit{Swift} X-ray telescope (XRT, \citealt{Burrows2005}), which covers the 0.3--10~keV energy range and typically starts observing  $\sim 100$~s after the trigger.  

Many GRBs exhibit a canonical X-ray light curve that consists of five parts: a steep decay, a shallow decay, a normal decay, a post-jet break component, as well as flares that are usually superposed on the earlier phases \citep{Zhang2006}. In this work we focus on the emission before $\sim 600$~s, which is often in the steep decay phase and sometimes contains flares that correlate with the gamma-ray emission. The emission at these times is thought to be due to a combination of late prompt emission from the central engine, including high-latitude emission from the jet \citep{Kumar2000,Liang2006,Zhang2009}, as well as the onset of the afterglow, with the relative contributions of these components varying for different GRBs (e.g., \citealt{Obrien2006}). Emission from a hot cocoon surrounding the jet \citep{meszaros2001} may also contribute, as discussed further below.  The spectra of the early X-ray emission are usually well fitted by power laws with typical photon indices ($\Gamma$) in range $1.5 \lesssim \Gamma \lesssim 3$ \citep{Racusin2009}, although it should be noted that flares sometimes exhibit more complex spectra \citep{Falcone2007, Peng2014}. 

There are a number of GRBs that deviate significantly from the canonical behaviour. For example, the low-luminosity (LL) GRB~060218 exhibits unusual properties, including a spectrum with a prominent blackbody component \citep{campana}. The intriguing properties of this GRB prompted the search for blackbody emission in other GRBs, and $\sim 20$ such cases have been reported to date \citep{page2011, Thone2011, starling2011, starling2012, sparre, mette, Nappo2016, Valan2018, Delia2018}.  However, with the exception of one more LL GRB (GRB~171205A), the properties of these GRBs differ significantly from those of GRB~060218. While the blackbody component in GRB~060218 cools from an initial value of $\sim 0.15$~keV on a time scale of several 1000~s, most of the other GRBs have higher temperatures, which decrease on a time scale of $\sim 100$~s or less. The luminosities of the blackbody components also differ dramatically, being $\sim10^{45}\ \rm{erg\ s^{-1}}$ in the LL GRBs, but at least two orders of magnitude higher in the other cases. The origin of the blackbody emission is still unclear, with proposed explanations including supernova shock breakout \citep{campana}, emission from a cocoon  \citep{meszaros2001, ghisellini2007, starling2012} and prompt emission of the jet itself \citep{mette, Nappo2016}. 

In order to investigate this phenomenon, we performed a time-resolved spectral analysis of 83 GRBs observed by XRT in \citet{Valan2018} (V18 from here on). Requiring a 3$\sigma$ detection in three consecutive time intervals, we identified nine GRBs with significant blackbody components, four of which had previously been reported in the literature. This detection criterion is stricter than most previous works, which reduces the number of detections, but also removes questionable cases and ensures that the spectral evolution can be studied. The results show that the shock-breakout scenario is  ruled out in all cases, with the possible exception of GRB~060218, while at least a part of the sample is consistent with cocoon emission. For some GRBs, prompt emission from the jet is strongly favoured. A larger sample is clearly needed to gain a more complete understanding of this emission component.

In this work we build on the methodology set up in V18 and analyse GRBs not covered in the previous analysis, more than doubling the total sample. The paper is structured as follows. Section 2 presents the sample, while Sections 3 and 4 describe the data reduction and spectral analysis, respectively. The results are presented in Section 5, followed by a discussion in Section 6 and conclusions in Section 7. In all our calculations we assume a flat Universe with $\rm{H_0} = 67.3\ \rm{km \ s^{-1} \ Mpc^{-1}}$, $\Omega_{M} = 0.315$ and  $\Omega_\Lambda = 0.685$ \citep{planck2014}. All uncertainties represent $90 \%$ confidence intervals. 

\section{Sample selection}
We analyse GRBs that meet the following selection criteria, which are the same as in V18:
\begin{itemize}
\item Known spectroscopic redshift ($z$)\footnote{Taken from: \url{http://www.mpe.mpg.de/~jcg/grbgen.html}.}
\item \textit{Swift} XRT windowed timing (WT) mode data available
\item Observed time-averaged WT flux ($F_{\rm{av, 0.3-10\ keV}}$) higher than $ 2 \times 10^{-10} \  \rm{erg \ cm^{-2} \ s^{-1}}$.
\end{itemize}
The analysis in V18 included all GRBs that meet these criteria observed between 2011 January 01 and 2015 December 31, as well as all GRBs observed prior to 2011 for which blackbody components had been reported in the literature \citep{starling2012, sparre, mette}. Here we analyse all GRBs observed before 2018 December 31 that were not already analysed in V18. 

The resulting sample (see Tab. \ref{sample}) consists of 116 GRBs, three of which are classified as short GRBs. The sample spans the redshift range 0.0368 -- 6.29 and the flux range $2 \times 10^{-10} \  \rm{erg \ cm^{-2} \ s^{-1}} \le F_{\rm{av,0.3-10\ keV}} \le 10.1 \times 10^{-8} \ \rm{erg \ cm^{-2} \ s^{-1}}$. The known redshifts are important for accurately constraining the absorption at low energies and also make it possible to infer the physical properties of the outflows from the blackbody components. The sample includes one GRB with a previously reported blackbody component; GRB~171205A \citep{Delia2018}. Together with the 83 GRBs presented in V18, the total number of GRBs analysed  is 199. 

\begin{table}
    \centering
    \caption{Sample of analysed GRBs with spectroscopic redshifts. The spectroscopic redshifts are usually reported without uncertainties. 83 GRBs not listed here were previously analysed in V18. Short GRBs are indicated with an "s" at the end of the name.} 
    \label{sample}
    \begin{tabular}{c c|c c|c c } % four columns, alignment for each
        \hline
        GRB  & $z$ & GRB & $z$ & GRB & $z$\\ 
        \hline
   050401 & 2.9  & 080430 & 0.767 & 161108A & 1.159\\
   050724s & 0.258  & 080603B & 2.69 & 161117A & 1.549\\
   050730 & 3.967 & 080604 & 1.416 & 161219B & 0.1475\\
   050814 & 5.77 &  080605 & 1.6398  & 170113A & 1.968\\
   050820 & 2.612 & 080721 & 2.591 &  170202A & 3.645\\
   050822A & 1.434 & 080805 & 1.505 & 170519A & 0.818\\
   050904 & 6.29 & 080810 & 3.35 & 170604A & 1.329\\
   050922C & 2.198 & 080905B & 2.374 & 170607A & 0.557\\
   051001 & 2.4296 & 080928 & 1.692 & 170705A & 2.01\\
   051109 & 2.346 & 081007 & 0.5295 & 170714A & 0.793\\
   060111 & 2.32 & 081008 & 1.9685 & 171205A & 0.0368\\
   060115 & 3.53 & 081028 & 3.038 & 171222A & 2.409\\
   060204B & 2.3393 & 081210 & 3.0631 & 180205A & 1.409 \\
   060210 & 3.91 & 081221 & 2.26  & 180325A & 2.248\\
   060502 & 1.51 & 081222 & 2.77  & 180329B & 1.998\\
   060510B & 4.94 & 090407 & 1.4485 & 180620B & 1.1175\\
   060526 & 3.221 & 090417B & 0.3415 & 180624A & 2.855    \\
   060604 & 2.1357 & 090418 & 1.608  & 180720B	& 0.654 \\
   060607 & 3.0749 & 090516 & 4.109 & 180728A & 0.117 \\
   060614 & 0.125 & 090529 & 2.625 & 181010A & 1.39 \\
   060708 & 1.92 & 090715B & 3 & 181020A	& 2.938 \\
   060714 & 2.711 & 090809 & 2.737  & 181110A & 1.505\\
   060719 & 1.532 & 090812 & 2.452 \\
   060729 & 0.54 & 090926B & 1.24\\
   060814 & 1.9229 & 091020 & 1.71\\
   060904B & 0.703 & 091029 & 2.752\\
   061202 & 2.2543 & 091127 & 0.49 \\
   061222A & 2.088 & 100302A & 4.813\\
   061222B & 3.355 & 100418A & 0.6235\\
   070110 & 2.352 & 100424A & 2.465 \\
   070129 & 2.3384 & 100425A & 1.755\\
   070223 & 1.6295 & 100615A & 1.398\\
   070306 & 1.4959 & 100728A & 1.567\\
   070318 & 0.836 & 100814A & 1.44\\
   070328 & 2.0627 & 100910A & 1.408\\
   070419A & 0.97 & 10906A & 1.727\\
   070419B & 1.9591 & 101213A & 0.414\\
   070721B & 3.626 & 160117B & 0.87\\
   071021 & 2.452 & 160131A & 0.972\\
   071025 & 5.2 & 160227A & 2.38\\
   071031 & 2.692 & 160110As & 1.717\\
   071112C & 0.823 & 160425A & 0.555\\
   071222 & 1.14 & 160804A & 0.736\\
   080210 & 2.641 & 160821Bs & 0.16\\
   080310 & 2.42 & 161014A & 2.823\\
   080319B & 1.1 & 161017A & 2.013 \\
   080325 & 1.78 & 161108A & 1.159\\
        \hline
    \end{tabular}
\end{table}

\section{Data reduction}\label{datared}

Our analysis is focused on XRT WT mode data, which has better temporal resolution than the photon counting (PC) mode. WT mode is used when the count rate is high and thus offers good quality spectra.
The XRT observations were downloaded from the UK {\it Swift} Science Data Centre XRT GRB repository\footnote{\url{http://www.swift.ac.uk/xrt_live_cat/}} and reduced using the automatic pipeline \citep{Evans}. As our analysis is time resolved, the light curves need to be binned in order to identify relevant start and stop times for the spectra. We performed the binning using Bayesian blocks \citep{scargle}, applying the routine \textsc{battblocks} with its default settings. We then used the resulting start and stop times to download time-resolved spectra from the {\it Swift} website. All spectra were grouped to a minimum of 20 counts per bin to enable the use of $\chi^2$ statistics. The resulting time-resolved spectra contain $\sim 200 - 4000$ counts each.

{\it Swift} XRT data are affected by several systematic effects. The most common one is pile-up, which occurs for count rates above $\sim$ 100 counts $\rm{s^{-1}}$ in WT mode. This is automatically corrected for when downloading reduced data from the UK {\it Swift} Science Data Centre. XRT data of moderately to highly absorbed sources (${N_{\rm H,intr}} \geq 5 \times 10^{21} \ \rm{cm^{-2}}$) may also be affected by redistribution issues at low energies ($ \lesssim 1 \ \rm{keV}$) as described in "XRT Calibration Digest".\footnote{\url{https://www.swift.ac.uk/analysis/xrt/digest_cal.php}} One of the possible problems is the appearance of a bump at low energies, which is seen when comparing spectra extracted using Grade 0-2  and Grade 0 events only, with the bump only being present in the former. We checked for this by comparing spectra extracted using the different Grade events and found no GRBs affected by this issue.
Another possible redistribution effect in moderately to highly absorbed sources is a ``turn-up" at energies $\lesssim 0.6~\rm{keV}$. GRB~170519A and GRB~180329B showed this issue and we therefore ignore all the counts below $0.6~\rm{keV}$ in the analysis of these bursts.  

We also reduced data from the BAT instrument, which sometimes provides a detection during the beginning of the XRT observations. We used the command \textsc{batbinevt} with start and stop times from the event files to produce spectra and lightcurves, and the \textsc{batdrmgen} command to produce associated response files.

\section{Data analysis}
\label{sec:analysis}
The data analysis follows V18 and we refer the reader to that paper for more details. The most important steps are summarized below. We used \textsc{xspec} version 12.10.1 \citep{xspec} to perform a time-resolved spectral analysis of all GRBs. The XRT data were fitted over 0.3--10~keV, except when there were no counts above an energy smaller than 10~keV, in which case the upper energy limit was set manually. In addition, the lower energy range was set to 0.6~keV for GRB~170519A and GRB~180329B as discussed in Section \ref{datared}. 

The primary goal of the analysis is to determine whether a blackbody component is present in the spectra. We use an absorbed power law as the baseline model (which is usually a good description of GRB spectra in the XRT energy range, e.g. \citealt{Racusin2009}) and investigate if there is a significant improvement in the fit statistic when a blackbody is added. In a small number of cases there is simultaneous BAT data available, covering the 15--150~keV energy range. In these cases we replace the power law with a cutoff power law in order to account for a likely cutoff at higher energies. All fit parameters were tied between the XRT and BAT apart from a cross normalisation constant that was allowed to vary in the range 0.9--1.1. We note that the BAT data usually have poor quality in the finely time-resolved spectra at these times ($\sim 100$~s after trigger) and therefore only have a minor impact on the results (discussed further below).

The spectra below $\sim 2~\rm{keV}$ are affected by Galactic and intrinsic absorption. To model this we use the \textit{tbabs} and \textit{ztbabs} models in XSPEC \citep{Wilms2000}. For each burst we determined the Galactic H column density ($N_{\rm{H,Gal}}$) using the  \textit{$N_{\rm{H,tot}}$} tool,\footnote{\url{http://www.swift.ac.uk/analysis/nhtot/index.php}} which accounts for contributions from both atomic and molecular H \citep{Willingale2013}.  The \textit{tbabs} model assumes that the molecular component accounts for $20 \ \%$ of the total column density. In cases where  the molecular component was $< 10 \ \%$ or $> 30 \ \%$ we used the \textit{tbvarbas} model to set the atomic and molecular contributions separately. The Galactic absorption was kept fixed in all fits. 

Absorption in excess of the Galactic value may originate in the host galaxy of the GRB as well as in the intergalactic medium (e.g., \citealt{starling2013}). Since we are not able to discriminate between these contributions we fit for the intrinsic absorption ($N_{\rm H,intr}$) using \textit{ztbabs} with the redshift fixed at the value of the host galaxy. As in V18 we determined $N_{\rm H,intr}$ by simultaneously fitting all time-resolved spectra with $N_{\rm H,intr}$ tied but all other parameters free to vary. This was done separately for the power-law and power-law + blackbody models since the $N_{\rm H,intr}$ determined from the power-law fits may be overestimated if a blackbody is present. In the case of GRB~171205A this method only returned an upper limit for $N_{\rm H,intr}$, and we therefore used the $N_{\rm H,intr}$ determined from late PC data when analysing this GRB. 

For nearly all GRBs with simultaneous BAT data available, the cutoff energy of the cutoff power law is well above the XRT energy range ($ 40 ~ {\rm keV} \le E_{\rm pk} \le 500 ~ {\rm keV}$) and often poorly constrained due to the low count statistics. The only exception to this is GRB~160227A, where the best-fit cutoff energy is in range $ 4 ~ {\rm keV} \le E_{\rm pk} \le 40 ~ {\rm keV}$. In this case we assessed the significance of the blackbody  based on the joint XRT+BAT fits. For all other GRBs we verified that including the BAT does not significantly affect the photon index in the XRT energy range or the results regarding the significance of the blackbody. In the following we will therefore only present the fits to the XRT data.

The significance of the blackbody component was assessed using Monte Carlo simulations. We simulated 10,000 spectra based on the best-fitting power-law model, using the response files of the original data. Background spectra were also simulated. Based on the experience from V18, we only performed simulations for bursts where the blackbody component showed a non-erratic evolution and improved the fit by $\Delta \chi^2 \geq 2$  in at least three consecutive bins (51 GRB in total). The simulated spectra were then analysed as described in V18. The significance of the added blackbody component was finally determined by comparing the resulting $\Delta \chi^2$ distribution with the $\Delta \chi^2$ obtained by fitting the real data with the two different models.

\section{Results}

\begin{figure*}
    \begin{subfigure}[b]{0.49\textwidth}
    \includegraphics[width=\columnwidth,  height = 6.66cm]{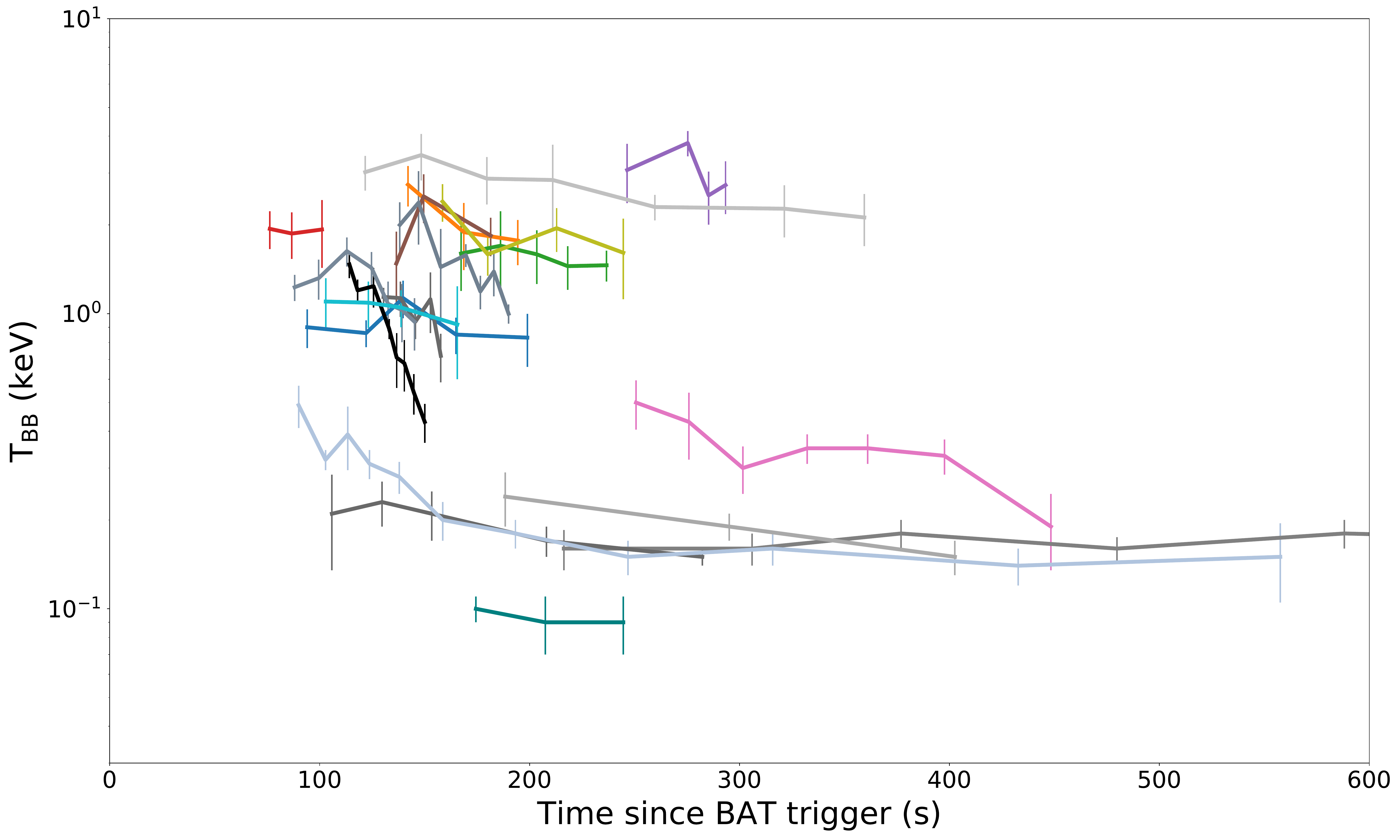}
    \end{subfigure}
    \begin{subfigure}[b]{0.49\textwidth}
     \includegraphics[width=\columnwidth, height = 6.66cm]{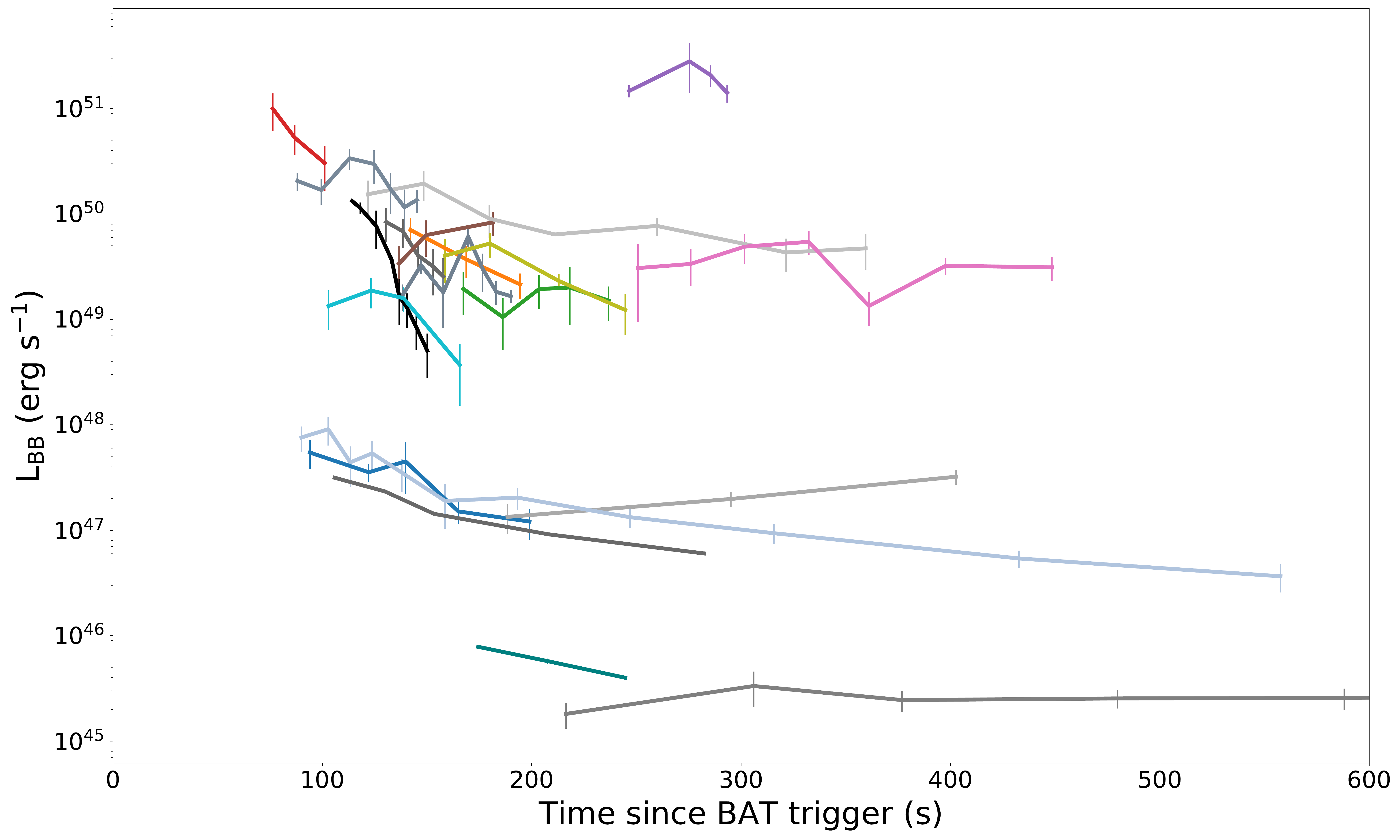}
    \end{subfigure}
     \begin{subfigure}[b]{0.49\textwidth}
     \includegraphics[width=\columnwidth, height = 6.66cm]{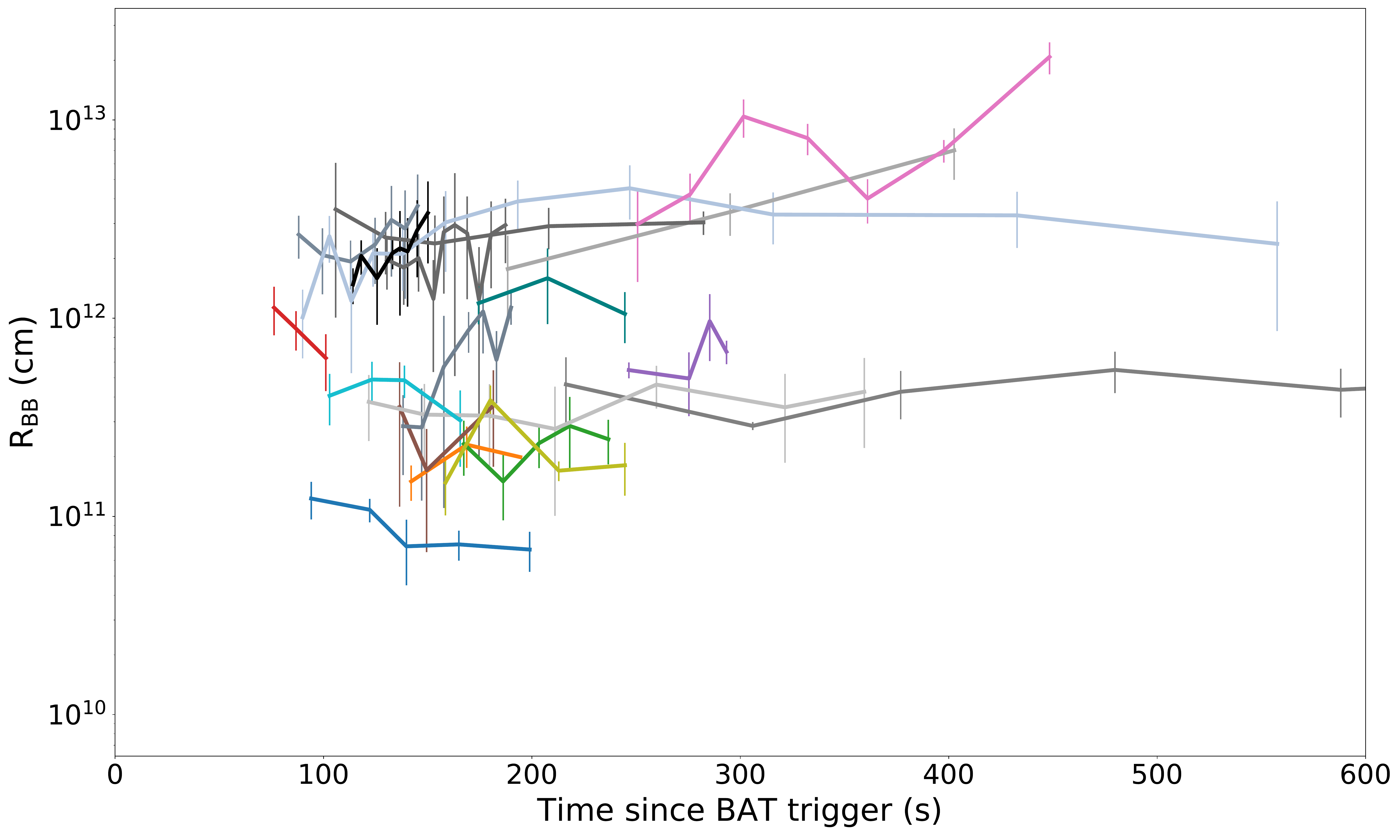}
    \end{subfigure}
    \begin{subfigure}[b]{0.49\textwidth}
    \includegraphics[width=\columnwidth,  height = 7cm]{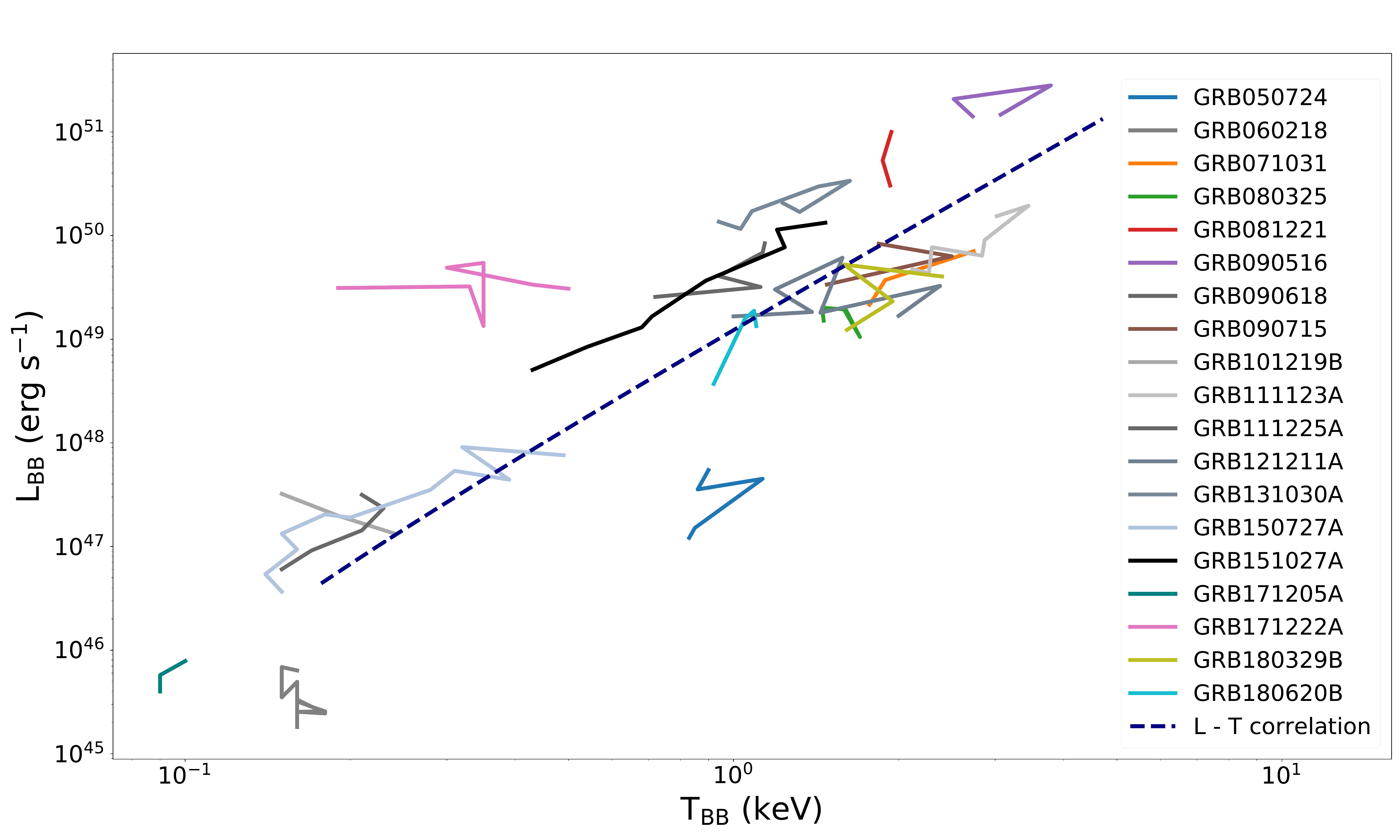}
    \end{subfigure}
    \caption{Best-fitting parameters of the blackbody components. The first three panels starting from the upper left show the time evolution of $T_{\rm BB}$, $L_{\rm BB}$ and $R_{\rm BB}$, respectively. The lower right panel shows the $L_{\rm BB} - T_{\rm BB}$ relation (where error bars are omitted for visual clarity). The gray-scale and black lines show GRBs that were presented in V18, while the more vivid colors show GRBs analysed in this paper. The dashed line represents the best-fit to the L-T relation; $L \propto T^{3.06 \pm 0.21}$.}
	\label{results}
\end{figure*}

We use the same criterion for a detection as in V18, i.e. the blackbody component must be significant at $ > 3 \sigma$ in at least three consecutive time bins. There are 10 GRBs that meet this criterion: GRB~050724, GRB~071031, GRB~080325, GRB~081221, GRB~090516, GRB~090715B, GRB~171205A, GRB~171222A, GRB~180329B, and GRB~180620B.  We also note that the blackbody components are significant in the time-integrated spectra of all these GRBs, though we base the following discussion exclusively on the time-resolved results. The best-fitting parameters for the time-resolved spectra of the 10 GRBs are provided in Table \ref{results}. In Table \ref{thermalwithz} we present absorption properties and blackbody energies for all GRBs with significant blackbodies, including the nine cases presented in V18.

\begin{table*}
    \centering
    \caption{Best-fitting parameters for the absorbed power-law+blackbody model for GRBs with significant blackbody components. The blackbody luminosity is the full unabsorbed luminosity, including parts of the blackbody that may fall outside the observed energy range. $\Gamma$ denotes the photon index. The full table, together with the best-fitting parameters for the absorbed power-law model, is available as online material.}
    \label{fits} 
    {\def\arraystretch{2}\tabcolsep=10pt

\begin{tabular}{c c c c c c c c}
\hline
\multicolumn{1}{p{2cm}}{\centering Time interval $\mathrm{(s)}$} & \multicolumn{1}{p{2cm}}{\centering $kT$  \\ $\mathrm{(keV)}$} & \multicolumn{1}{p{2cm}}{\centering $L_{\mathrm{BB}}$  \\ $\mathrm{(erg \ s^{-1})}$} & \centering $\Gamma$  &  \multicolumn{1}{p{2cm}}{\centering $F_{\mathrm{obs,0.3-10\ keV}}$   \\ $(10^{-9} \ \mathrm{erg \ cm^{-2} \ s^{-1})}$} & \multicolumn{1}{p{2cm}}{\centering $F_{\mathrm{BB, obs}}/ F_{\mathrm{tot, obs}}$ \\ $(\%)$ } & $\chi^2/ \mathrm{dof}$\\
\hline

GRB 050724 & & & & & & \\
80 - 109 & $0.90_{-0.10}^{+0.17}$ &  $5.46\pm 1.67 \times 10^{47}$ &  $0.73_{-0.99}^{+0.33}$ &  $2.44 \pm 0.86$ & $27\pm11$  & 62.76/65 \\
109 - 136 &  $0.86_{-0.08}^{+0.10}$ &  $3.55\pm 0.68\times 10^{47} $ &  $1.04_{-0.26}^{+0.19}$ &  $1.57_{-0.52}^{+0.53}$ &  $31 \pm 5$ & 113.44/108 \\
136 - 144 & $1.13_{-0.19}^{+0.14}$ &  $4.50\pm 2.31 \times 10^{47}$ &  $2.20_{-0.74}^{+1.13}$ &  $2.11_{-0.68}^{+0.12}$ &  $69 \pm 21$ & 28.46/25 \\
144 - 186 & $0.85^{+0.11}_{-0.13}$ &  $1.51\pm 0.37 \times 10^{47}$ &  $1.56_{-0.05}^{+0.05}$ &  $0.66_{-0.30}^{+0.30}$ & $27 \pm 5$  & 72.03/93 \\
186 - 212 & $0.83_{-0.15}^{+0.19}$ &  $1.21\pm 0.40 \times 10^{47}$ &  $1.50_{-0.36}^{+0.39}$ &  $0.53_{-0.27}^{+0.21}$ &  $32 \pm 2$ & 23.58/40\\
 \hline
\end{tabular}
}
\end{table*}

\begin{table*}
    \centering
    \caption{All GRBs with significant blackbody emission, including cases identified in V18. The table lists the redshifts, absorption properties (with $N_{\rm H,intr}$ derived by fitting an absorbed power-law + blackbody), as well as the total energy of the blackbody components. \\
    * The energy of the blackbody for GRB~060218 is a lower limit from the XRT WT mode data only.}
    \label{thermalwithz}
    {\def\arraystretch{2}\tabcolsep=10pt
    \begin{tabular}{c c c c c} % four columns, alignment for each
        \hline
        GRB  & $z$  & \multicolumn{1}{p{2cm}}{\centering $N_{\mathrm{H,Gal}}$ \\ $(10^{22}\  \mathrm{cm^{-2})}$} & \multicolumn{1}{p{2cm}}{\centering $N_{\mathrm{H,intr}}$  \\ $(10^{22}\  \mathrm{cm^{-2})}$} & \multicolumn{1}{p{2cm}}{\centering $E_{\mathrm{BB,tot}}$ \\ $\mathrm{(erg)}$ }\\ 
        \hline
   
		050724s & 0.258 & 0.277 & $0.16 \pm 0.10$ & $3.07 \times 10^{49}$ \\ 
		060218 & 0.0331  & 0.142 & $0.60 \pm 0.01$ & $ 1.10 \times 10^{49 *} $ \\
		071031 & 2.692 & 0.013 & $1.43 \pm 0.25$ & $ 1.15 \times 10^{51}$ \\
		080325 & 1.78 & 0.046 & $3.28 \pm 0.30$ & $5.29 \times 10^{50}$\\
		081221 & 2.26 & 0.022 & $5.04 \pm 0.49$ & $5.86 \times 10^{51}$ \\
		090516 & 4.109 & 0.053 & $4.90 \pm 0.65$ & $ 2.57 \times 10^{52}$ \\
        090618A & 0.54 & 0.076 & $0.28 \pm 0.03$ & $ 1.41 \times 10^{51}$ \\
        090715B & 3 & 0.014 & $ 2.41 \pm 0.33$ & $ 1.35 \times 10^{51}$\\
        101219B & 0.5519 & 0.033 & $0.16 \pm 0.11$ & $ 4.56 \times 10^{49}$ \\ 
        111123A & 3.1516 & 0.069 &  $4.79 \pm 0.36$ & $ 6.68 \times 10^{51}$ \\
        111225A	& 0.297  & 0.275 & $ 0.1 7 \pm 0.17$ & $2.30 \times 10^{49}$ \\
        121211A	& 1.023  & 0.009 & $ 0.58 \pm 0.17 $ & $1.04 \times 10^{51}$ \\
        131030A & 1.295  & 0.056  &  $0.42 \pm 0.11$ & $7.24 \times 10^{51}$ \\
        150727A	& 0.313 & 0.098 & $ 0.08 \pm 0.05$ & $ 5.94 \times 10^{49}$ \\
	    151027A	& 0.810 & 0.038 & $0.33 \pm 0.06$ & $ 1.40 \times 10^{51}$ \\
	    171205A & 0.0368 & 0.059  & $0.029 \pm 0.001$ &  $4.67 \times 10^{47}$ \\
	    171222A & 2.409 & 0.011 & $1.62 \pm 0.20$ & $2.45 \times 10^{51}$ \\
	    180329B & 1.998 & 0.025 & $0.06 \pm 0.06$ & $1.08 \times 10^{51}$ \\
	    180620B & 1.1175 & 0.014 & $ 1.00 \pm 0.10$ & $ 4.83 \times 10^{50}$ \\
        \hline 
    \end{tabular}
}
\end{table*}

\begin{figure*}
    \begin{subfigure}[b]{0.49\textwidth}
    \includegraphics[width=\columnwidth,  height = 4.5cm]{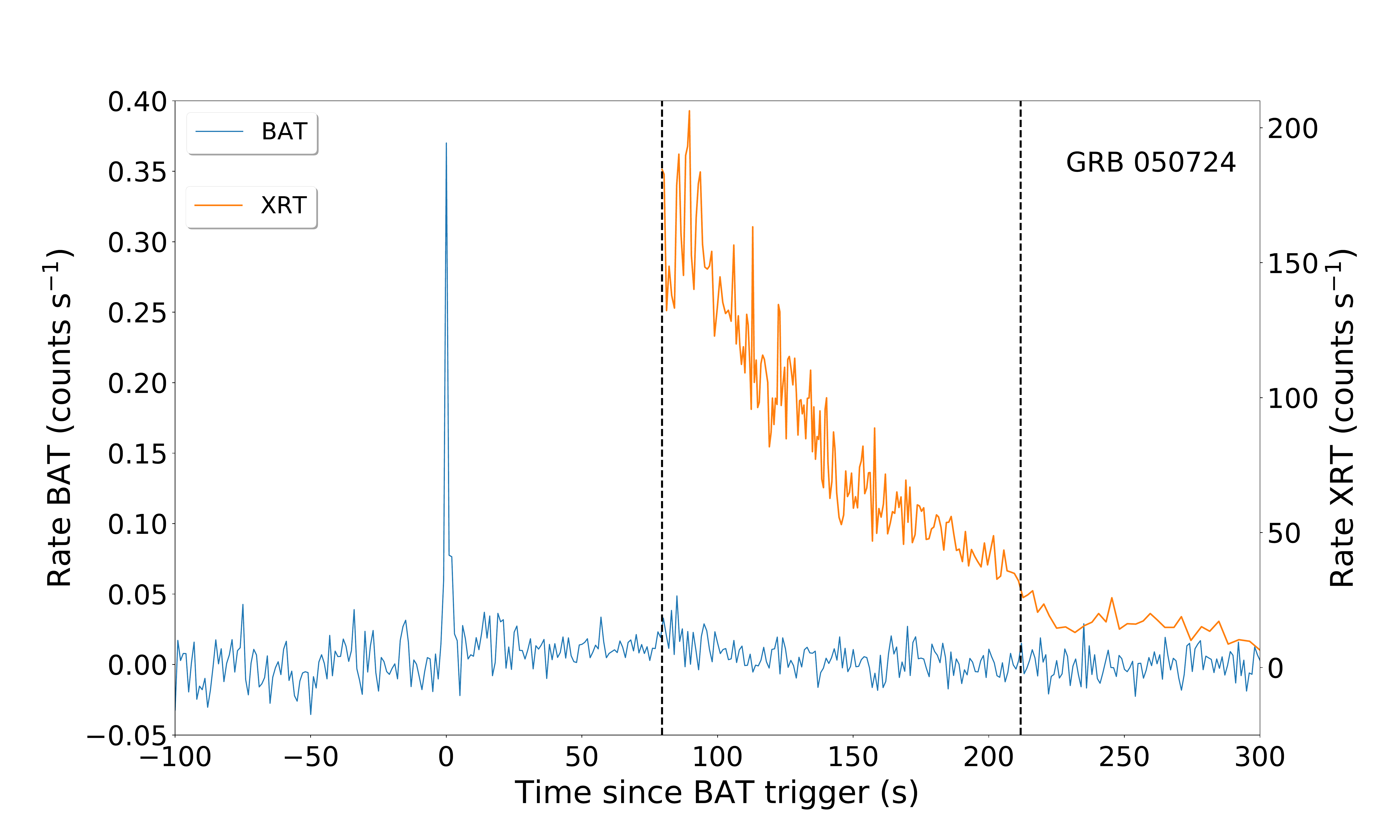}
    \end{subfigure}
    \begin{subfigure}[b]{0.49\textwidth}
     \includegraphics[width=\columnwidth, height = 4.5cm]{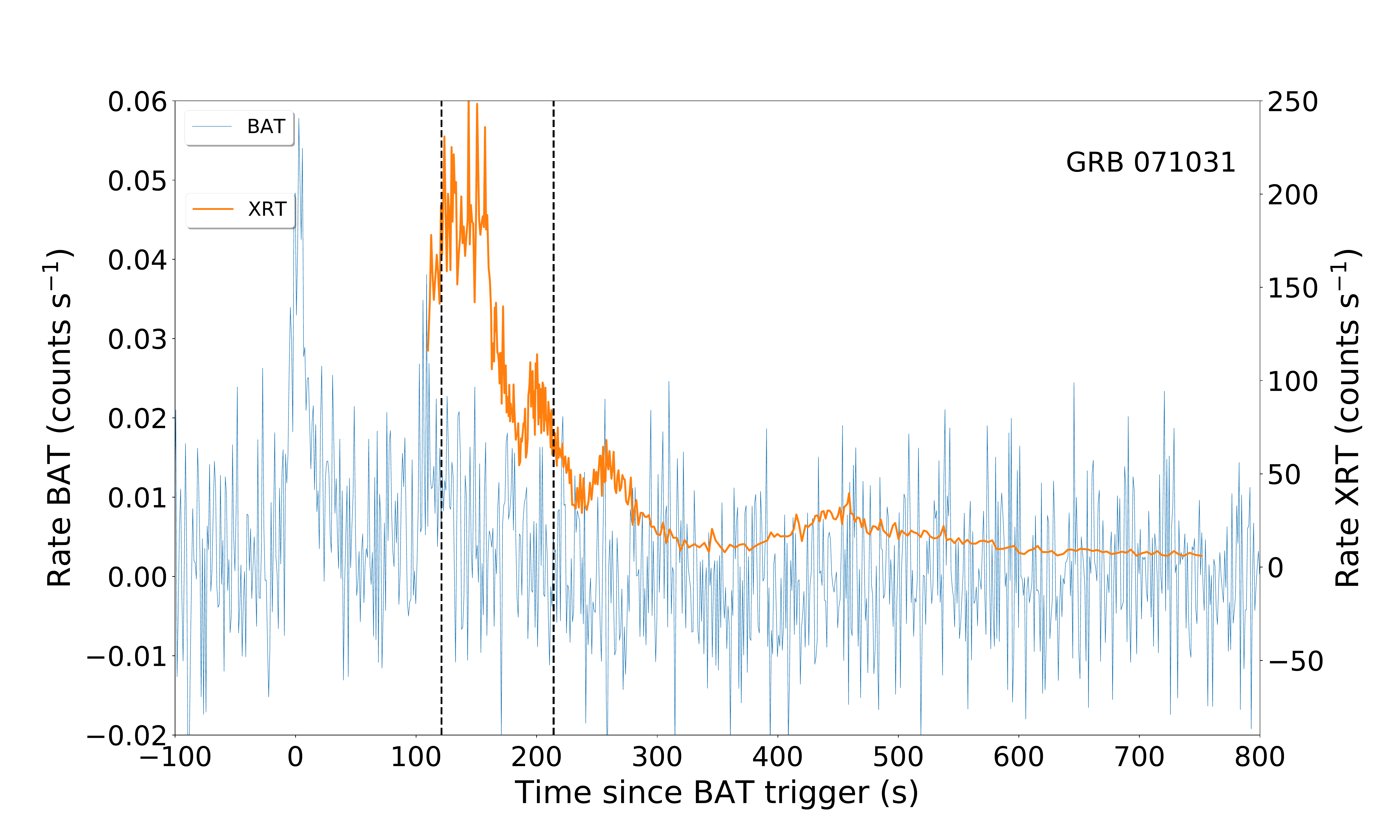}
    \end{subfigure}
        \begin{subfigure}[b]{0.49\textwidth}
     \includegraphics[width=\columnwidth, height = 4.5cm]{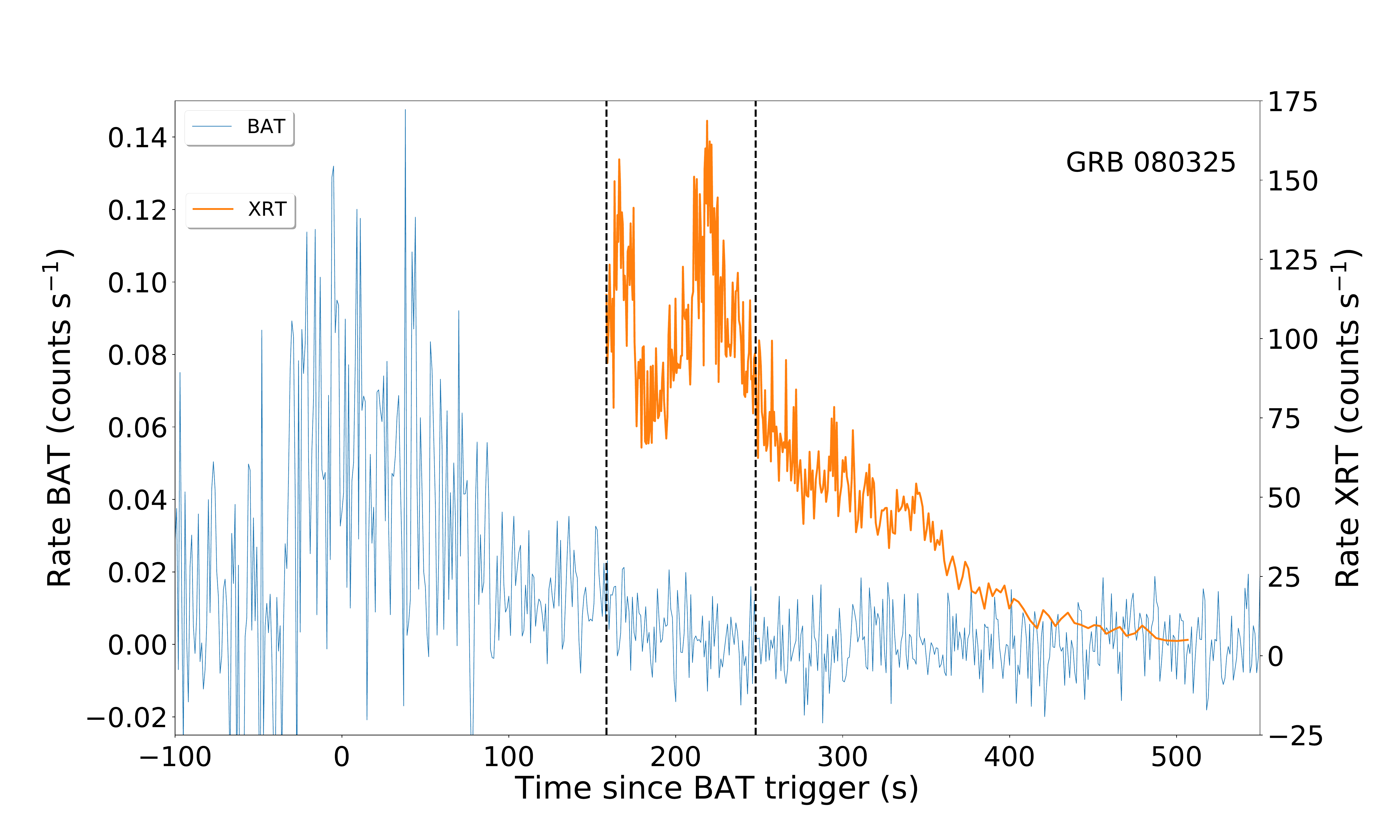}
    \end{subfigure}
        \begin{subfigure}[b]{0.49\textwidth}
     \includegraphics[width=\columnwidth, height = 4.5cm]{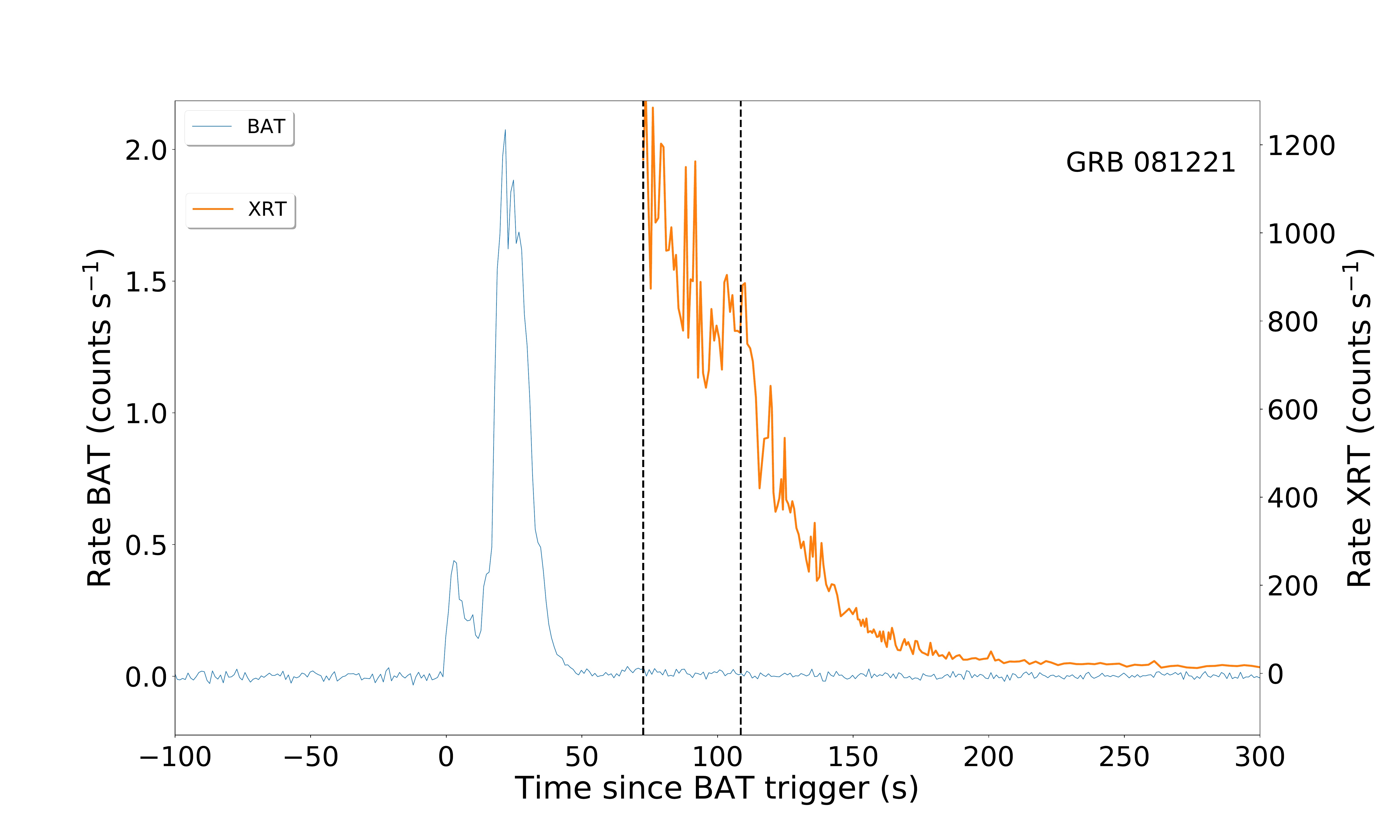}
    \end{subfigure}
        \begin{subfigure}[b]{0.49\textwidth}
     \includegraphics[width=\columnwidth, height = 4.5cm]{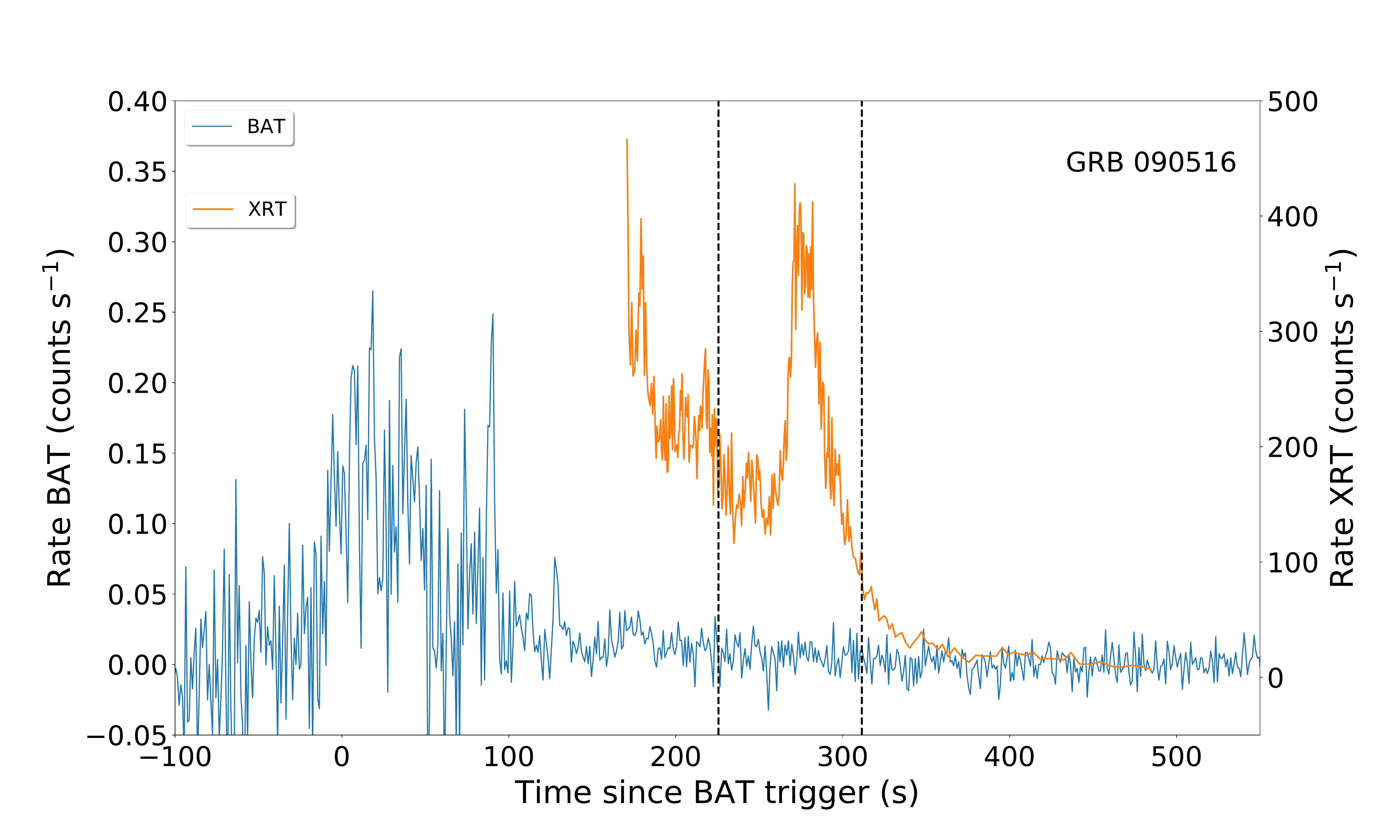}
    \end{subfigure}
        \begin{subfigure}[b]{0.49\textwidth}
     \includegraphics[width=\columnwidth, height = 4.5cm]{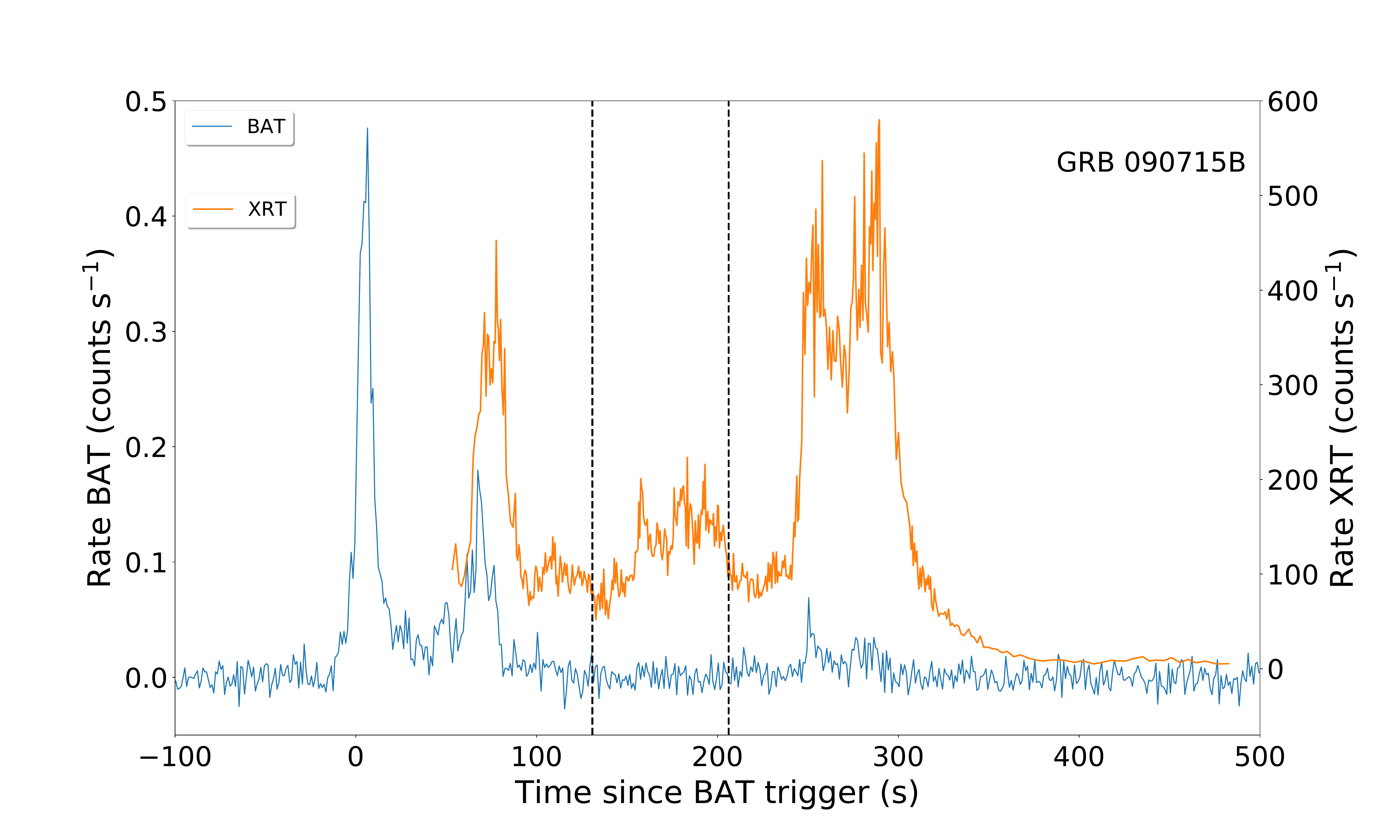}
    \end{subfigure}
     \begin{subfigure}[b]{0.49\textwidth}
     \includegraphics[width=\columnwidth, height = 4.5cm]{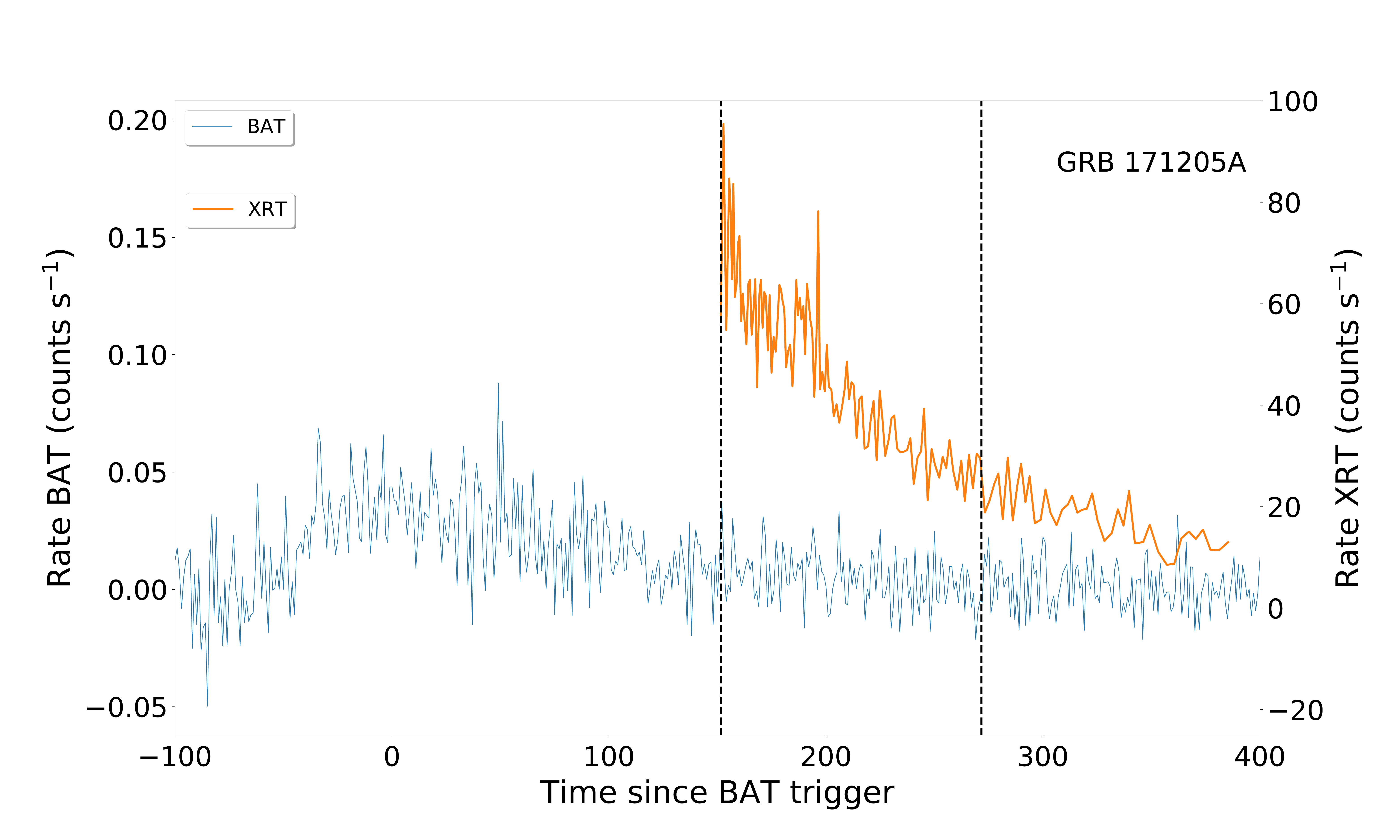}
    \end{subfigure}
        \begin{subfigure}[b]{0.49\textwidth}
     \includegraphics[width=\columnwidth, height = 4.5cm]{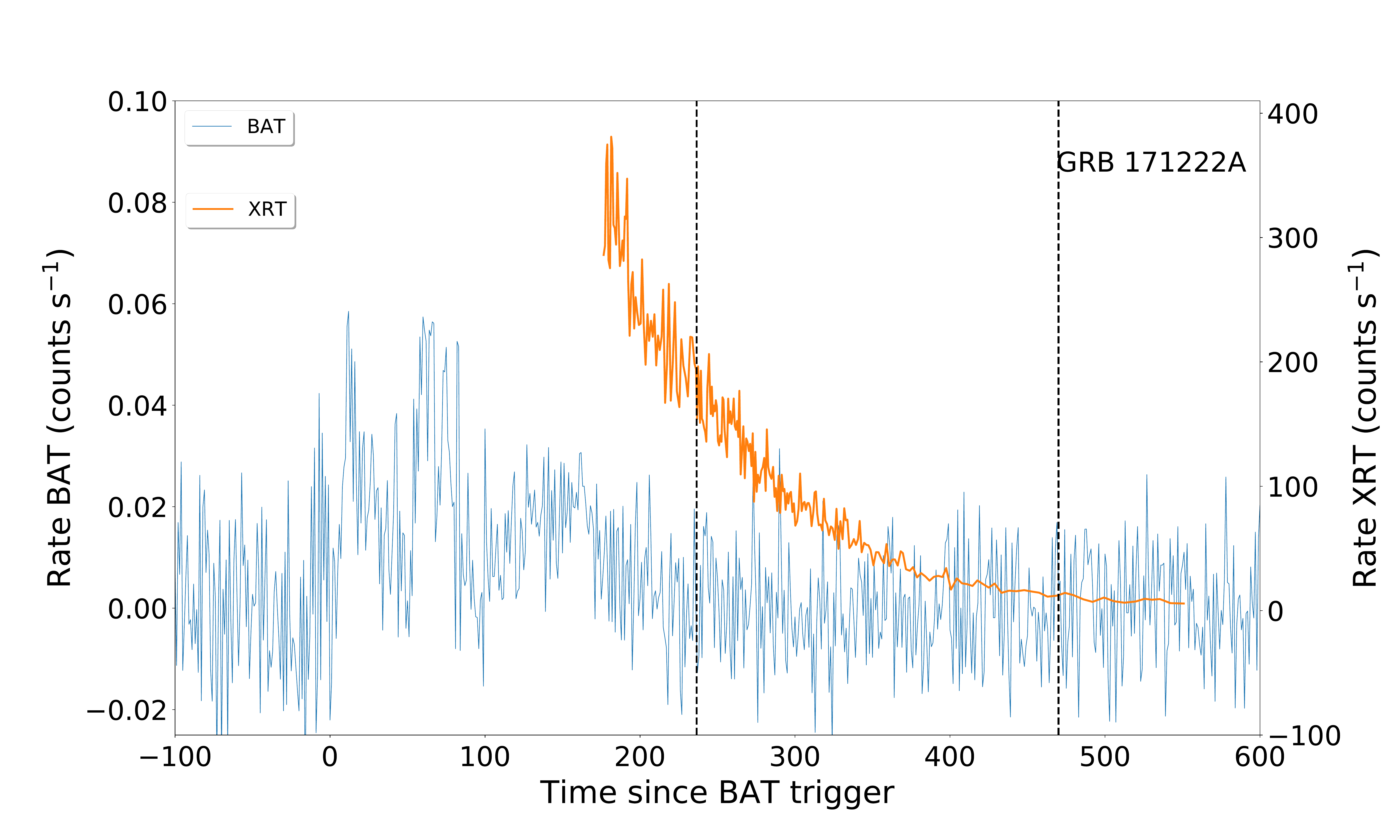}
    \end{subfigure}
        \begin{subfigure}[b]{0.49\textwidth}
     \includegraphics[width=\columnwidth, height = 4.5cm]{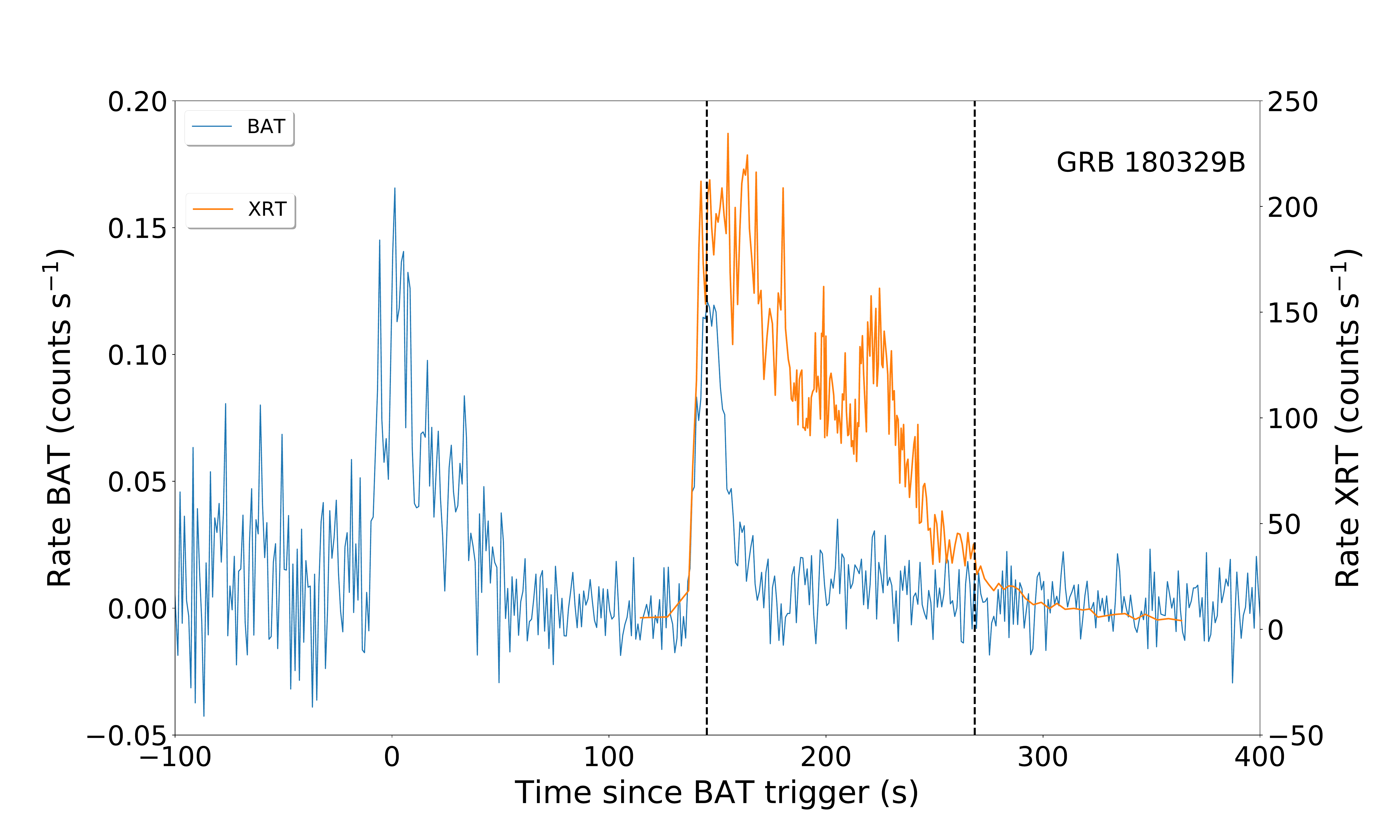}'
    \end{subfigure}
        \begin{subfigure}[b]{0.49\textwidth}
     \includegraphics[width=\columnwidth, height = 4.5cm]{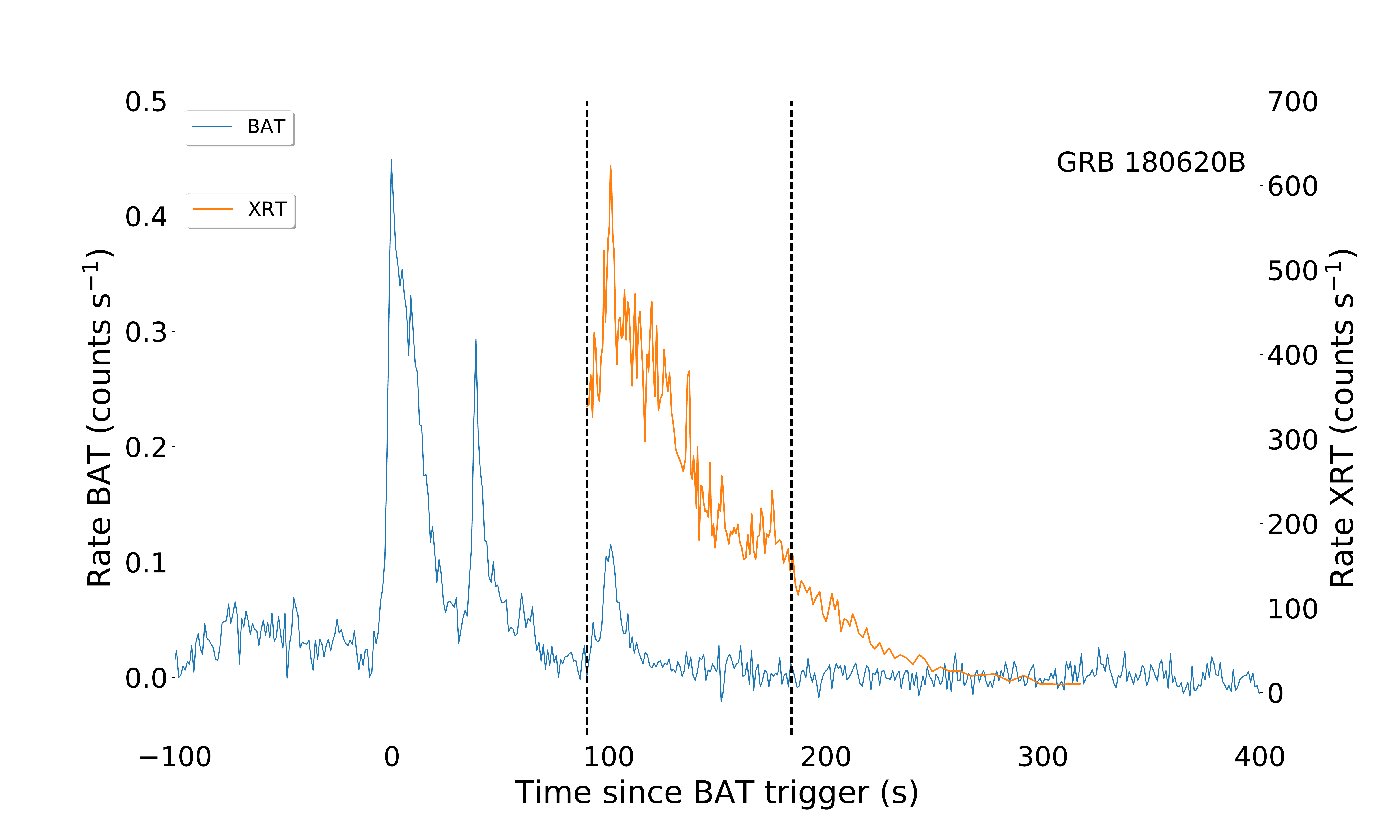}
    \end{subfigure}
    \caption{BAT and XRT light curves for GRBs with significant blackbody components. The black dashed lines mark the time interval where the blackbody is significant for each GRB.}
	\label{lc}
\end{figure*}

The BAT+XRT light curves of the 10 GRBs are plotted in Fig. \ref{lc}, while Fig.~\ref{results} shows the time evolution of the blackbody temperatures ($T_{\rm BB}$), luminosities ($L_{\rm BB}$) and radii ($R_{\rm BB}$\footnote{Obtained from the classical Stefan-Boltzmann equation}), as well as the relation between $L_{\rm BB}$ and $T_{\rm BB}$. Fig.~\ref{results} also includes the nine GRBs with blackbody components previously reported in V18. Plots of the time evolution of the best-fitting parameters for each individual GRB are provided in  Appendix \ref{grbscomplete}. This appendix also includes plots of spectral fits for each GRB in the time interval where the blackbody is most significant. Note that the spectra are plotted in the observer frame, while the reported $T_{\rm BB}$ are in the rest frame of the GRBs.

\section{Discussion}

In this paper we have identified 10 new GRBs with significant blackbody components in the early X-ray spectra. This brings the total number of GRBs with  blackbody components to 19 out of 199 GRBs analysed, considering the previous results in V18. We stress that these results rely on the assumption that the underlying spectrum is an absorbed power law.  Furthermore, we have not made any a-priori selection on light curves in order to select which parts of the emission to analyse. Below we discuss the properties of the sample in Section~\ref{sec:disc-sample}, derive fireball parameters in Section~\ref{sec:disc-fireball} and finally discuss the origin of the emission in Section~\ref{sec:disc-origin}. 

\subsection{Sample properties}
\label{sec:disc-sample}
  
 Fig.~\ref{results} shows that $T_{\rm BB}$ of the sample span 0.1--3.8~keV and that $L_{\rm BB}$ cover a wide range of $\sim 10^{45} - 10^{51} \ \rm{erg \ s^{-1}}$. A trend of decreasing $T_{\rm BB}$ and $L_{\rm BB}$ with time is seen in the majority of GRBs, although there are some exceptions, such as GRB~080325 and GRB~171205A, where the temperature is consistent with being constant. As already noted for the smaller sample in V18, there are distinct groups in the luminosity distribution. Apart from the LL GRB~060218 and GRB~171205A, there is one group at  $L_{\rm BB} \sim 10^{47} - 10^{48} \ \rm{erg \ s^{-1}}$ and one with $L_{\rm BB}  >10^{49} \ \rm{erg \ s^{-1}}$, with the latter group also having higher temperatures. Most of the new detections added in this paper fall in the more luminous group, with the notable exceptions of GRB~171205A and GRB~050724. The latter has a luminosity of $L_{\rm BB} \sim 10^{47} \ \rm{erg \ s^{-1}}$ and is the only short GRB in the sample.

Because of the relation between $L_{\rm BB}$ and $T_{\rm BB}$ (lower, right panel in Fig.~\ref{results}), the blackbody radii span a relatively narrow range, as previously noted in V18. This may point to similar progenitors and environments for these GRBs. Fitting the relation with a power-law function gives $L_{\rm{BB}} \propto T^{3.06 \pm 0.21}_{\rm{BB}}$.
We stress that the radii in Fig.~\ref{results} were obtained from the classical Stefan-Boltzmann equation. The Lorentz factors and impact of relativistic effects are discussed in Section \ref{sec:disc-fireball}. At the same time as the blackbodies cool with time, the photon indices of the power-law components are often seen to soften, reaching typical values of $\sim 2$  (see Appendix~\ref{grbscomplete}, V18 and Table~\ref{fits}).  This is compatible with the distribution of photon indices for the early X-ray emission in large samples of GRBs (\citealt{Evans,Racusin2009})

\begin{figure*}
    \begin{subfigure}[b]{0.49\textwidth}
    \includegraphics[width=\columnwidth,  height = 6.66cm]{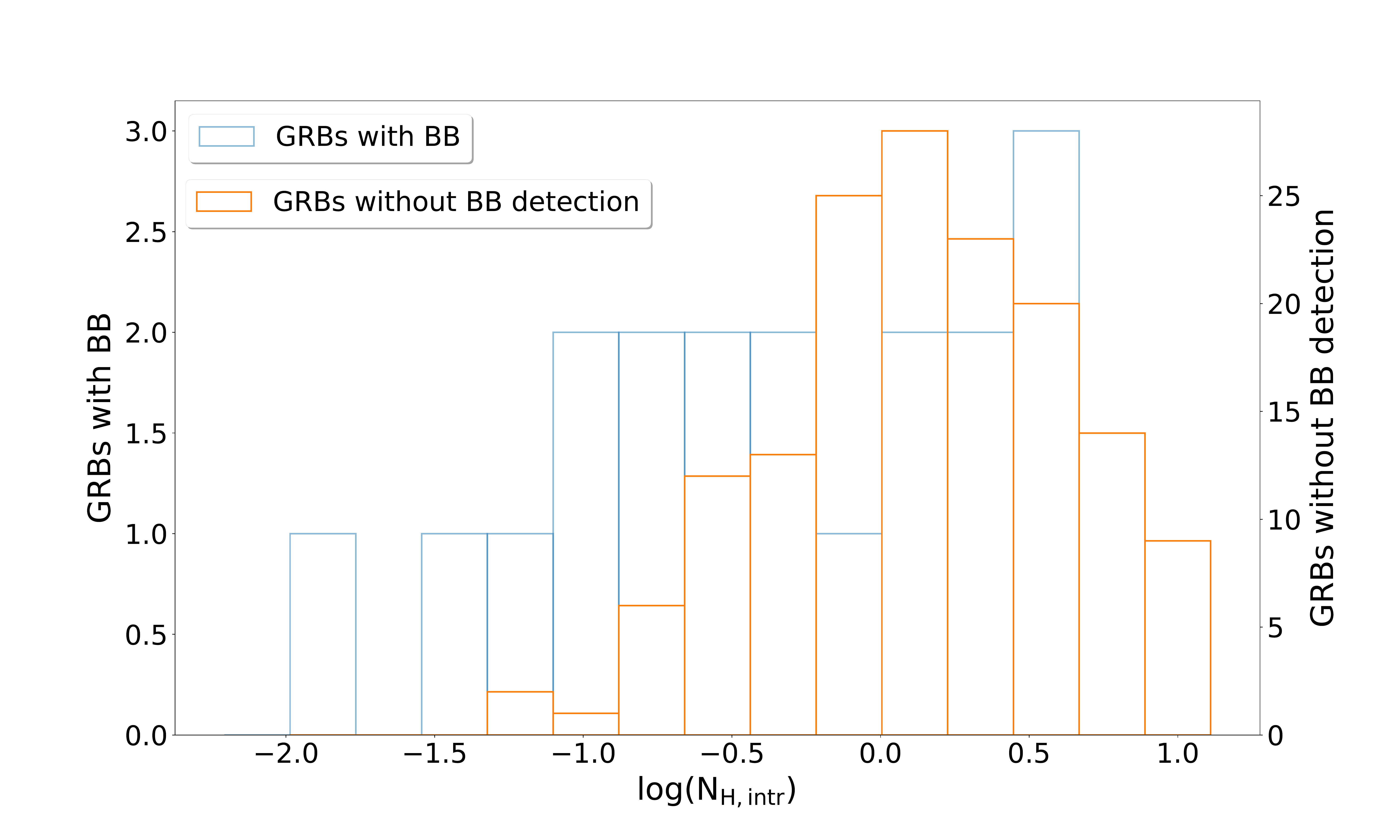}
    \end{subfigure}
    \begin{subfigure}[b]{0.49\textwidth}
     \includegraphics[width=\columnwidth, height = 6.66cm]{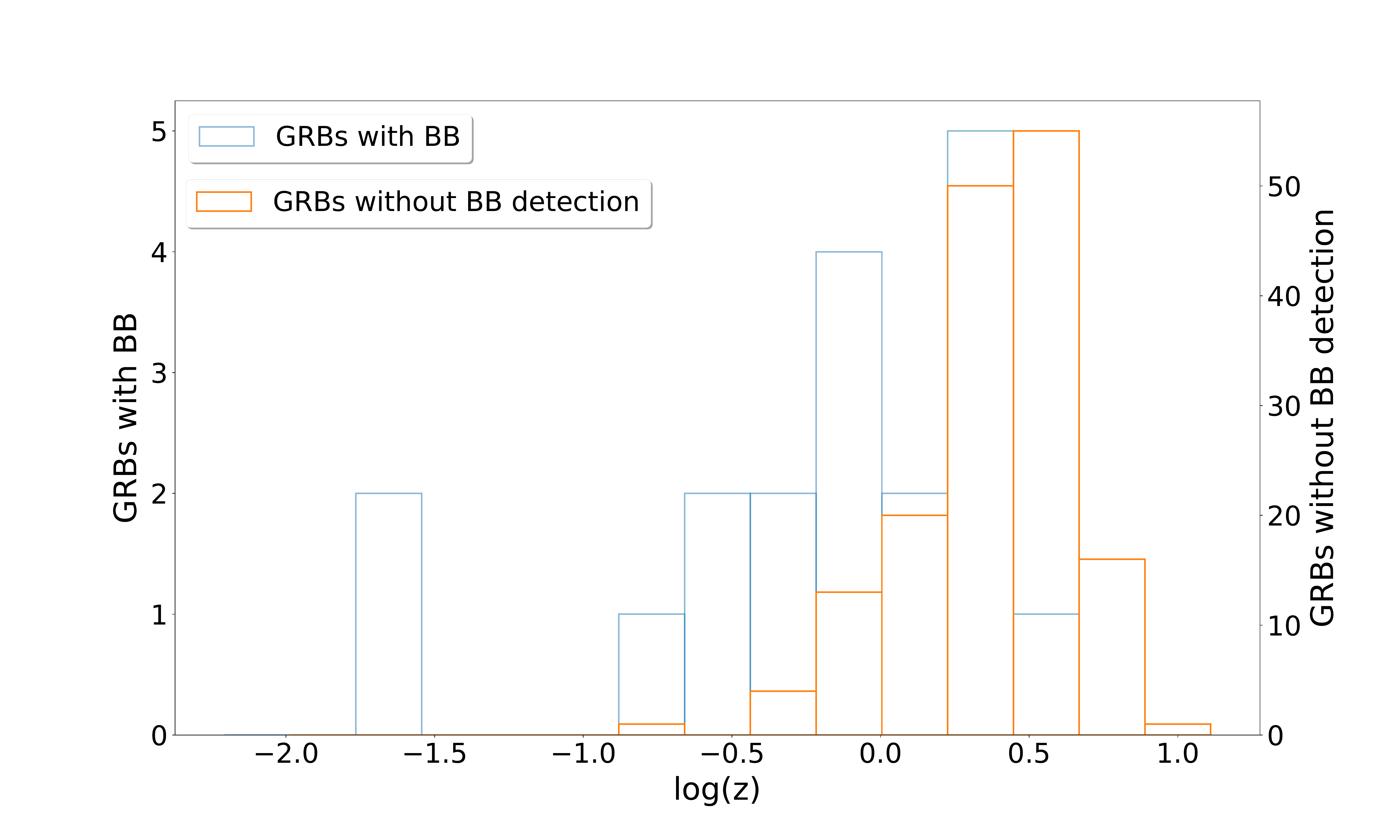}
    \end{subfigure}
    \caption{Histograms of $N_{\rm H,intr}$ (left) and redshift (right) for GRBs with and without significant blackbody components.}
	\label{nh}
\end{figure*}

The larger fraction of GRBs with high luminosities and temperatures identified in this work is also reflected in the distributions of $N_{\rm H,intr}$ and $z$, which now extend to higher values than in V18 (Fig.~\ref{nh} and Table~\ref{thermalwithz}).  In particular, there are now a number of GRBs with $N_{\rm H,intr} > 1 \times 10^{22} \ \rm{cm^2}$ and  $z >1$, with the highest $z$ being close to 4.  
%%%
The observed flux fraction of the blackbody in the 19 GRBs is mainly below $50\%$ (typically $\sim 30 \%$). However, there are some cases where the blackbody dominates the spectra in certain time intervals, including GRB~050724, GRB~081221 and GRB~151027A. In the latter case the blackbody constitutes nearly $100\%$ of the flux in some bins. We also find that the observed and unabsorbed flux fractions differ significantly in some cases. The most notable example is GRB~171222A, where the blackbody contributes on average $\sim 80\%$ of the unabsorbed flux, but only $\sim 20\%$ of the observed flux. 

The light curves of the sample exhibit a wide range of morphologies (see Fig. \ref{lc}). It is clear that some XRT light curves resemble typical afterglows (e.g., GRB~081221), while others catch late, flaring prompt emission (e.g., GRB~090715B). For completeness, we have plotted the XRT + BAT light curves of the GRBs presented in V18 in Fig. \ref{lcv18}. The variety in light curve morphology indicates that the origin of the blackbody component is different in different bursts, as discussed further below. 

There are some systematic effects in the analysis that are expected to have a small impact on the results, including some degeneracies between the parameters $kT$, $N_{\rm H,intr}$ and $\Gamma$. All uncertainties are discussed in section 7.2 of V18, and we refer the reader to that paper for details. The most important caveat is that the blackbody+power-law model is unlikely to provide unique descriptions of the spectra of all the GRBs in the sample. In particular, we have not tested whether other models, or combinations of models, can provide similar or better fits to the data.  The only exception to this is the blackbody + cutoff power-law model used for GRBs with simultaneous BAT data (Section~\ref{sec:analysis}). Below we focus on the interpretation that the blackbody components in the spectral fits represent ``real" thermal emission, but also point to some alternative interpretations.  

\subsection{Fireball parameters}
\label{sec:disc-fireball}

To gain further insight into the properties of the GRBs, we use the method described in   \citet{Peer2007}  to determine the properties of the fireball from the blackbody components. We focus on the coasting Lorentz factor ($\Gamma_{\rm{jet}}$, not to be confused with $\Gamma$ which is used to denote photon index), the photospheric radius ($r_{\rm{ph}}$) and the nozzle radius of the jet  ($r_0$, where the fireball starts expanding adiabatically). The derivations rely on the scaling relations of the standard fireball model and assumes on-axis emission with $\Gamma_{\rm{jet}}$ much larger than the inverse of the jet opening angle. We stress that these assumptions may not be valid for all the GRBs in the sample, considering that two LL~GRBs are included and that the observations probe emission 100s of seconds after the initial triggers. In addition, the calculations do not account for the effect of radiative diffusion, which is expected to be relevant for cocoons \citep{Vereshchagin2020}. The results obtained below should therefore only be seen as indicative.  

In addition to the properties of the blackbodies, the fireball parameters depend on the the high-energy flux of any non-thermal component in the prompt emission.  This introduces an uncertainty since the the origin of the power-law component in our fits is not always clear. For simplicity, we assume that the power-law is due to prompt emission only when rapid variability is observed  and/or when there is a simultaneous gamma-ray detection with BAT (see further Section~\ref{sec:disc-origin-jet}).  Another source of uncertainty is the parameter $Y$, which is the inverse of the radiative efficiency, defined as $Y =  (E_{\rm{\gamma,iso}} + E_{\rm{K}})/E_{\rm{\gamma,iso}}$ where $E_{\rm{\gamma,iso}}$ is the isotropic energy of the prompt phase and $E_{\rm{K}}$ is the kinetic energy of the jet. These parameters can be estimated from the BAT and XRT data, given a number of assumptions. 

We determine $E_{\rm{\gamma,iso}}$ in the 10--10 000~keV rest frame energy interval based on the flux obtained from fitting the time-averaged BAT spectra with a cutoff power law. The parameters of the cutoff power-law model must  be constrained for this estimate to be reliable, which means that we disregard GRBs with low-quality BAT spectra and/or high cutoff energies. 

 $E_{\rm{K}}$ can be estimated from the X-ray afterglow, following the method of \cite{Zhang2007}. We estimate $E_{\rm{K}}$ only for GRBs where the normal decay phase can be clearly defined based on the light curve fits in the {\it Swift}~XRT GRB Catalogue \citep{Evans}. We further require that there are at least 200 counts during this phase. For these GRBs, we fit the full PC spectrum during the normal phase, with  $N_{\rm H,  Gal}$ and $N_{\rm H,intr}$ fixed at the values from Table \ref{thermalwithz}, to obtain the photon index and unabsorbed flux needed to estimate $E_{\rm{K}}$.  We assume typical values for the fraction of energy in electrons and magnetic fields, $\epsilon_e = 0.1$ and $\epsilon_b = 0.01$, respectively, as well as a number density of $n = 1 \ \rm{cm^{-3}}$. We also neglect inverse Compton emission. In order to asses if the X-ray spectrum is above or below the synchrotron cooling frequency, we check the so-called closure relations for the case of slow cooling synchrotron emission, constant-density interstellar medium, and no energy injection  (see e.g.,  \citealt{Zhang2006}).

\begin{figure*}
    \begin{subfigure}[b]{0.49\textwidth}
    \includegraphics[width=\columnwidth,  height = 6.66cm]{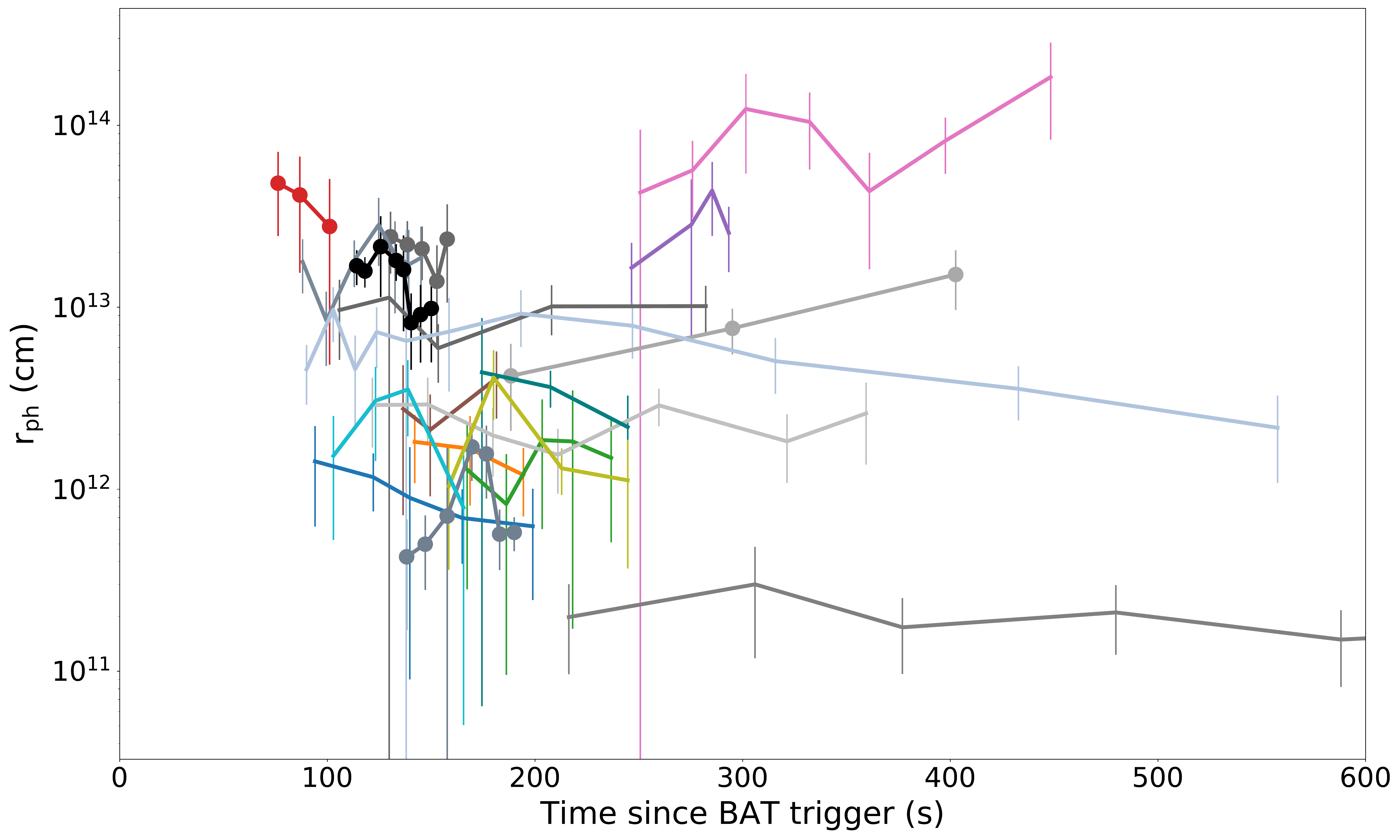}
    \end{subfigure}
    \begin{subfigure}[b]{0.49\textwidth}
     \includegraphics[width=\columnwidth, height = 6.66cm]{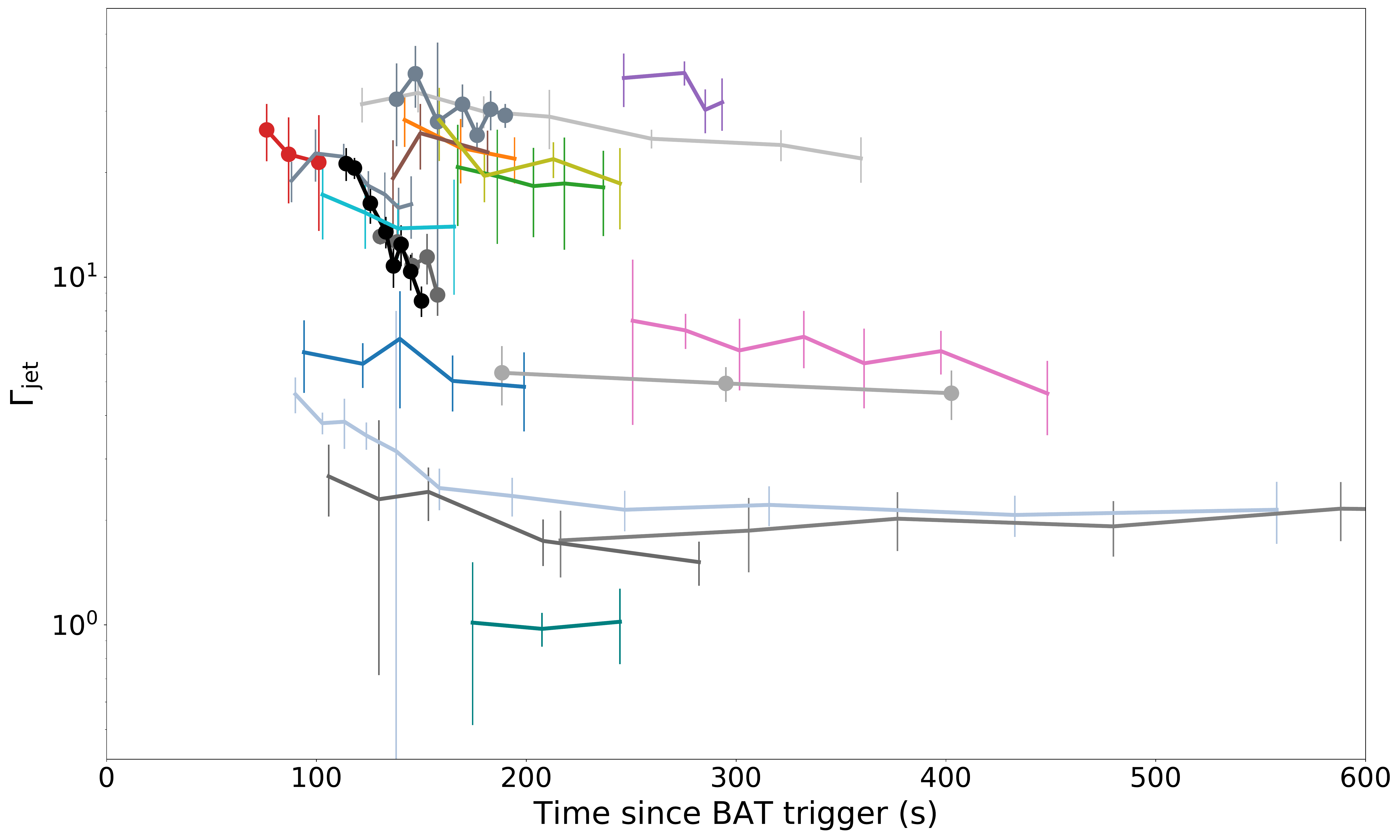}
    \end{subfigure}
     \begin{subfigure}[b]{0.49\textwidth}
     \includegraphics[width=\columnwidth, height = 6.5cm]{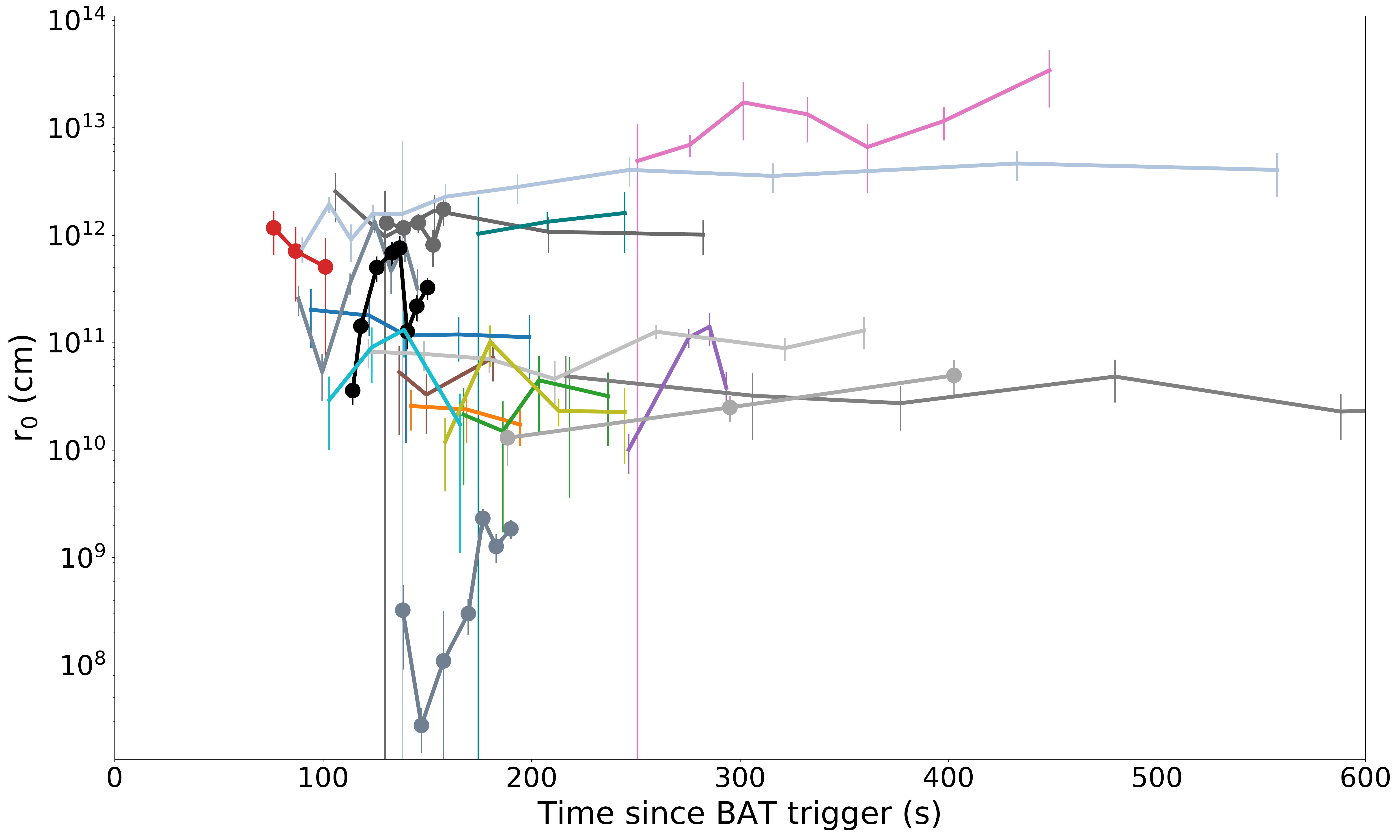}
    \end{subfigure}
    \begin{subfigure}[b]{0.49\textwidth}
    \includegraphics[width=\columnwidth,  height = 7cm]{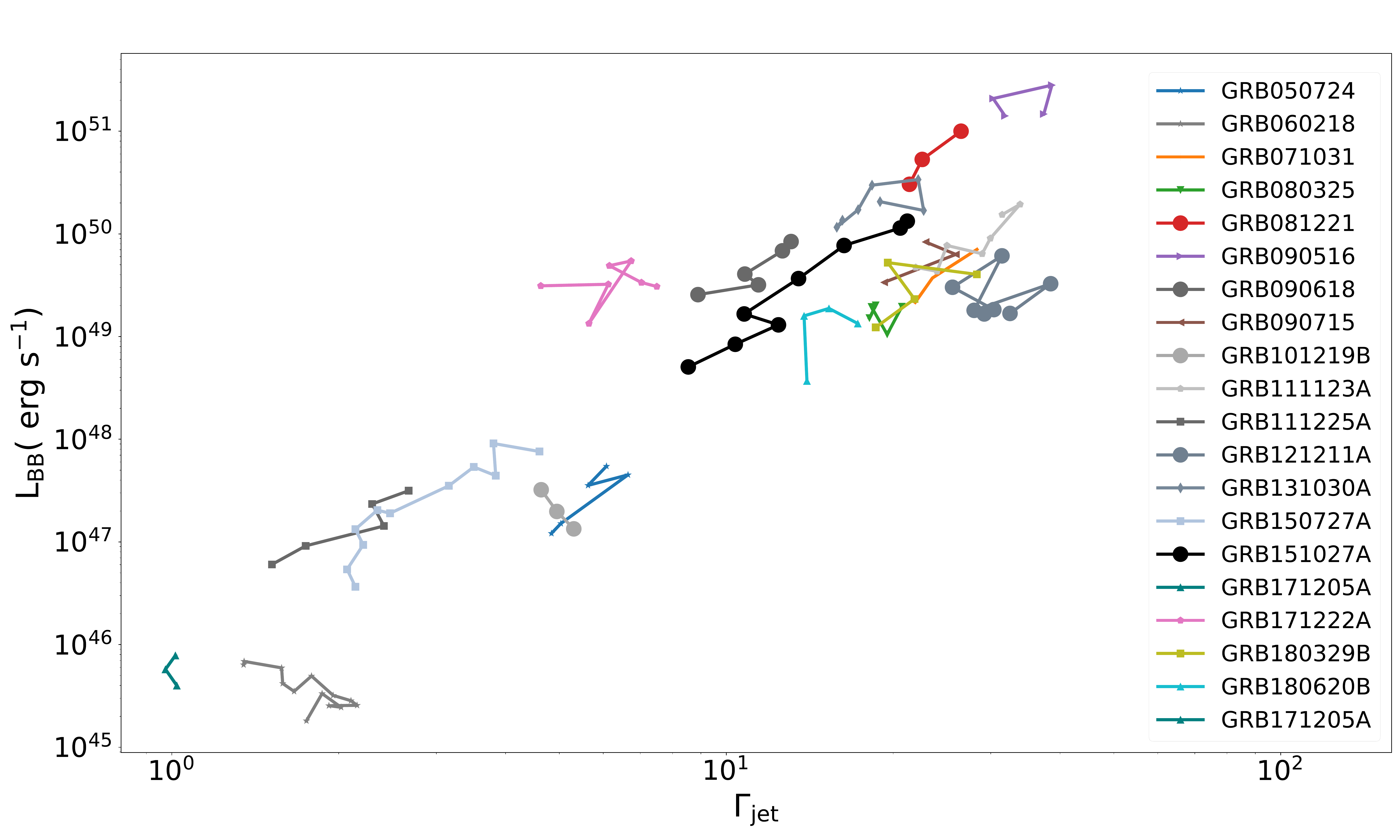}
    \end{subfigure}
    \caption{Inferred fireball parameters for the GRBs with blackbody components. The upper panels show the time evolution of $r_{\rm{ph}}$ and $\Gamma_{\rm{jet}}$. The evolution of $r_{\rm{0}}$ and the relation between $L_{\rm BB}$--$\Gamma_{\rm{jet}}$ are plotted in the lower panels. In the $L_{\rm BB}$--$\Gamma_{\rm{jet}}$ panel we have omitted error bars for clarity. GRBs plotted with large points have estimates of $Y$ from
    Table~\ref{tab:fireball}. In the other cases $Y=1$ has been assumed. Note that $\Gamma_{\rm{jet}}$ and $r_{\rm{ph}}$ scale as  $Y^{1/4}$, while $r_0 \propto Y^{-3/2}$.}
	\label{fireball}
\end{figure*}

Only five GRBs have both BAT data of good enough quality for constraining $E_{\rm{\gamma,iso}}$ and a clearly defined normal decay phase in the X-ray light curve with more than 200 counts: GRB~081221, GRB~090618, GRB~101219B, GRB~121211A and GRB~151027A. Values of $E_{\rm{\gamma,iso}}$,  $E_{\rm{K}}$ and $Y$ for these GRBs are presented in Table \ref{tab:fireball}. For the remaining GRBs we simply set $Y=1$ when calculating the fireball parameters. The parameter that depends most strongly on $Y$ is the nozzle radius ($r_0 \propto Y^{-3/2}$), whereas both $\Gamma_{\rm{jet}}$ and $r_{\rm{ph}}$ scale as  $Y^{1/4}$.  We plot the time evolution of $r_{\rm{ph}}$, $\Gamma_{\rm{jet}}$ and $r_{\rm{0}}$ in Fig. \ref{fireball}, together with the relation between $L_{\rm BB}$ and $\Gamma_{\rm{jet}}$.

The figure shows that $r_{\rm{ph}}$ is clustered around $10^{12} - 10^{13} \ \rm{cm}$, which is well outside  the typical size of the presumed WR star progenitors ($\sim 10^{11} \ \rm{cm}$, \citealt{Sander2012}). The $r_{\rm{ph}}$  are slightly larger than the classical blackbody radii (Fig.~\ref{results}) as expected, but the small spread of radii from Fig.~\ref{results} is preserved. It should also  be noted that there is typically little evolution in $r_{\rm{ph}}$ with time, considering the parameter uncertainties. The clearest evidence of expansion is seen for GRB~101219B and  GRB~171222A. In the case of $r_0$, the values are mainly within $10^{10} - 10^{12} \ \rm{cm}$, although we note the stronger dependence on $Y$ for this parameter, which means that the GRBs that lack efficiency estimates may in reality have $r_0$ about an order of magnitude lower. However, this is still clearly larger than the $\sim 10^{7} \ \rm{cm}$ that would be expected for a fireball created near a stellar-mass black hole. This may indicate that the jet does not start expanding freely until it has reached the surface of the star. 

The values of $\Gamma_{\rm{jet}}$ span a wide range from just over 1 to $\sim 40$. These relatively low values are expected given the low temperatures. Much higher values of $\gtrsim 100$ are typically inferred during the prompt gamma-ray phase (e.g., \citealt{Ghirlanda2018}). There are some groups in the distribution of $\Gamma_{\rm{jet}}$, which may be connected to different origins of the blackbody component in different GRBs, as discussed in Section~\ref{sec:disc-origin}. We note that some of the error bars on $\Gamma_{\rm{jet}}$ in Fig.~\ref{fireball} extend below 1. This is clearly not physical, but simply reflects the statistical uncertainties in the measurements. As noted above, the assumptions in the calculations may also be inappropriate for some GRBs, which is likely to particularly affect the low end of the $\Gamma_{\rm{jet}}$ distribution.

%\begin{table}
%    \centering
%    \caption{Values of $E_{\rm{\gamma,iso}}$, $E_{\rm{K}}$ and efficiencies $\eta$ for GRBs in which we could estimate efficiencies. }
%    \label{tab:fireball}
%  \begin{tabular}{c c c c } % four columns, alignment for each
%        \hline
%         GRB & $E_{\rm{\gamma,iso}}$ (ergs)  & $E_{\rm{K}}$ (ergs)& $\eta$ ($\%$)\\ 
%        \hline
%   081221 & $1.11 \pm 0.2 \times 10^{54}$ &  $2.07 \pm0.54 \times 10^{54}$ & $65 \pm 13$ \\
%   090618 & $2.03 \pm 1.1 \times 10^{53}$ & $2.33 \pm 1.17 \times 10^{52}$ & $90 \pm 45$ \\
%   101219B & $3.4 \pm 0.2 \times 10^{51}$ & $6.4 \pm 3.5 \times 10^{52}$ & $5 \pm 2$\\
%   121211A & $1.77 \pm 0.22 \times 10^{51}$ & $1.12 \pm 0.13 \times 10^{52}$ & $14 \pm 2$ \\
%   151027A & $1.18 \pm 0.11 \times 10^{53}$ & $12.69 \pm 0.13 \times 10^{52}$ & $48 \pm 5$ \\
%        \hline
%    \end{tabular}
%\end{table}

%%Version in terms of Y
\begin{table}
    \centering
    \caption{Values of $E_{\rm{\gamma,iso}}$, $E_{\rm{K}}$ and $Y$ for GRBs where all quantities could be estimated. }
    \label{tab:fireball}
  \begin{tabular}{c c c c } % four columns, alignment for each
        \hline
         GRB & $E_{\rm{\gamma,iso}}$   & $E_{\rm{K}}$ & $Y$\\ 
         & (erg) & (erg) & \vspace{1mm} \\
        \hline
  081221 & $1.11 \pm 0.22 \times 10^{54}$ &  $2.07 \pm0.54 \times 10^{54}$ & $1.5 \pm 0.3$ \\
  090618 & $2.03 \pm 1.10 \times 10^{53}$ & $2.33 \pm 1.17 \times 10^{52}$ & $1.1 \pm 0.6$ \\
  101219B & $3.4 \pm 0.2 \times 10^{51}$ & $6.4 \pm 3.5 \times 10^{52}$ & $20 \pm 8$\\
  121211A & $1.77 \pm 0.22 \times 10^{51}$ & $1.12 \pm 0.13 \times 10^{52}$ & $7 \pm 1$ \\
  151027A & $1.18 \pm 0.11 \times 10^{53}$ & $12.69 \pm 0.13 \times 10^{52}$ & $2.1 \pm 0.2$ \\
        \hline
    \end{tabular}
\end{table}

The lower, right panel of Fig.~\ref{fireball} reveals a correlation between $L_{\rm BB}$ and $\Gamma_{\rm{jet}}$. A correlation between these parameters (or more generally, $L$ rather than  $L_{\rm BB}$) has previously been noted in several papers, which use different methods to estimate the Lorentz factors \citep{Lu2012,Peng2014,Ghirlanda2018,Ahlgren2019A}. Fitting the relation in Fig.~\ref{fireball}  gives $\Gamma_{\rm{jet}} \propto L_{\rm BB}^{0.33 \pm 0.21}$, which is in line with the expectations from the photospheric model when $r_0$ is approximately constant \citep{Peng2014}.  We also note that this relation drives the $L_{\rm BB}-T_{\rm BB}$ correlation in Fig.~\ref{results} and that there is no correlation between $L_{\rm BB}$ and the comoving temperature ($T'_{\rm BB} \propto T_{\rm BB}/\Gamma_{\rm{jet}}$). 

These results may support a photospheric origin of the emission (see also \citealt{Fan2012}), though we note that the observed correlations are likely affected by selection effects. The  $L_{\rm BB}-T_{\rm BB}$ and $L_{\rm BB} - \Gamma_{\rm{jet}}$ relations are similar to the well-known spectral energy correlations for the prompt emission \citep{Amati2002,Yonetoku2004,Ghirlanda2010,Frontera2012}, which are also observed in X-ray flares  \citep{Margutti2010,Peng2014}. While it is known that these correlations are affected by selection effects, conclusions differ on whether the selection effects are the main driver for the correlations or simply serve to make real physical relations somewhat tighter \citep{Nava2008,Krimm2009,Butler2010, Shahmoradi2011,Collazzi2012, Kocevski2012}. The most important selection effect is that the detection of faint bursts with high peak energies is limited by detector sensitivity. The lack of luminous GRBs with low peak energies can be attributed to the fact that rare, bright GRBs are preferentially detected at high redshifts. Possible low peak energies for these GRBs would be redshifted outside the observed energy range. Although investigations of these selection effects have focused on the gamma-ray emission,  we expect expect similar biases to operate in the X-ray band.

\subsection{Possible origins of the blackbody component}
\label{sec:disc-origin}

\begin{figure*}
    \begin{subfigure}[b]{0.49\textwidth}
    \includegraphics[width=\columnwidth,  height = 6.66cm]{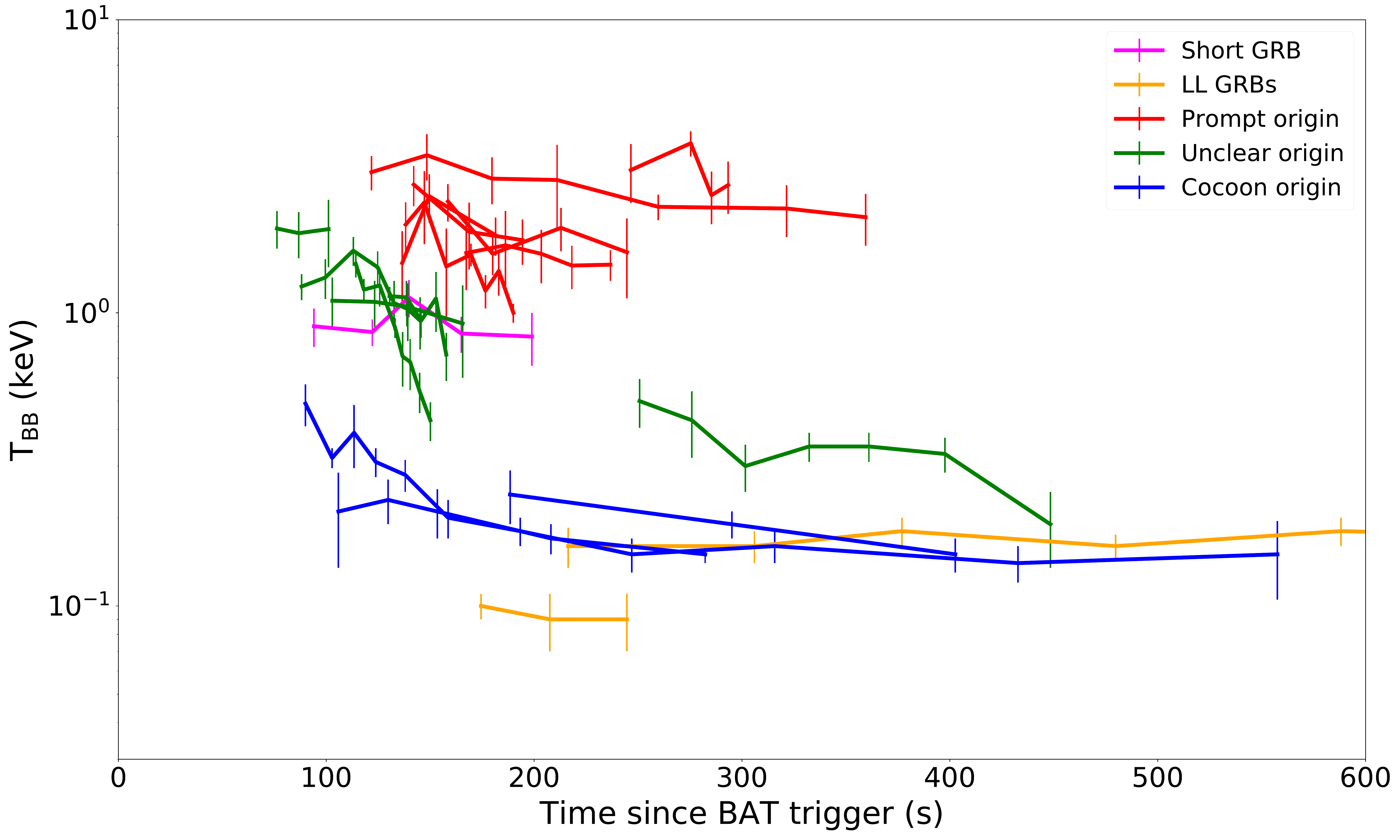}
    \end{subfigure}
    \begin{subfigure}[b]{0.49\textwidth}
     \includegraphics[width=\columnwidth, height = 6.66cm]{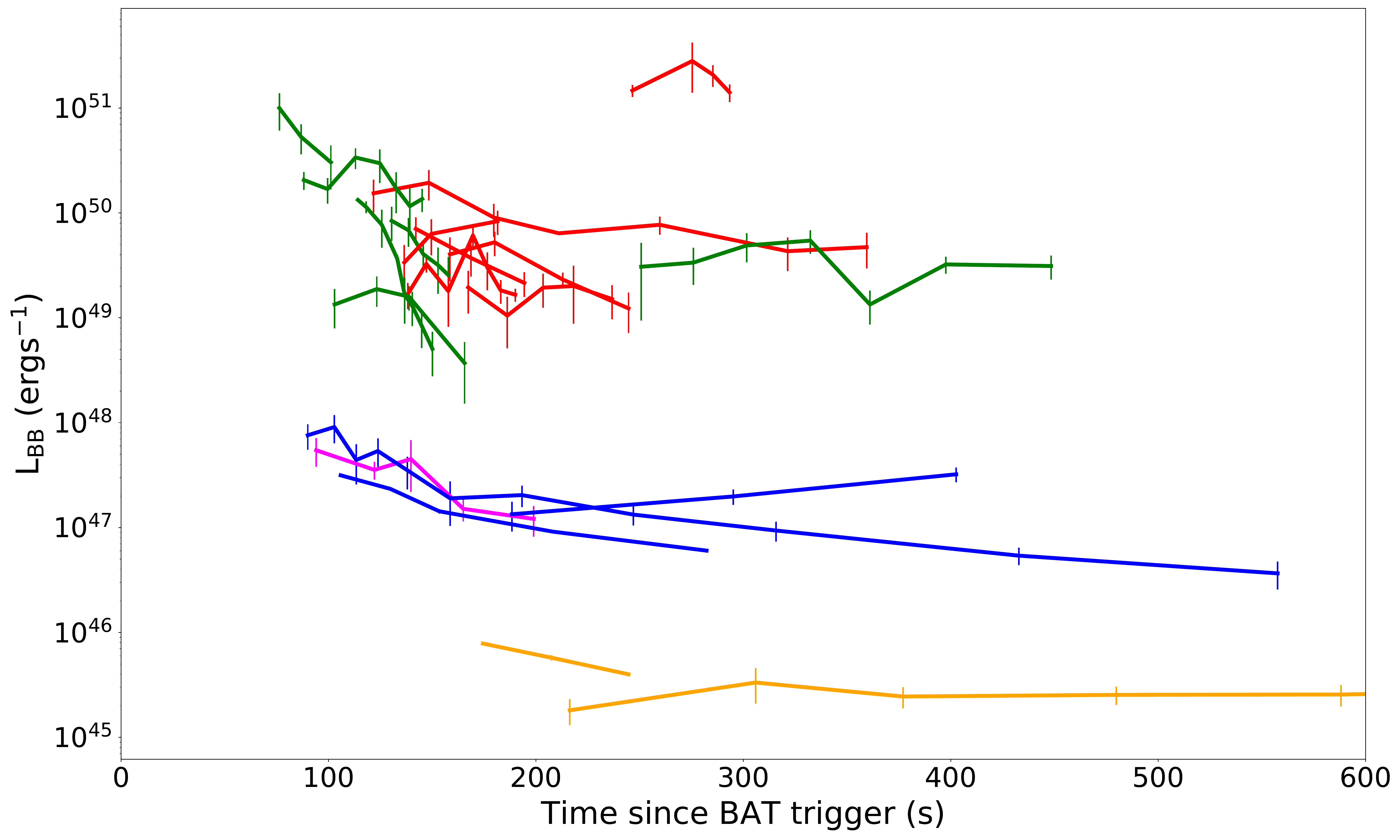}
    \end{subfigure}
     \begin{subfigure}[b]{0.49\textwidth}
     \includegraphics[width=\columnwidth, height = 6.66cm]{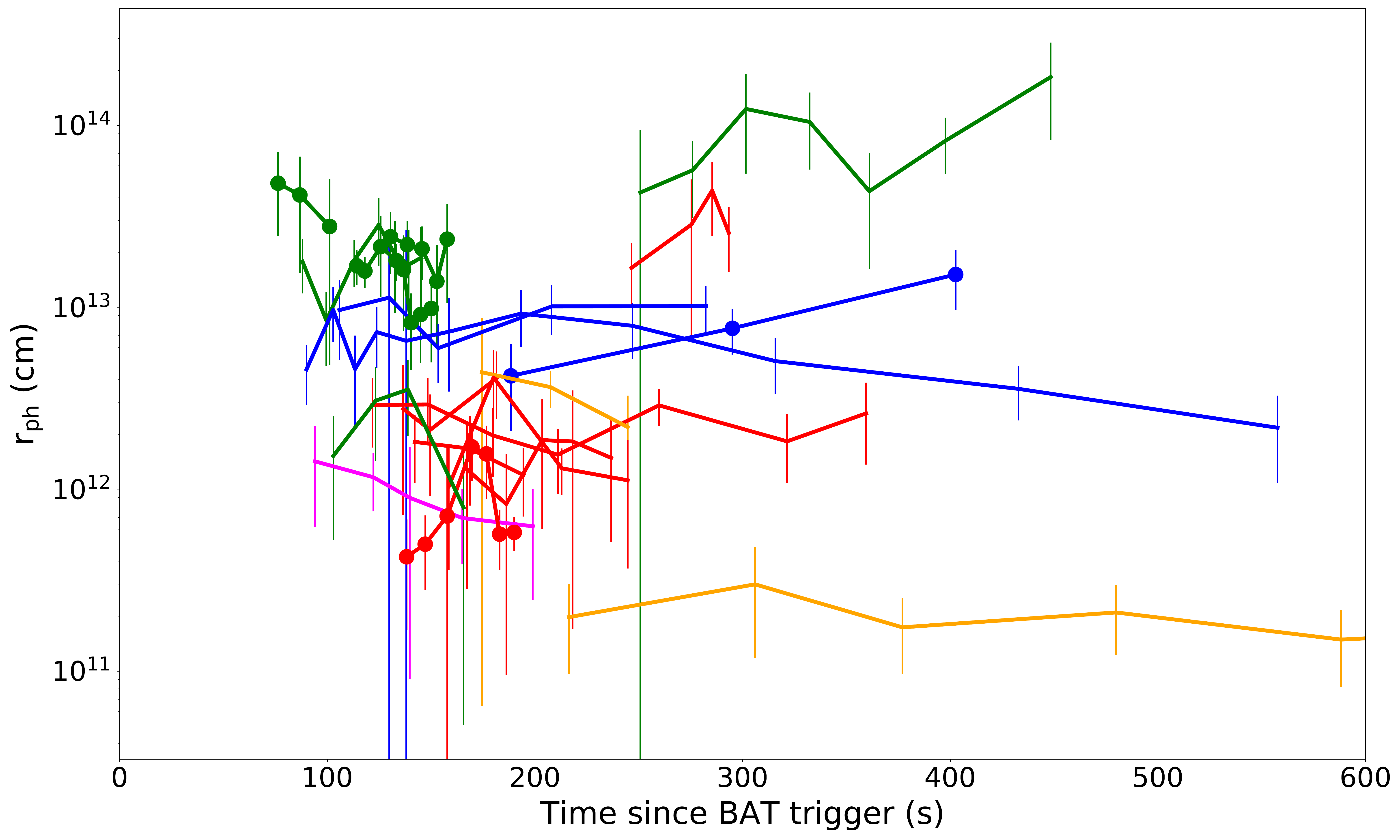}
    \end{subfigure}
    \begin{subfigure}[b]{0.49\textwidth}
    \includegraphics[width=\columnwidth,  height = 6.66cm]{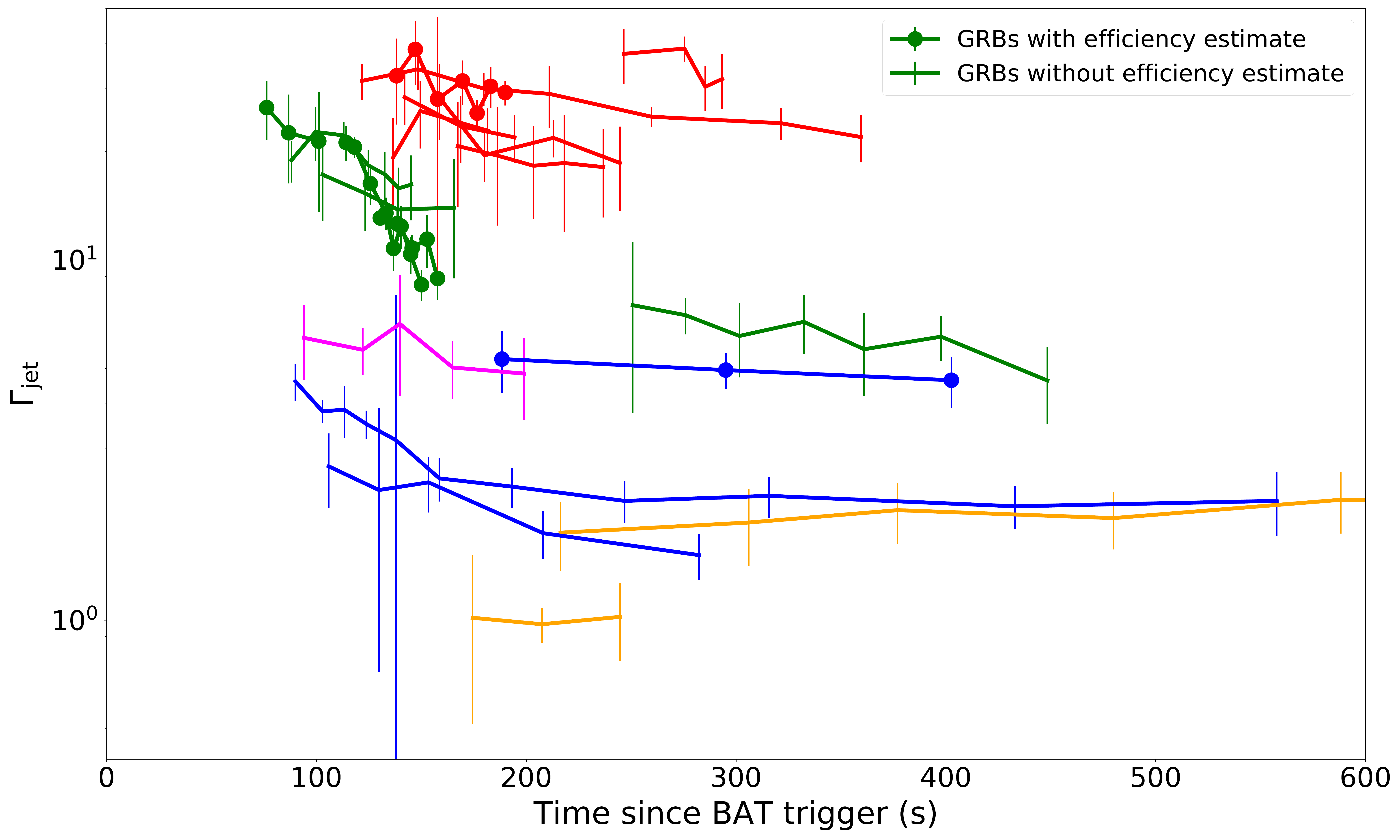}
    \end{subfigure}
    \caption{Same as the upper panels of Fig.~\ref{results} (upper row) and Fig.~\ref{fireball} (bottom row), but with colour coding that represent different populations and presumed origins of the blackbody components.}
	\label{results2}
\end{figure*}

The main explanations that have been proposed for the origin of the blackbody emission in the X-ray spectra are late-time prompt emission, emission from a cocoon surrounding the jet, and shock breakout from the SN. As discussed in V18, the shock breakout scenario is ruled out by the high luminosities in all cases (with the possible exceptions of GRB~060218 and GRB~171205A) and this will therefore not be discussed further. Below we discuss the other two scenarios, as well as GRBs where the interpretation is unclear. We find that different mechanisms are likely responsible for different parts of the sample. To aid the discussion, we have plotted the top panels of Figs~\ref{results} and \ref{fireball} in Fig.~\ref{results2}, with different colours representing different preferred explanations and/or populations.

\subsubsection{Prompt emission}
\label{sec:disc-origin-jet}

The light curves of the GRBs give important information about the origin of the emission.  As opposed to emission from a cocoon, prompt emission originating from a jet is characterised  by variability on short times scales of seconds or less (e.g., \cite{Golkhou2014}). 
The light curves during the prompt phase commonly display an erratic evolution with multiple peaks. The early observations by {\it Swift}~XRT sometimes catch the tail of this emission, as seen from a similar evolution of the X- and gamma-ray light curves (e.g., \cite{Oganesyan_2018,Ahlgren2019}). This is clearly the situation for three GRBs in our sample: GRB~090715B, GRB~111123A and GRB~121211A (see light curves in Figs.~ \ref{lc} and \ref{lcv18}).

In addition to these cases, there are some GRBs in the sample that display strong flares superposed on a smoothly decaying light curve. Such flares, which are observed in about one third of all GRBs, exhibit properties similar to the prompt emission and are thought to be due to late central-engine activity \citep{Chincarini2010, Margutti2010}. 
In GRB~071031, GRB~080325, GRB~090516, and GRB~180329B (Fig.~\ref{lc} and Appendix~\ref{grbscomplete}), it is clear that the blackbody components are associated with these flares rather than the smoothly decaying component. This gives a total of seven GRBs where the blackbodies are unambiguously connected to prompt emission. 
These GRBs (indicated in red in Fig.~\ref{results2}) all have high temperatures ($\gtrsim 1$~keV) and luminosities ($\sim 10^{49}-10^{51} \rm{erg\ s^{-1}}$), and also tend to show a strong evolution of the photon indices (Appendix~\ref{grbscomplete} and V18). In addition, these GRBs have the highest Lorentz factors in the sample ($\Gamma_{jet} \ge 20$). The fact that these values are lower than typically inferred from gamma-ray observations is in line with studies that show the Lorentz factor decreasing with time during the prompt phase, possibly indicating increased pollution by baryons \citep{Iyyani2013, Iyyani2015}.

The most straight forward interpretation of the blackbody component detected in these GRBs is that it is due to the photosphere of the jet, while the power law can be interpreted as a non-thermal component originating in the optically thin region. Similar models, comprising a  blackbody and a non-thermal component, have previously been found to provide good fits to both X-ray flares \citep{Peng2014} and prompt gamma-ray spectra of some GRBs \citep{Ryde2004, Axelsson2012, Burgess2014}. In a related scenario, subphotospheric dissipation can produce spectra that resemble a blackbody + power law over a limited energy range \citep{Ahlgren2015, Ahlgren2019A}. In this case the power-law component is due to Comptonisation of the thermal photons by a power-law distribution of electrons that was accelerated below the photosphere by, e.g., internal shocks. Finally, it is possible that the blackbody + power-law model is capturing the spectral shape of another emission process, such as synchrotron emission. In fact, the X- and gamma-ray spectra of GRB~090715B, GRB~111123A and GRB~121211A  have been interpreted as synchrotron emission by \cite{Oganesyan_2017,Oganesyan_2018}, based on fits to a Band function (or broken power law) with an additional exponential cutoff.

\subsubsection{Cocoon emission}
\label{sec:disc-origin-cocoon}

When a relativistic jet propagates through a star, it shocks the surrounding medium,  producing a hot cocoon  \citep{meszaros2001,Ramirez2002}. The cocoon is mildly relativistic and starts expanding spherically when it emerges from the star. Predictions for light curves and spectra of cocoons have been obtained by post-processing hydrodynamical simulations \citep{Suzuki2013, DeColle2017}. While the details of the signal depend on the properties of the jets, progenitor and circumstellar medium, the basic picture is a light curve that rapidly rises to a a maximum luminosity of $\sim 10^{47} \ \rm{erg\ s^{-1}}$  and then decays slowly on a time-scale of several $~ 100$~s. The spectra are dominated by thermal emission with temperatures in the soft X-ray range. 

Here we focus on the GRBs that are best explained by this model, while more ambiguous cases are noted in Section~\ref{sec:disc-origin-pec}. There are three long GRBs in the sample that are clearly consistent with the fading emission from a cocoon: GRB~101219B, GRB~111225A and GRB~150727A. These are indicated in blue in Fig.~\ref{results2}, which shows that they evolve very similarly, with low temperatures ($\sim 0.1-0.2$~keV) and luminosities ($\sim 10^{47} - 10^{48} \ \rm{erg s^{-1}}$). The Lorentz factors of $\Gamma_{\rm jet} \lesssim 6$ are also consistent with cocoon models. Interestingly, these GRBs form a distinct group in the luminosity distribution together with GRB~050724 (magenta in Fig.~\ref{results2}). 

GRB~050724 is one of three short GRBs in our full sample and the only one with a significant blackbody component. The blackbody in this GRB was previously noted by \citet{sparre}, but not considered further due to the high temperature of $kT \sim 0.9$~keV and the fact that it was a short GRB.\footnote{This paper was focused on GRB~101219B-like blackbody components.} 
$\Gamma_{\rm jet}$ for this GRB is similar to the long GRBs with presumed cocoon emission, while the $R_{\rm BB}$ and $r_{\rm phot}$ are lower (Figs.~\ref{results} and \ref{results2}).

\citet{Nagakura2014} showed that a cocoon is expected to be produced also in short GRBs. The emission properties of such cocoons have been explored by \citet{Lazzati2017}. Although the expected signal is a sensitive function of the viewing angle, the results show that cocoons of short GRBs can produce high temperatures and a wide range of luminosities compatible with the observations of GRB~050724. However, the models for cocoons in short GRBs are more uncertain than for long ones, making it difficult to draw firm conclusions. We also note that the model for short GRB cocoons by  \citet{Nakar2017} predict a much less energetic signal. 

The reason why emission components consistent with cocoons are identified in such a small number of GRBs is most likely that they are difficult to detect. \citet{sparre} simulated GRB~101219A-like events and concluded that they can only be detected at relatively low redshifts ($z < 1.5$) and in systems with $N_{\rm H, intr} < 4 \times 10^{21}\ {\rm cm}^{-2}$. In addition, we note that the average (WT mode) X-ray luminosities of GRB 101219B, GRB 111225A and GRB 150727A are in the lower $25 \%$ of the distribution of the full sample analysed (see also V18). It is clear that similar cocoon emission would be impossible to detect in the GRBs discussed in the previous section, which have luminosities that are several orders of magnitudes higher, as well as higher redshifts and $N_{\rm H, intr}$. In the case of short GRBs, we cannot draw any conclusions regarding the detectability due to very small sample analysed.

\subsubsection{Peculiar and unclear cases}
\label{sec:disc-origin-pec}

The origin of the blackbody components in the remaining GRBs is less clear, including the two LL~GRBs marked in yellow in Fig.~\ref{results2}, as well as the six GRBs marked in green in the same figure. We discuss these cases below, highlighting individual GRBs that display peculiar behaviour.  

\paragraph*{GRB~060218 and GRB~171205A.} The X-ray spectra of these LL~GRBs reveal blackbody components with low temperatures, first reported by \cite{campana} and \cite{Delia2018}, respectively. Both events have been extensively followed with multiwavelength observations, which showed the blackbody component cooling into the UV/optical range over time \citep{campana,Izzo2019}. Our X-ray analysis confirms the previous results with a higher time resolution. The origin of the thermal emission in these GRBs is debated. Recently proposed models for GRB~060218 include a low-power jet \citep{Irwin2016} as well as mildly relativistic shock breakout powered by a jet that is choked in an extended stellar envelope \citep{Nakar2015}. In the case of GRB~171205A, \citet{Izzo2019} interpret the blackbody component as emission from a cocoon, while \cite{Suzuki2019} propose that the full multiwavelength emission is due to interaction between relativistic ejecta and the circumstellar medium. 

\paragraph*{GRB~081221, GRB 090618, GRB~131030A and GRB~180620B.} The blackbodies in these GRBs have  luminosities that are comparable to the cases of prompt emission discussed above (Section~\ref{sec:disc-origin-jet}), but somewhat lower temperatures (see Fig.~\ref{results2}). At the same time these GRBs have values of the Lorentz factor in between the populations with presumed prompt and cocoon emission.  In all four GRBs, the blackbodies are primarily detected in the smooth, steep decay phase of the X-ray light curves. GRB~081221 and GRB~180620B display some small flares during this phase, but the blackbody components do not appear to be associated with those flares. In the case of GRB~131030A, the light curve rises smoothly before the steep decay begins, and the blackbody is detected throughout this evolution. The steep decay phase is commonly attributed to high-latitude emission from the jet, which arrives later than the on-axis prompt emission due to a longer travel path \citep{Kumar2000}. While the simplest model for high-latitude emission is inconsistent with observations \citep{Obrien2006,Zhang2007a}, several authors have considered more realistic scenarios that give better agreement \citep{Yamazaki2006,Takami2007,Zhang2009,Genet2009}. Given the association with the steep phase, we consider it likely that the blackbody components in GRB~081221, GRB~090618, GRB~131030A and GRB~180620B have an origin in high-latitude emission.

An alternative interpretation is that the spectra are due to energetic emission from a cocoon. In particular, \cite{Peer2006} suggested that cocoon emission can explain the steep decay phase. A key difference between this model and the cocoon models discussed in Section~\ref{sec:disc-origin-cocoon} is that part of the cocoon kinetic energy is assumed to be dissipated in a similar way as for the jet. Assuming that the dissipation occurs where the cocoon is optically thick, the emission will diffuse out during the adiabatic expansion, leading to a time delay and a quasi-thermal spectrum, which is compatible with our observations. The model also predicts a rising part of the light curve, similar to what we see in GRB~131030A, but which may be drowned out by the end of the prompt emission or occur before the start of the XRT observations in other cases.

\paragraph*{GRB~171222A} shows several unusual properties. While the blackbody luminosity of $\sim 3 \times 10^{49} \ \rm{erg \ s^{-1}}$ and the smoothly decaying, steep light curve is similar to the cases discussed above, the temperature is significantly lower (cooling from $\sim 0.5$ to $\sim 0.2\ \rm{keV}$). It also shows evidence for an expanding $r_{\rm phot}$ as well as the largest $r_{\rm phot}$ in the whole sample (bottom left panel of Fig. \ref{fireball}). The contribution from the blackbody component reaches as high as  $\sim 80\%$ of the unabsorbed X-ray flux. The inferred Lorentz factor of $\sim 6$ is compatible with expectations for a cocoon, though we note that the luminosity is significantly higher than for the cocoon models discussed in Section~\ref{sec:disc-origin-cocoon}.   

\paragraph*{GRB~151027A.} This GRB was first reported to have a blackbody component by \citet{Nappo2016}. The blackbody is associated with a single $\sim 100$~s flare in the XRT light curve, the first part of which has some associated high-energy emission in the BAT (see Fig.~\ref{lcv18}). This is the only GRB in the sample where the blackbody completely dominates the observed flux, reaching 97$\%$ of the total flux near the peak of the flare (V18). The temperature and luminosity of the blackbody is similar to the GRBs where the blackbodies are detected in the steep light curve phase (see Figs.~\ref{results} and \ref{results2}). \citet{Nappo2016} suggested that the the unusual properties of this GRB can be explained with a re-born fireball, where the interaction between fast ejecta with slower material causes dissipation of energy and a re-acceleration. 

\subsection{Detectability}

We have identified blackbody components in 19 out of the 199 analysed GRBs, which corresponds to $10 \%$ of the sample. In order to investigate if there are systematic differences between the  GRBs with and without blackbodies, we have run non-parametric Anderson-Darling tests on the following parameters: $z$, $N_{\rm{H,intr}}$, $L_{\rm{av}}$, $T_{90}$ and photon index $\Gamma$. This shows that the GRBs with blackbodies  are consistent with being drawn from the full sample of analysed GRBs (see also distributions of $z$ and $N_{\rm{H,intr}}$ in Fig.~\ref{nh}). This is expected given the diverse properties of the blackbody components discussed above.

At the same time, the different groups of GRBs identified in Section~\ref{sec:disc-origin} clearly share common properties. The probable cocoon cases all have low $N_{\rm{H,intr}}$, $z$ and X-ray luminosities, while the opposite is true for the GRBs where the blackbodies are identified as prompt emission.  As discussed for the cocoons in Section~\ref{sec:disc-origin-cocoon}, this is likely due to issues of detectability rather than the intrinsic differences between the GRBs.  However, the small number of GRBs in each group prevents us from drawing firm statistical conclusions regarding different subpopulations based on the current sample. 

\section{Conclusions}

In this paper we have presented a search for blackbody components in the X-ray spectra of 116 GRBs observed by {\it Swift}~XRT. Together with our previous work in V18, the full analysed sample comprises 199 GRBs with known $z$ observed between 2005 Apr 01 and 2018 Dec 31. The sample covers a wide range of light curve morphologies, including flares as well as smoothly decaying light curves. We perform finely time-resolved spectroscopy of the early WT mode data (starting $\sim 100$~s after trigger) and assess the significance of the blackbody with respect to an absorbed power-law model.   

Ten GRBs are found to have a blackbody component that is significant at $>3\sigma$ in at least three consecutive time bins: GRB~050724, GRB~071031, GRB~080325, GRB~081221, GRB~090516, GRB~090715B, GRB~171205A, GRB~171222A, GRB~180329B and GRB~180620B. Together with the results from V18, the total number of GRBs with blackbody components is 19, corresponding to $\sim 10\%$ of the analysed sample. 

We focus on the interpretation that the blackbodies represent genuine thermal emission components, but stress that it is possible that other models may be able to describe many of the spectra equally well.  The 19 GRBs with detections exhibit diverse characteristics, both in terms of light curves and the properties of the blackbodies. This includes distinct groups in the distribution of $L_{\rm BB}$, already noted in the smaller sample in V18, as well as a large spread of $T_{\rm BB}$ and the inferred $\Gamma_{\rm jet}$. These diverse properties point to different origins of the blackbody components, as summarised below. 

\begin{itemize}
\item The blackbody emission in $\sim 1/3$ of the GRBs is clearly associated with late prompt emission from the jet, as evidenced by rapid, flaring variability of the light curves. The blackbodies also have high temperatures ($\gtrsim 1$~keV) and luminosities ($\sim 10^{49}-10^{51}\ \rm{erg\ s^{-1}}$), which translates to relatively high $\Gamma_{\rm jet} \sim 20 - 40$.  

\item Three long GRBs (GRB~101219B, GRB~111225A and GRB~150727A) are fully consistent with the expectations for the smoothly decaying emission from a cocoon. These GRBs have $T_{\rm BB}\sim 0.1-0.2$ keV,  $L_{\rm BB} \sim 10^{47} - 10^{48} \ \rm{erg s^{-1}}$ and $\Gamma_{\rm jet} \sim 2 - 6$. The short GRB~050724 has similar properties, apart from a higher temperature of $\sim 0.9$~keV, making it a promising candidate for cocoon emission from a short GRB. 

\item Another $\sim 1/3$ of the sample  have blackbody properties intermediate between the two groups discussed above. They are mostly associated with the steep decay phase of the X-ray light curve and may thus have an origin in high-latitude emission from the jet. An alternative interpretation is that they are due to more energetic cocoon emission. GRB~171222A is a particular promising candidate for the latter, showing strongly dominant blackbody emission, with  $T_{\rm BB}$ and $\Gamma_{\rm jet}$ similar to the likely cocoon cases, but a higher $L_{\rm BB}$ of  $\sim 3 \times 10^{49} \ \rm{erg s^{-1}}$. 

\item Two LL~GRBs with previously reported blackbody components are also included in our sample of detections; GRB~060218 and GRB~171205A. Various interpretations have previously been discussed for these events, including emission from a low-luminosity jet and a cocoon. 
\end{itemize}

The results tentatively support a scenario where thermal emission is associated with all parts of the jet, including on-axis prompt emission, high-latitude delayed emission, as well as emission from the surrounding cocoon. The sample also shows correlations between $L_{\rm BB}-T_{\rm BB}$ and $L_{\rm BB}-\Gamma_{\rm jet}$. These may be related to the spectral-energy correlations most commonly observed in gamma-rays and thus subject to similar selection effects.  The correlations imply a relatively narrow range of observed radii (classical $R_{\rm BB} \sim 10^{11}-10^{13}$~cm and somewhat larger inferred photospheric radii of $r_{\rm{ph}}\sim 10^{12}-10^{14}$~cm), which are clearly larger than the presumed WR progenitors. In the cocoon interpretation, the large radii may indicate breakout from thick winds that surround the progenitors.   

The blackbodies with low temperatures and luminosities are preferentially detected at low $z$ and $N_{\rm H,intr}$, but there are no such selection effects for the sample as a whole, with many of the luminous blackbodies being identified in GRBs at high $z$. Given the relatively low detection rate, we expect {\it Swift}~XRT to observe 1-2 GRBs with significant blackbody components every year, meaning that the sample will grow very slowly. To gain further insight into the nature of the blackbody emission it would be beneficial to also study the prompt gamma-ray phase and subsequent UV/optical emission, which may reveal the connection between different emission components and/or show the blackbody evolving through the different bands with time. A better understanding of the range of emission properties compatible with cocoons would also aid the interpretation of the observations. 

\section*{Acknowledgements}
We thank Bj\"{o}rn Ahlgren for valuable discussions during the preparation of this manuscript. This work was supported by the Knut and Alice Wallenberg Foundation. This work uses services provided by the UK Swift Science Data Centre at the University of Leicester.

\section*{Data availability}

The data underlying this article were accessed from the UK {\it Swift} Science Data Centre XRT GRB repository \url{http://www.swift.ac.uk/xrt_live_cat/} and are publicly available. 

%%%%%%%%%%%%%%%%%%%%%%%%%%%%%%%%%%%%%%%%%%%%%%%%%%



\bibliographystyle{mnras}
\bibliography{thermal_component} % if your bibtex file is called example.bib

% % Alternatively you could enter them by hand, like this:
% % This method is tedious and prone to error if you have lots of references
% \begin{thebibliography}{99}
% \bibitem[\protect\citeauthoryear{Author}{2012}]{Author2012}
% Author A.~N., 2013, Journal of Improbable Astronomy, 1, 1
% \bibitem[\protect\citeauthoryear{Others}{2013}]{Others2013}
% Others S., 2012, Journal of Interesting Stuff, 17, 198
% \end{thebibliography}

%%%%%%%%%%%%%%%%%%%%%%%%%%%%%%%%%%%%%%%%%%%%%%%%%%

%%%%%%%%%%%%%%%%% APPENDICES %%%%%%%%%%%%%%%%%%%%%

\appendix

\section{Details of GRBs with significant blackbody components} \label{grbscomplete}

\begin{figure*}
    \begin{subfigure}[b]{0.49\textwidth}
    \includegraphics[width=\columnwidth,  height = 6cm]{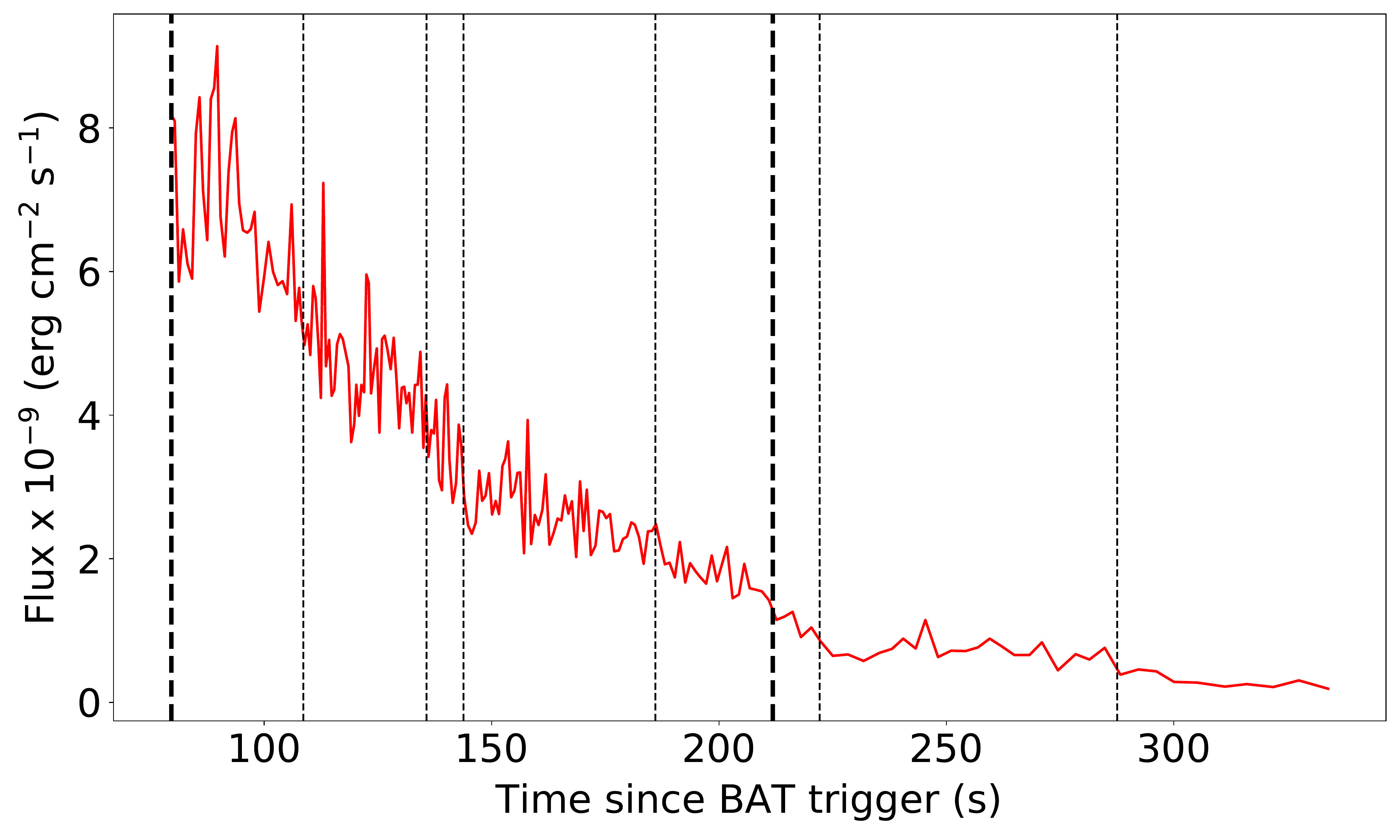}
    \end{subfigure}
    \begin{subfigure}[b]{0.49\textwidth}
    \includegraphics[width=\columnwidth,  height = 6cm]{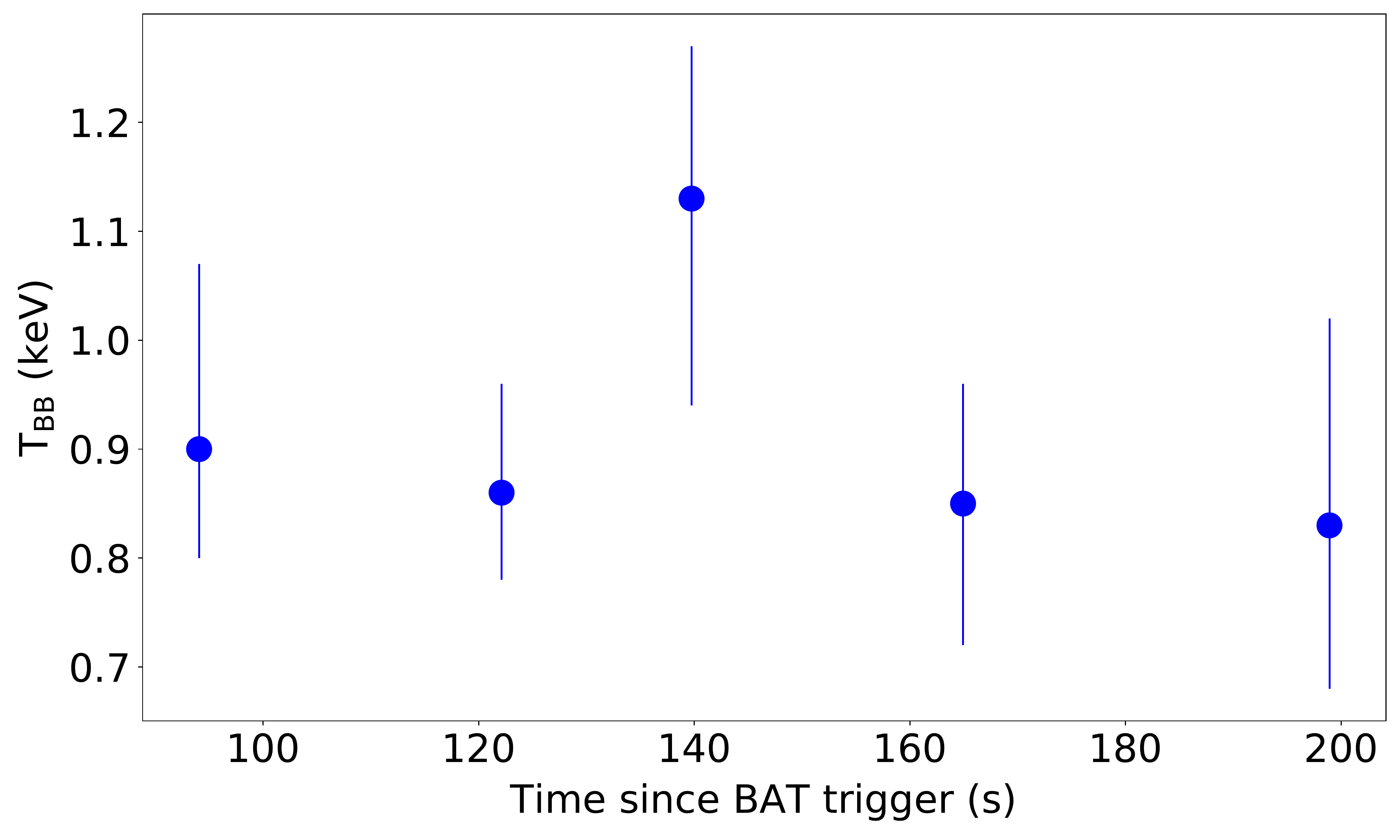}
    \end{subfigure}
    \begin{subfigure}[b]{0.49\textwidth}
     \includegraphics[width=\columnwidth, height = 6cm]{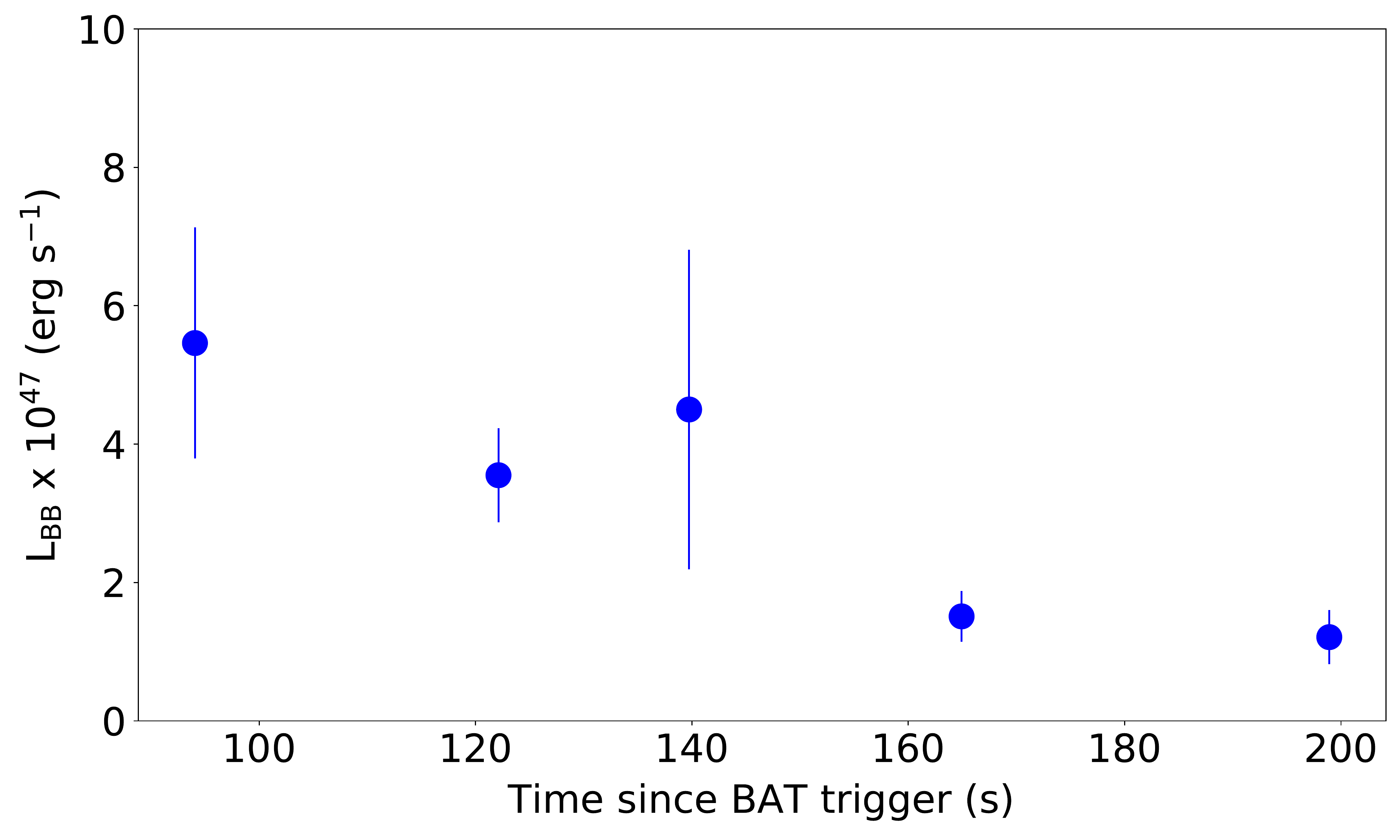}
    \end{subfigure}
    \begin{subfigure}[b]{0.49\textwidth}
     \includegraphics[width=\columnwidth, height = 6cm]{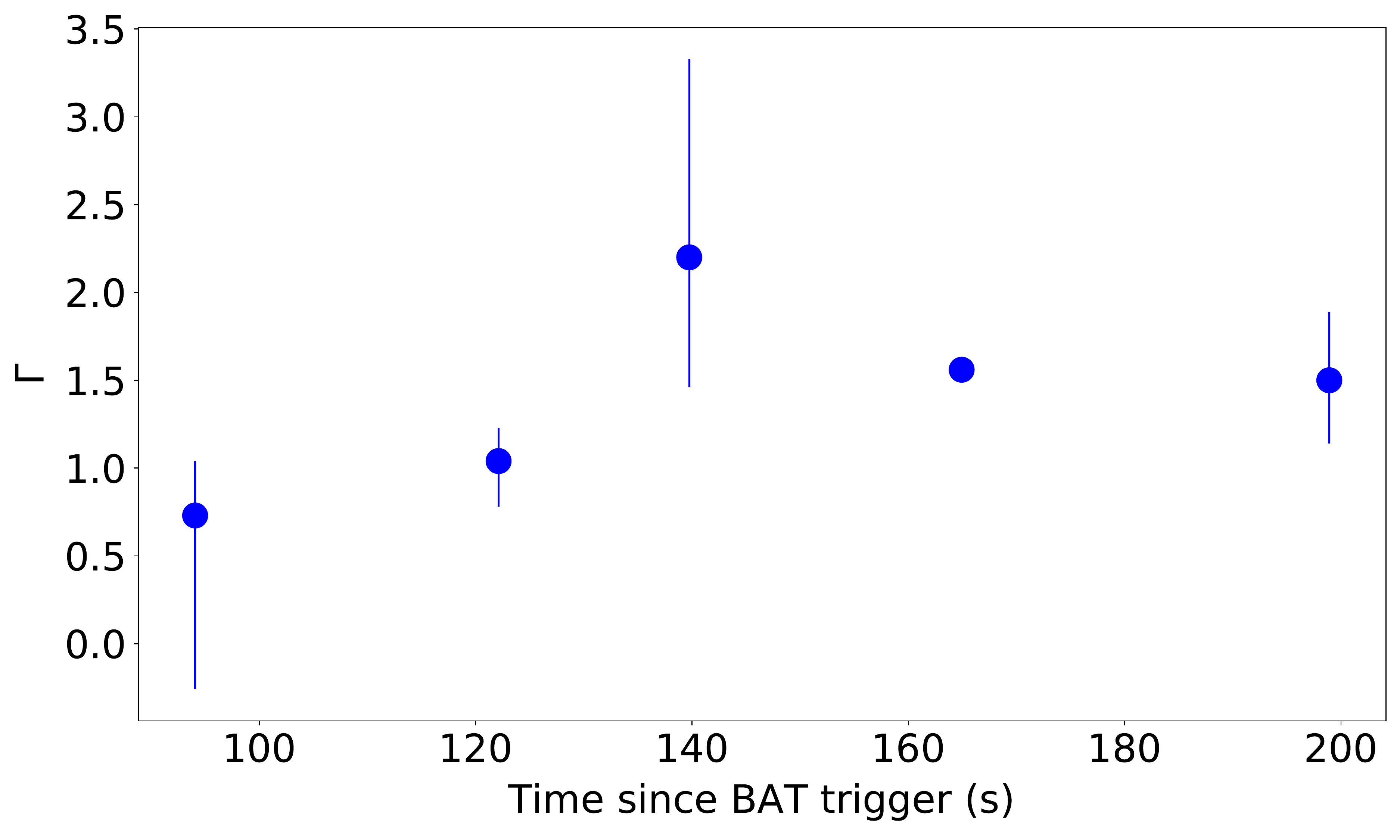}
    \end{subfigure}
    \caption{Light curve and fit results for GRB 050724. The Bayesian blocks used for the spectral analysis are marked by thin dashed lines on the light curve in the upper left panel. The thick dashed lines indicate the time interval where the blackbody component is significant. The next three panels show the time evolution of the rest-frame blackbody temperature, blackbody luminosity and photon index, respectively. }
	\label{050724res}
\end{figure*}

\begin{figure*}
    \begin{subfigure}[b]{0.49\textwidth}
    \includegraphics[width=\columnwidth, height = 7.6cm]{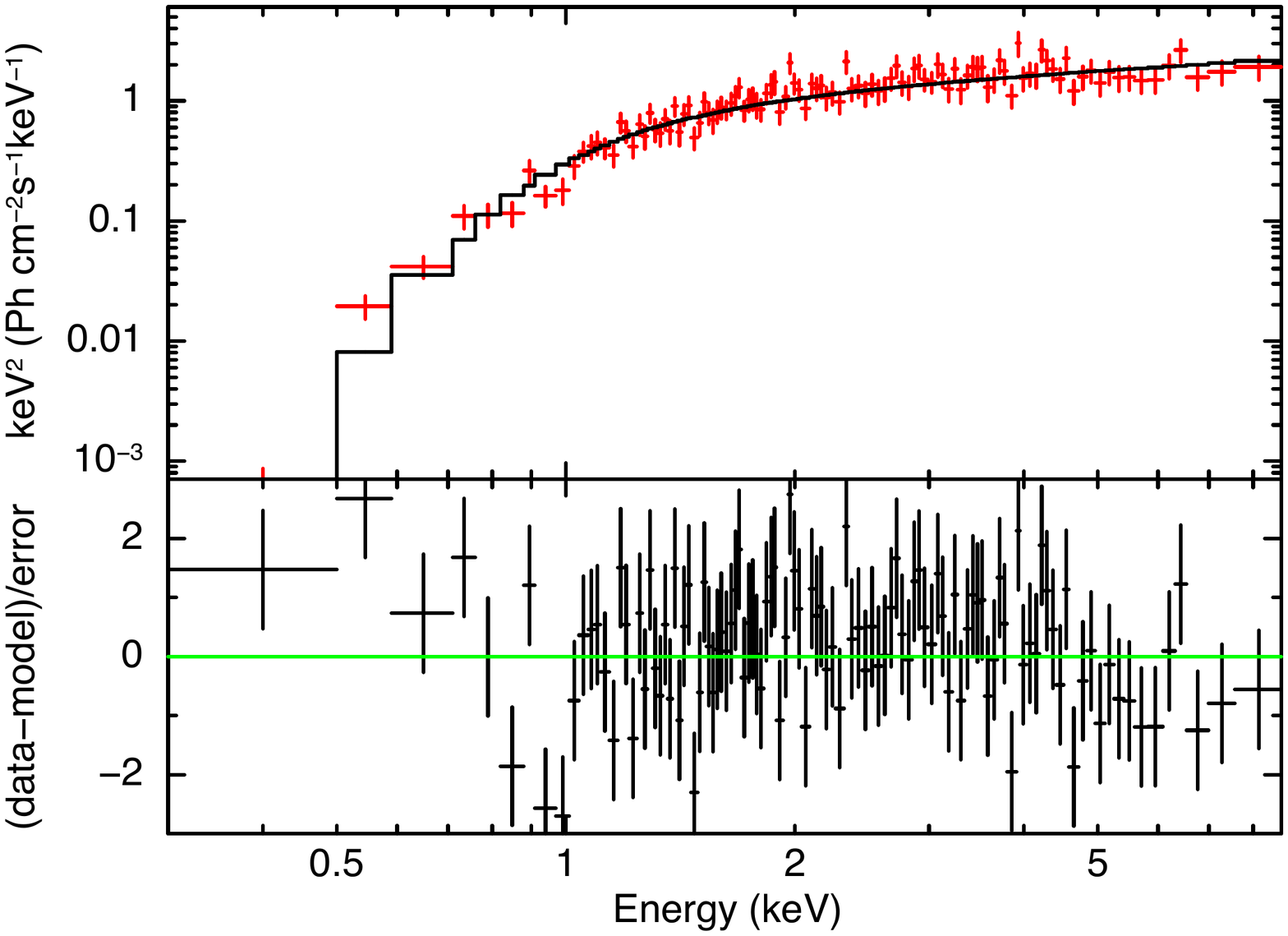}
    \end{subfigure}
    \begin{subfigure}[b]{0.49\textwidth}
     \includegraphics[width=1.1\columnwidth, height = 7.6cm]{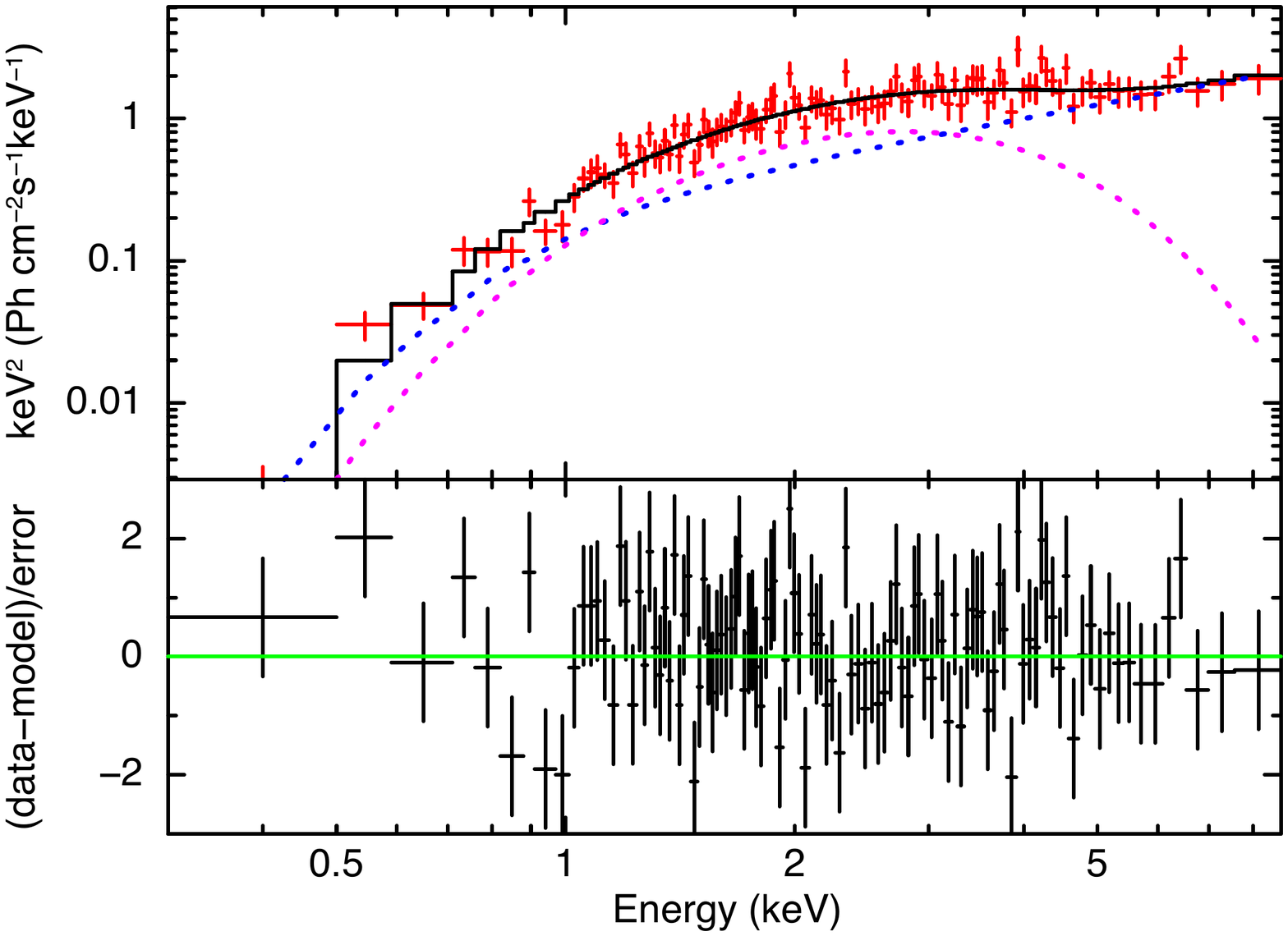}
    \end{subfigure}
    \caption{Fits to the spectrum of GRB 050724 in the time interval 109 - 136 s. Left panel: fit to the absorbed power-law model, where the lower panel shows residuals to the fit. Right panel: fit to the absorbed power-law + blackbody model. The model components are shown in magenta (blackbody) and blue (power law) while the total model is shown in black. The residuals to the fit are shown in the lower panel.}
	\label{050724xspec}
\end{figure*}

\begin{figure*}
    \begin{subfigure}[b]{0.49\textwidth}
    \includegraphics[width=\columnwidth,  height = 5cm]{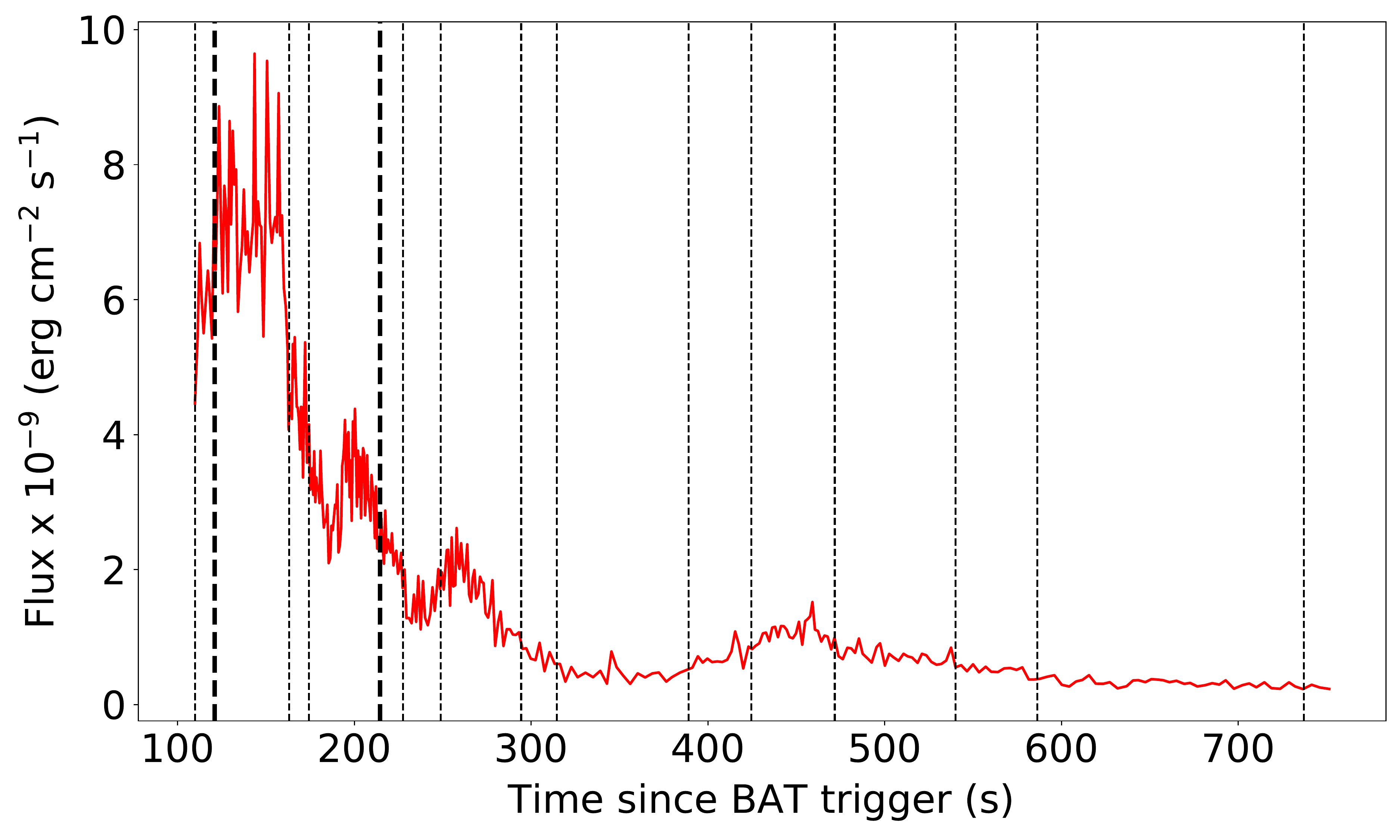}
    \end{subfigure}
    \begin{subfigure}[b]{0.49\textwidth}
    \includegraphics[width=\columnwidth,  height = 5cm]{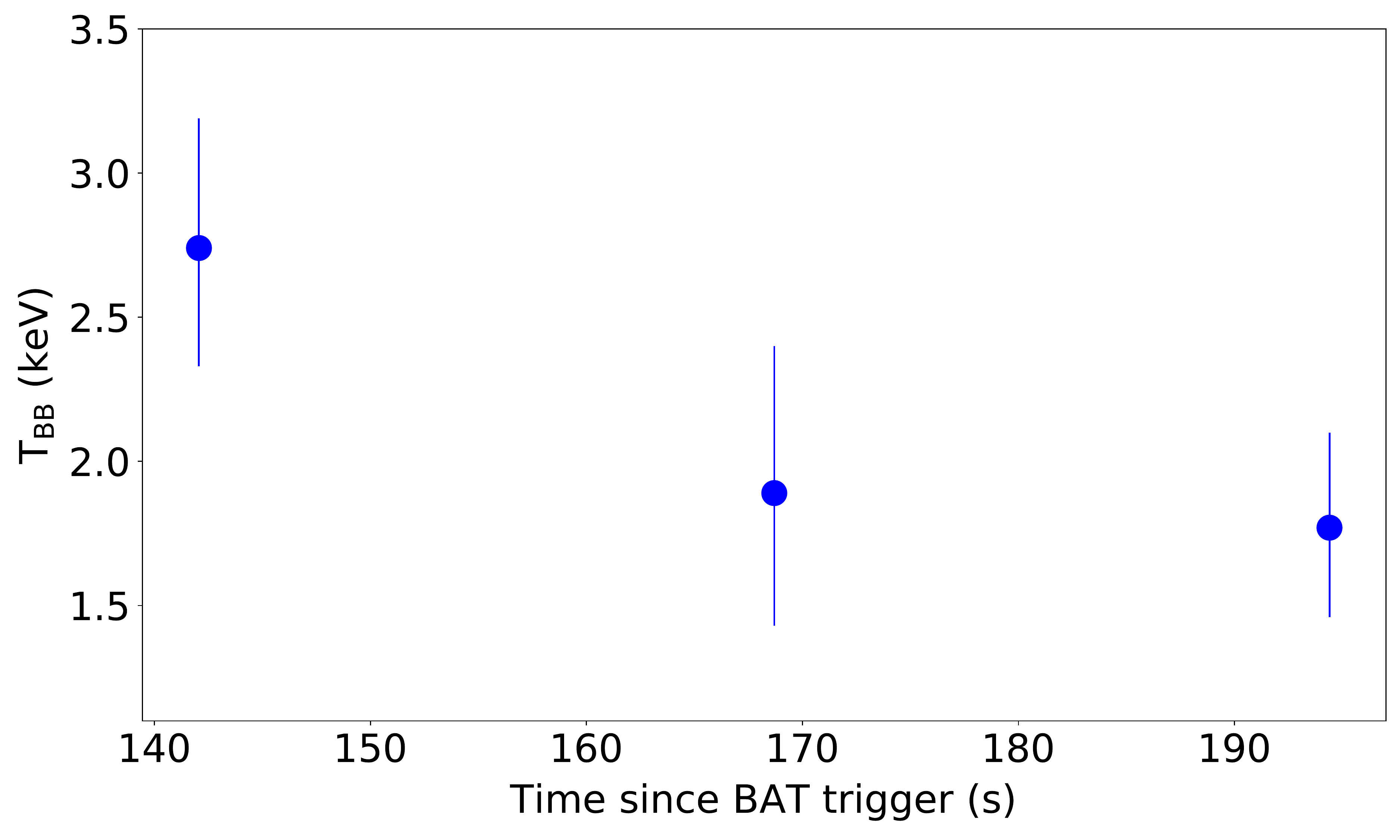}
    \end{subfigure}
    \begin{subfigure}[b]{0.48\textwidth}
     \includegraphics[width=\columnwidth, height = 5cm]{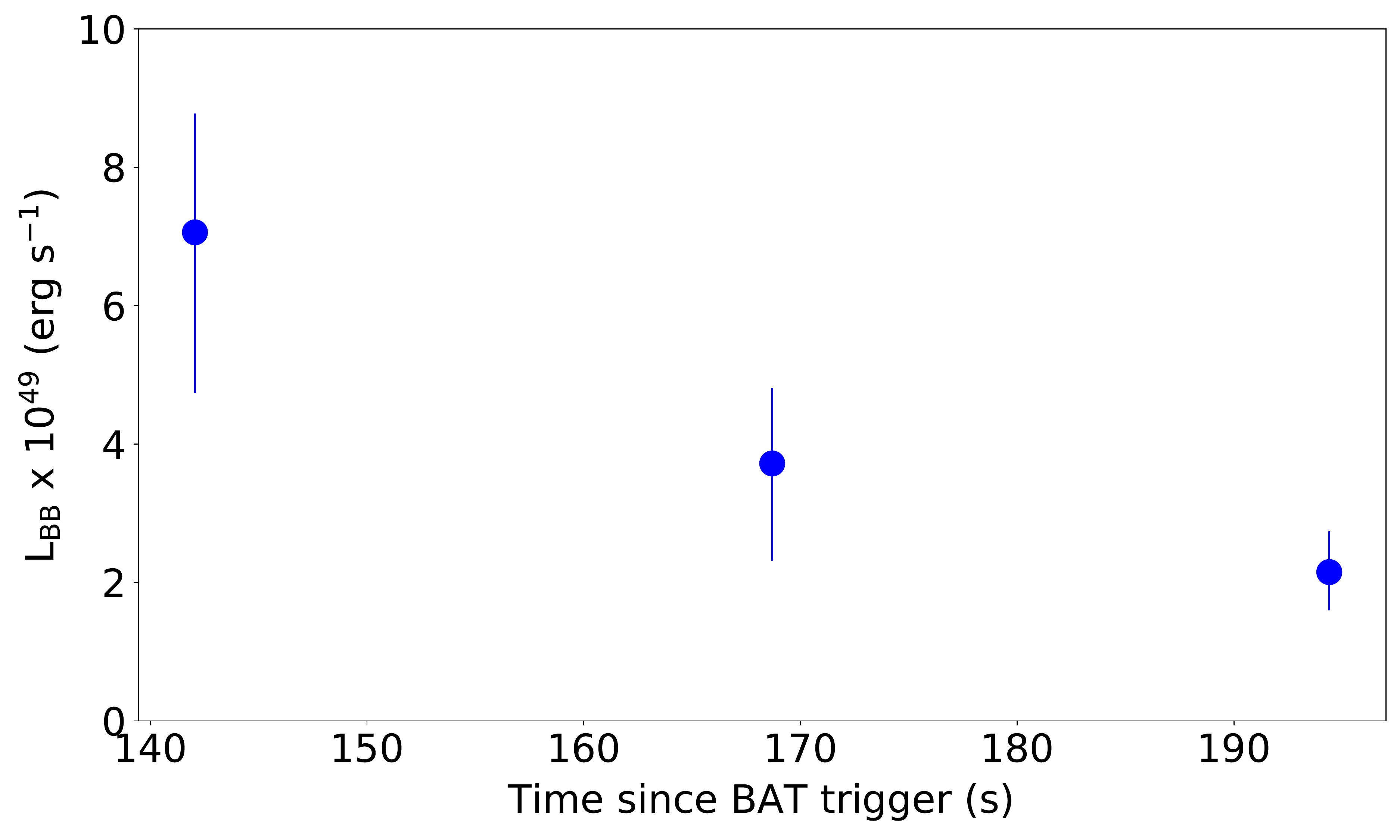}
    \end{subfigure}
    \begin{subfigure}[b]{0.49\textwidth}
     \includegraphics[width=\columnwidth, height = 5cm]{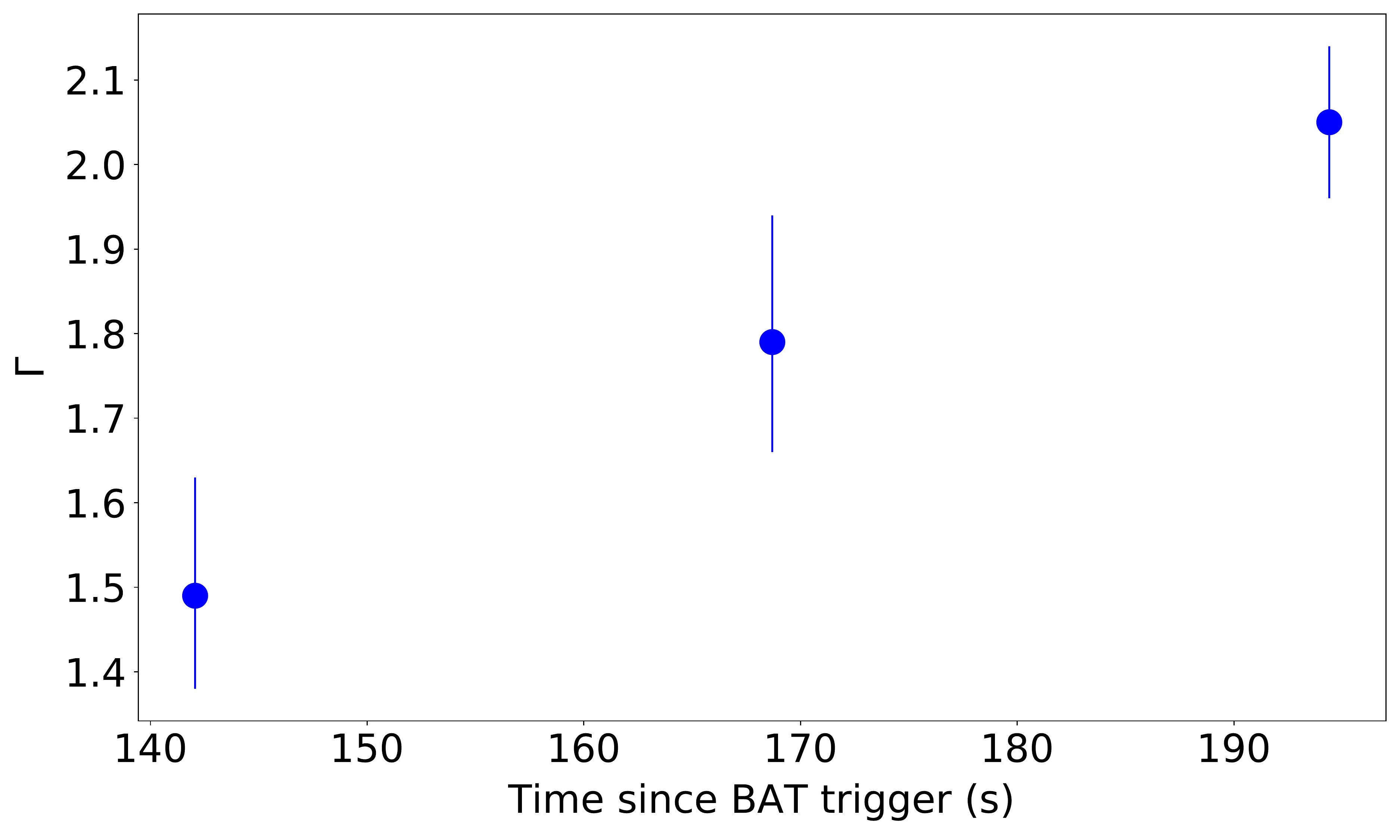}
    \end{subfigure}
    \caption{Same as in Fig.\ref{050724res}, but for GRB~071031.}
	\label{071031res}
\end{figure*}

\begin{figure*}
    \begin{subfigure}[b]{0.49\textwidth}
    \includegraphics[width=\columnwidth, height = 7.6cm]{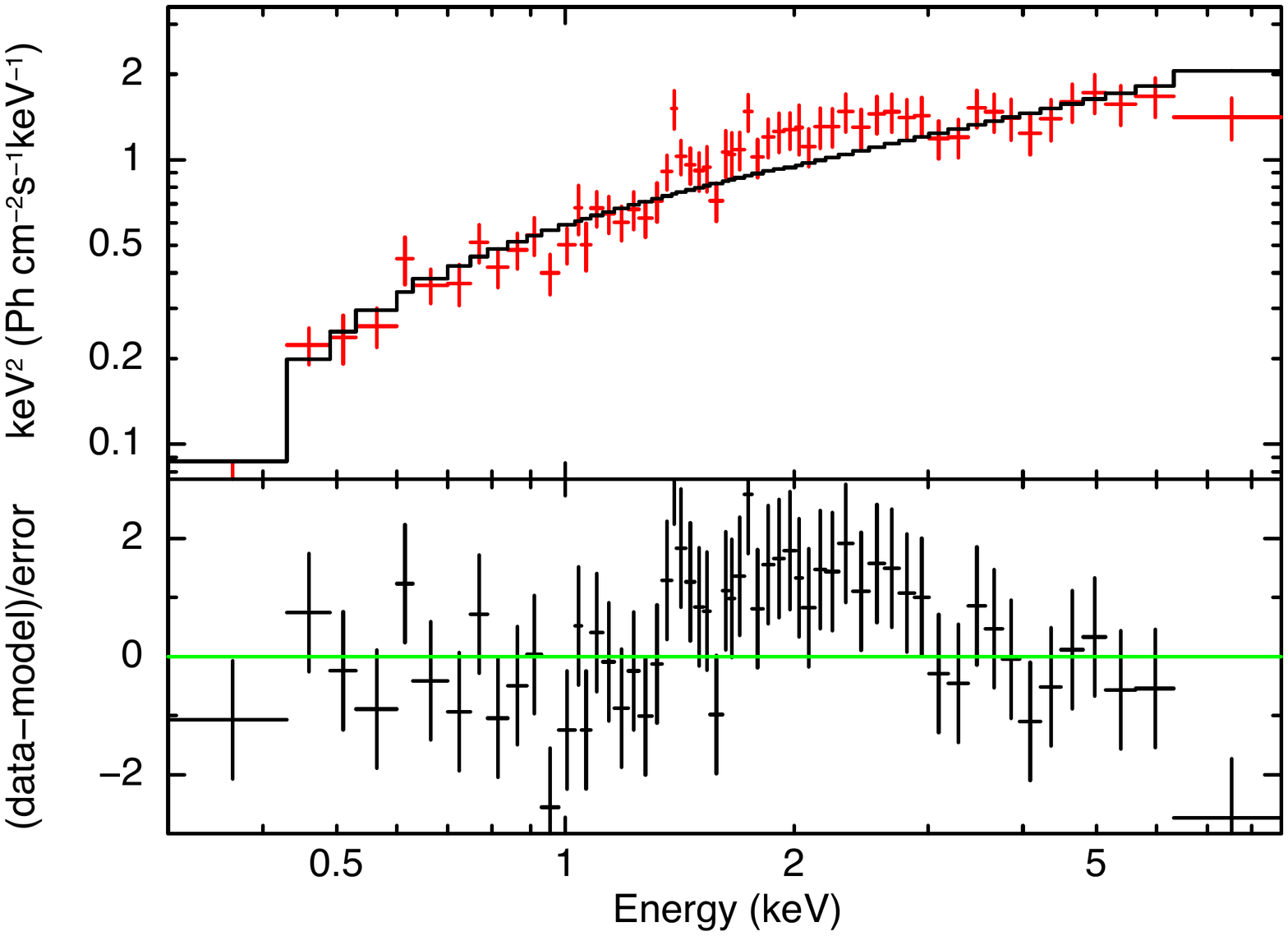}
    \end{subfigure}
    \begin{subfigure}[b]{0.49\textwidth}
     \includegraphics[width=\columnwidth, height = 7.6cm]{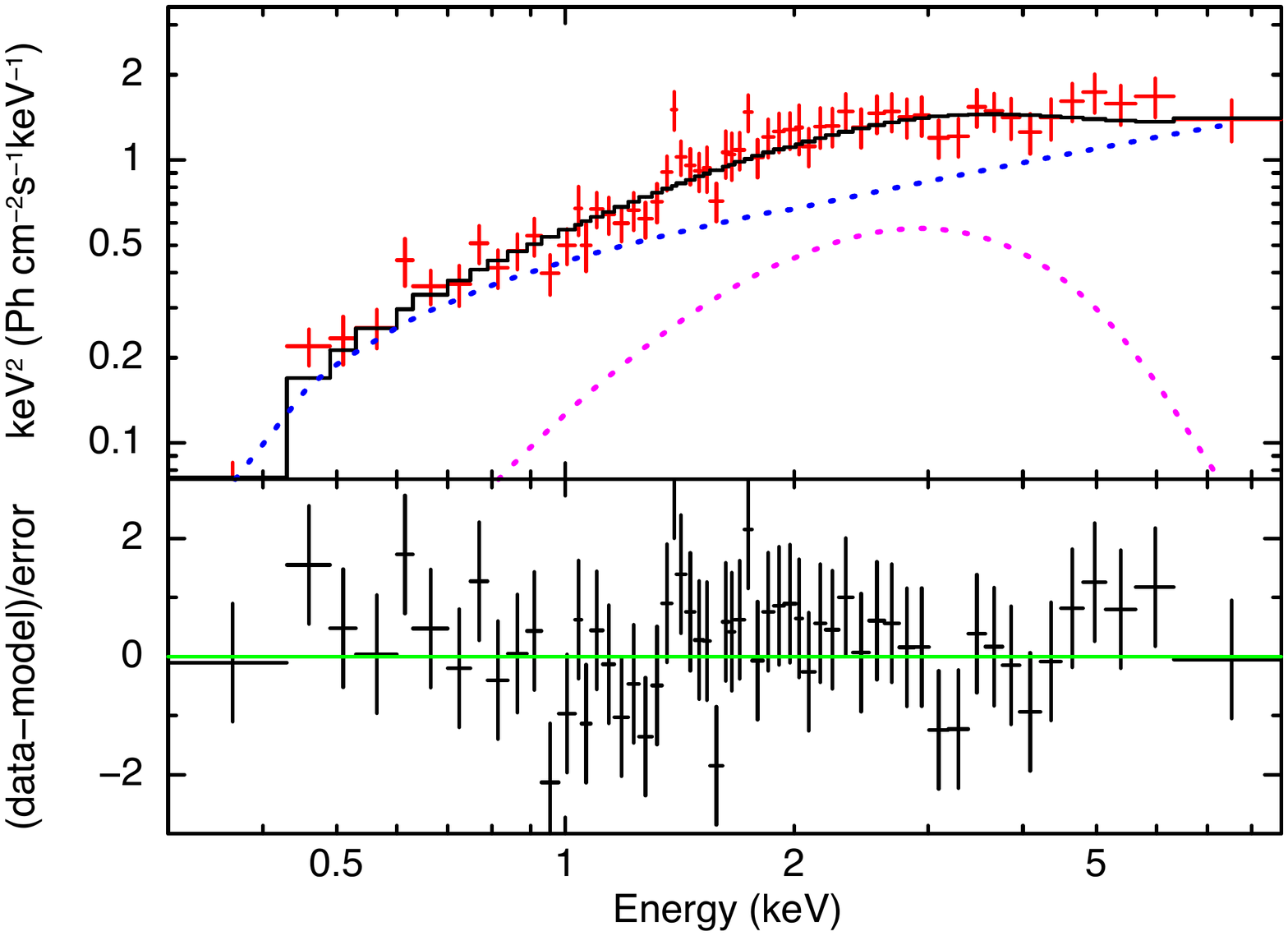}
    \end{subfigure}
    \caption{Same as in Fig. \ref{050724xspec} but for GRB~071031 in the time interval 163 - 174 s. The data have been rebinned for visual clarity.}
	\label{071031xspec}
\end{figure*}

\begin{figure*}
    \begin{subfigure}[b]{0.49\textwidth}
    \includegraphics[width=\columnwidth,  height = 5cm]{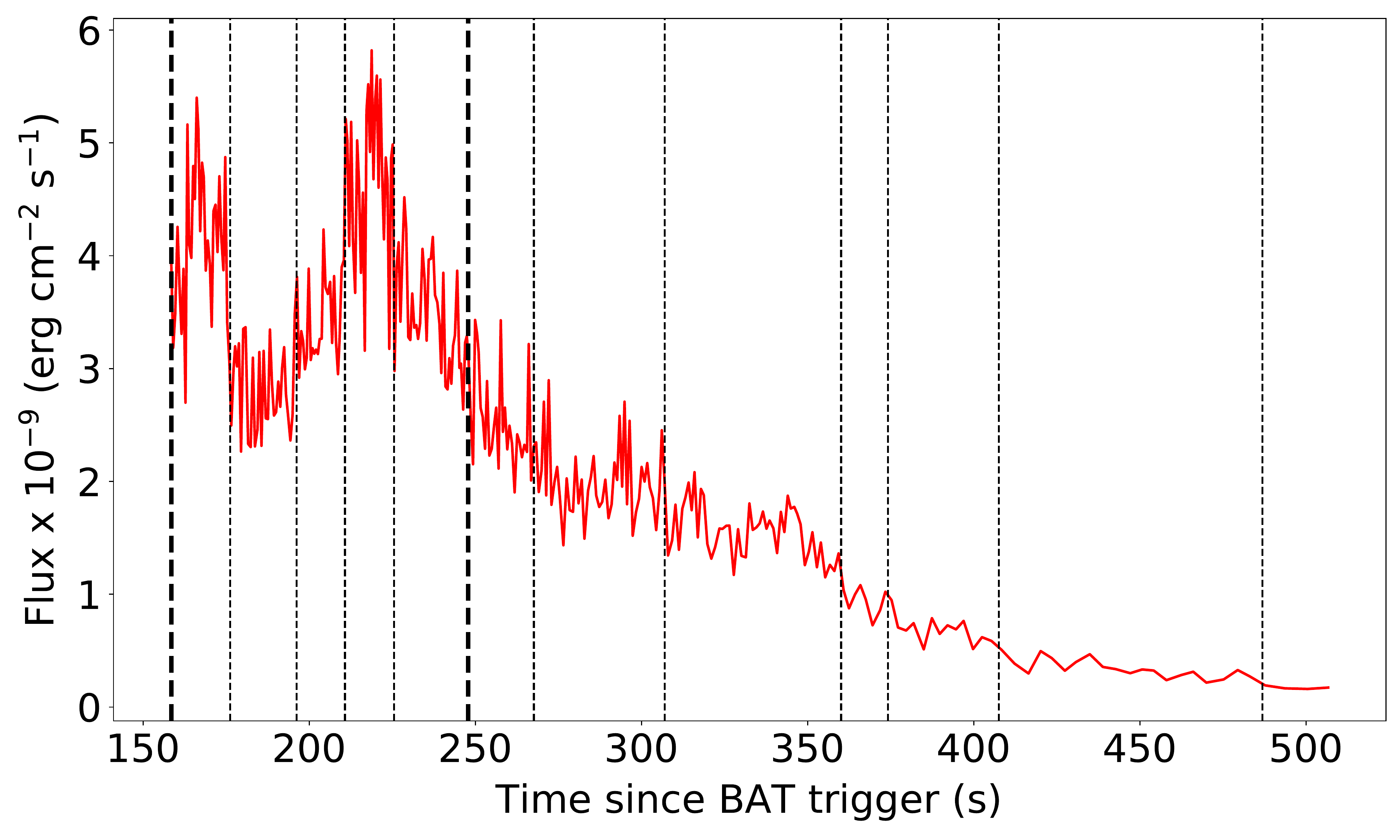}
    \end{subfigure}
    \begin{subfigure}[b]{0.49\textwidth}
    \includegraphics[width=\columnwidth,  height = 5cm]{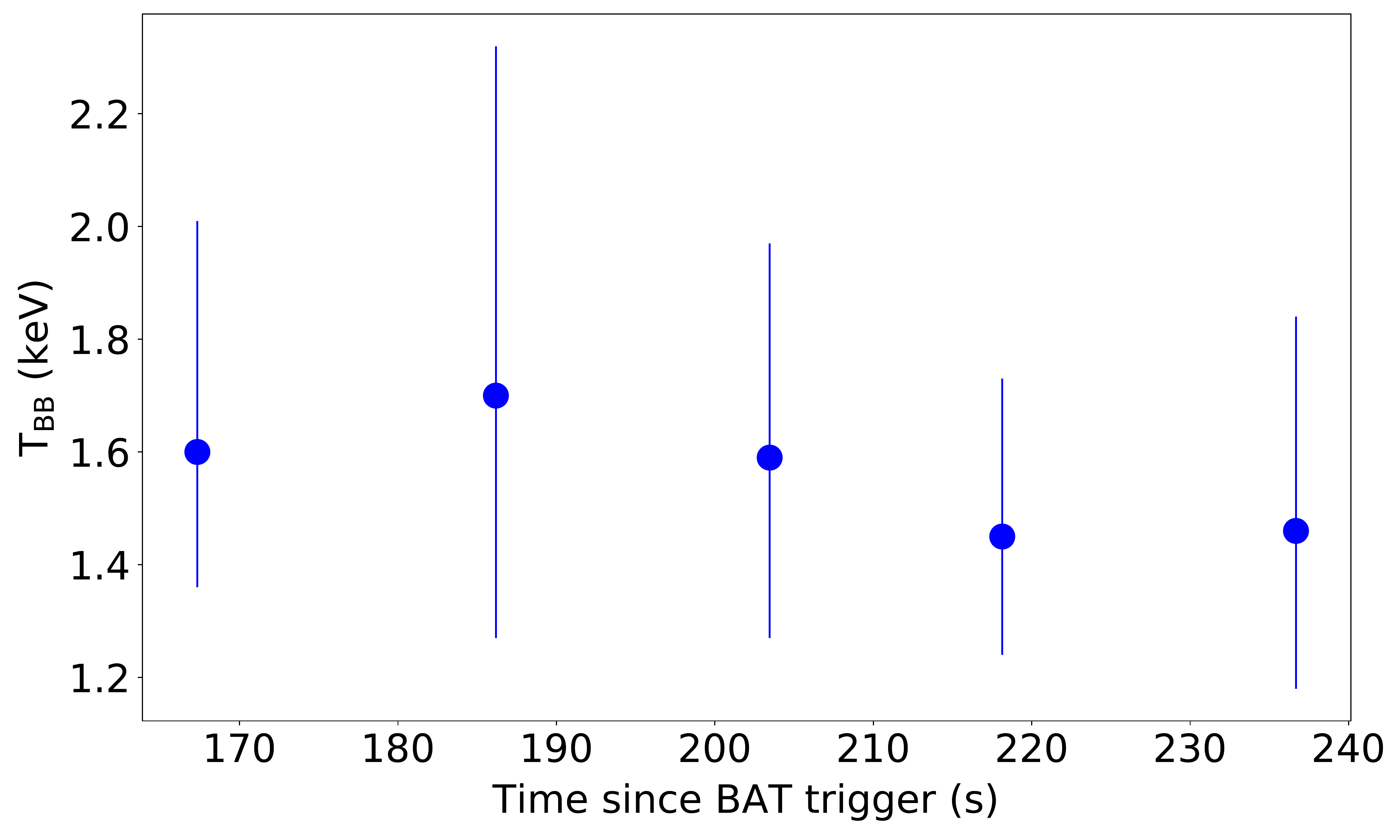}
    \end{subfigure}
    \begin{subfigure}[b]{0.49\textwidth}
     \includegraphics[width=\columnwidth, height = 5cm]{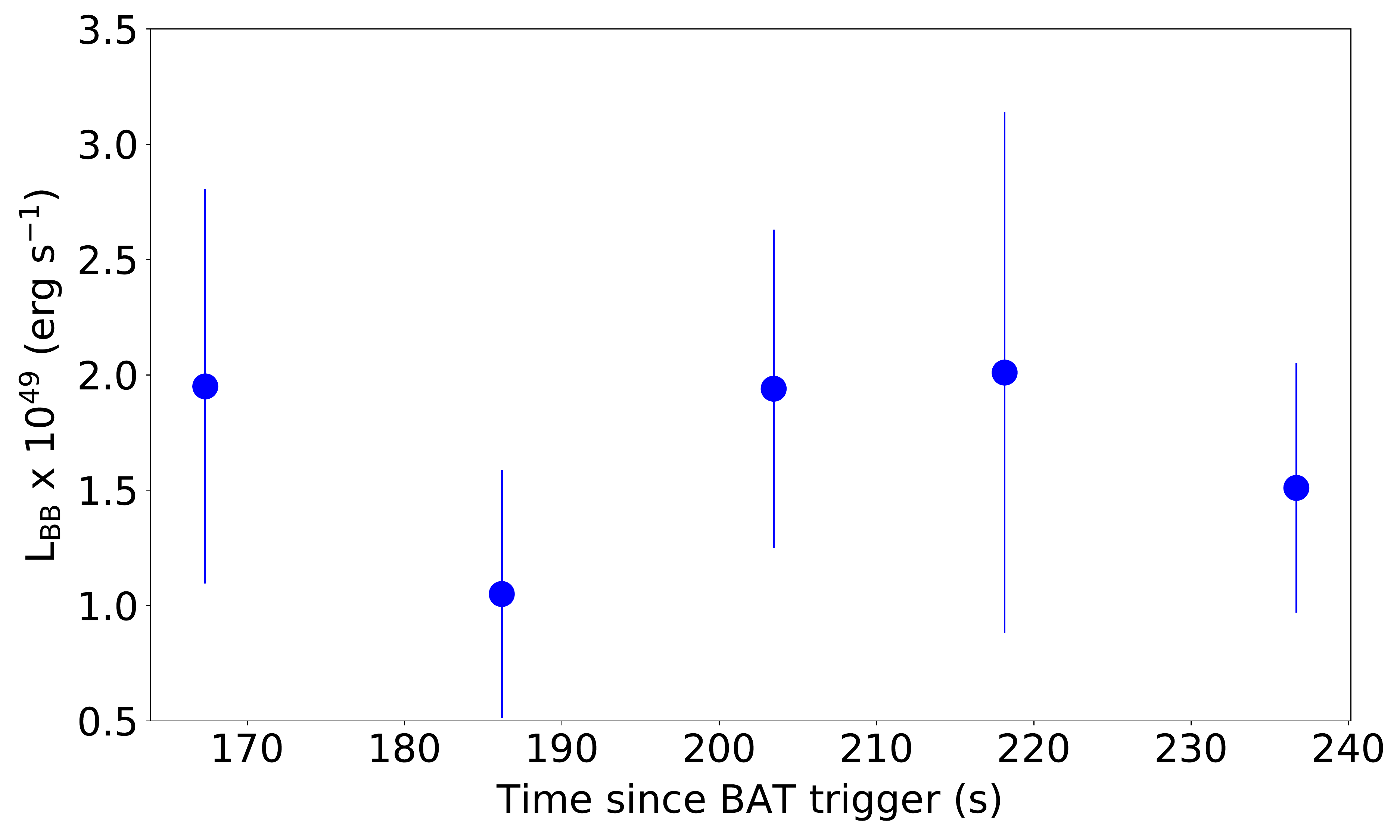}
    \end{subfigure}
    \begin{subfigure}[b]{0.49\textwidth}
     \includegraphics[width=\columnwidth, height = 5cm]{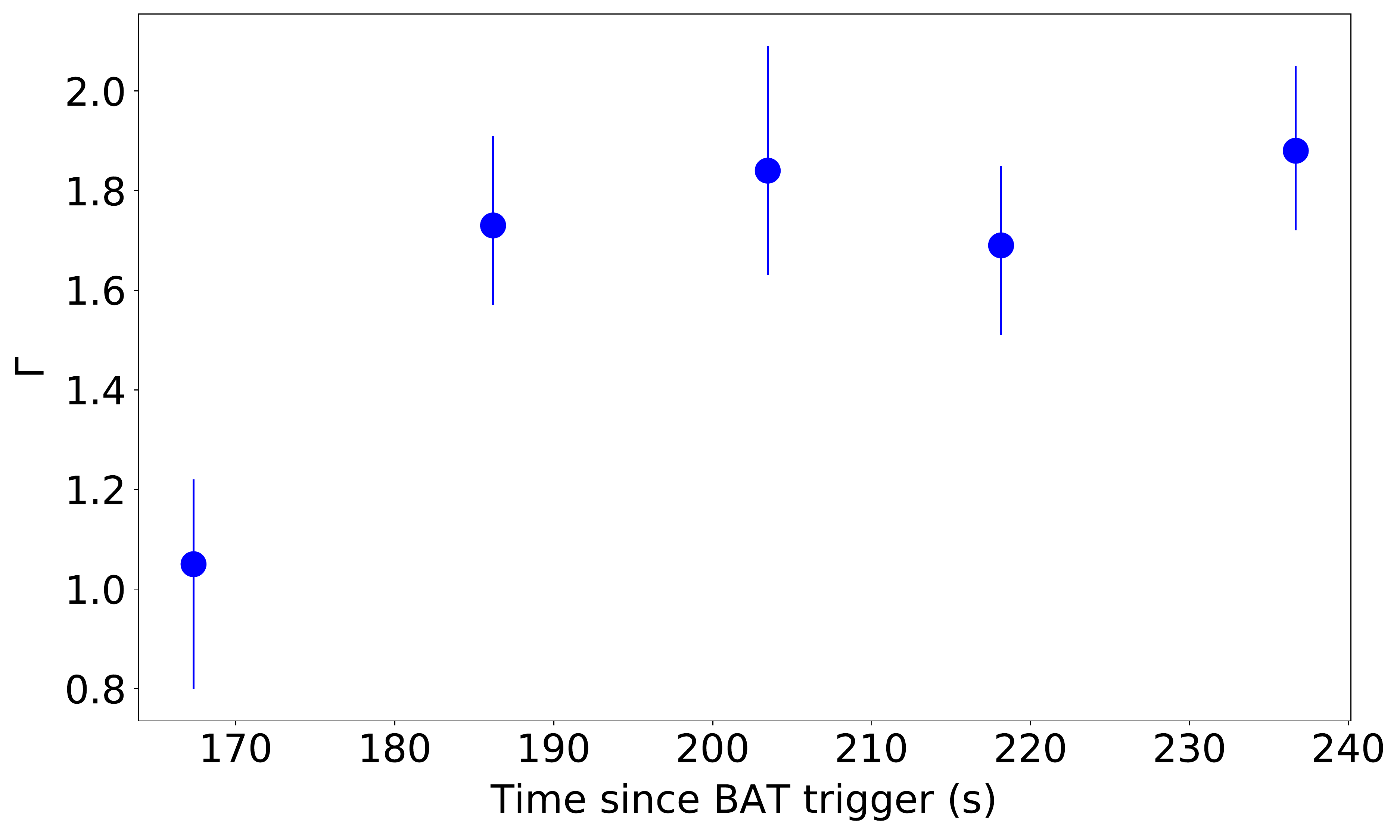}
    \end{subfigure}
    \caption{Same as in Fig. \ref{050724res} but for GRB~080325.}
	\label{080325res}
\end{figure*}

\begin{figure*}
    \begin{subfigure}[b]{0.49\textwidth}
    \includegraphics[width=\columnwidth, height = 7.6cm]{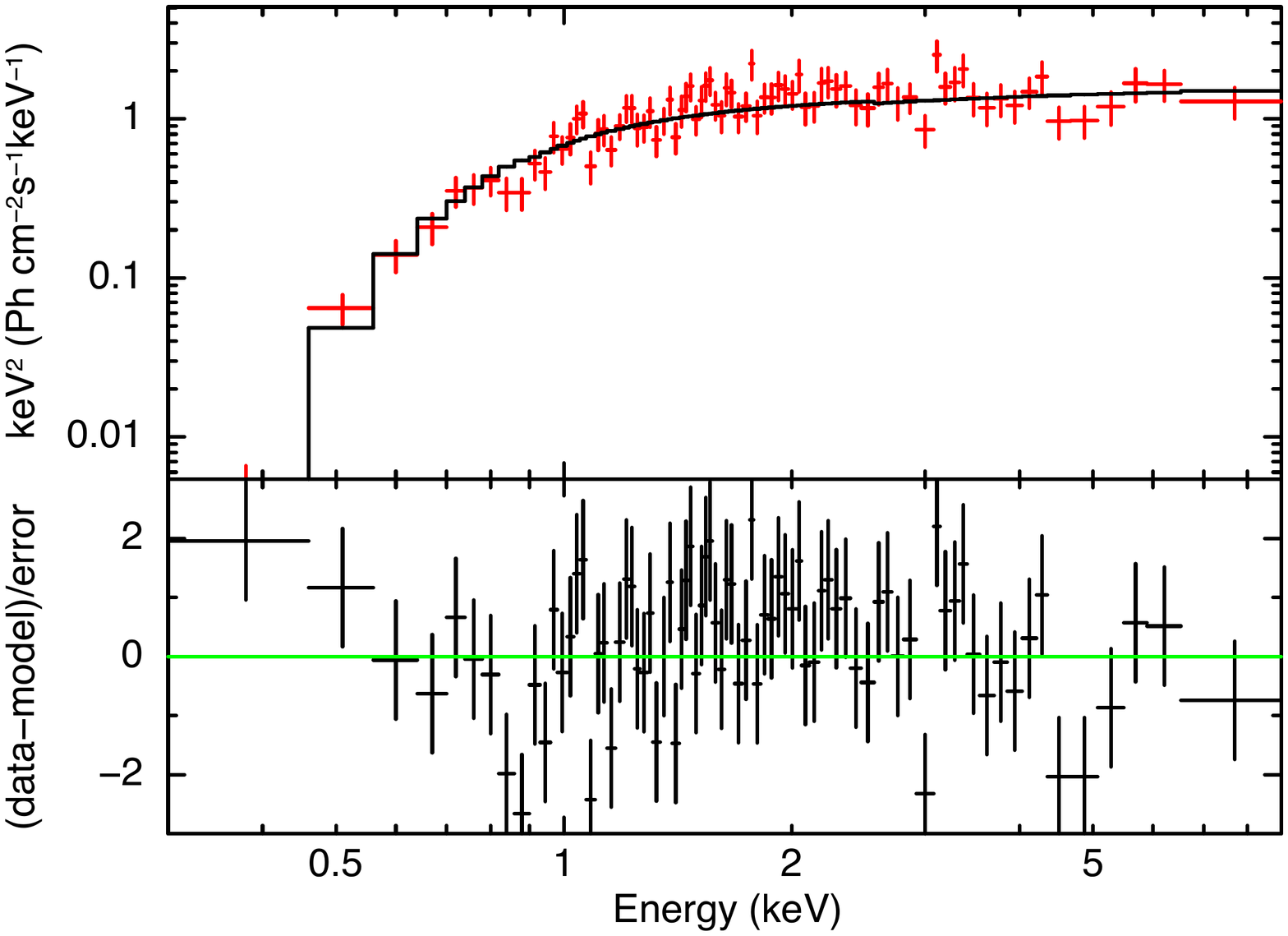}
    \end{subfigure}
    \begin{subfigure}[b]{0.49\textwidth}
     \includegraphics[width=\columnwidth, height = 7.6cm]{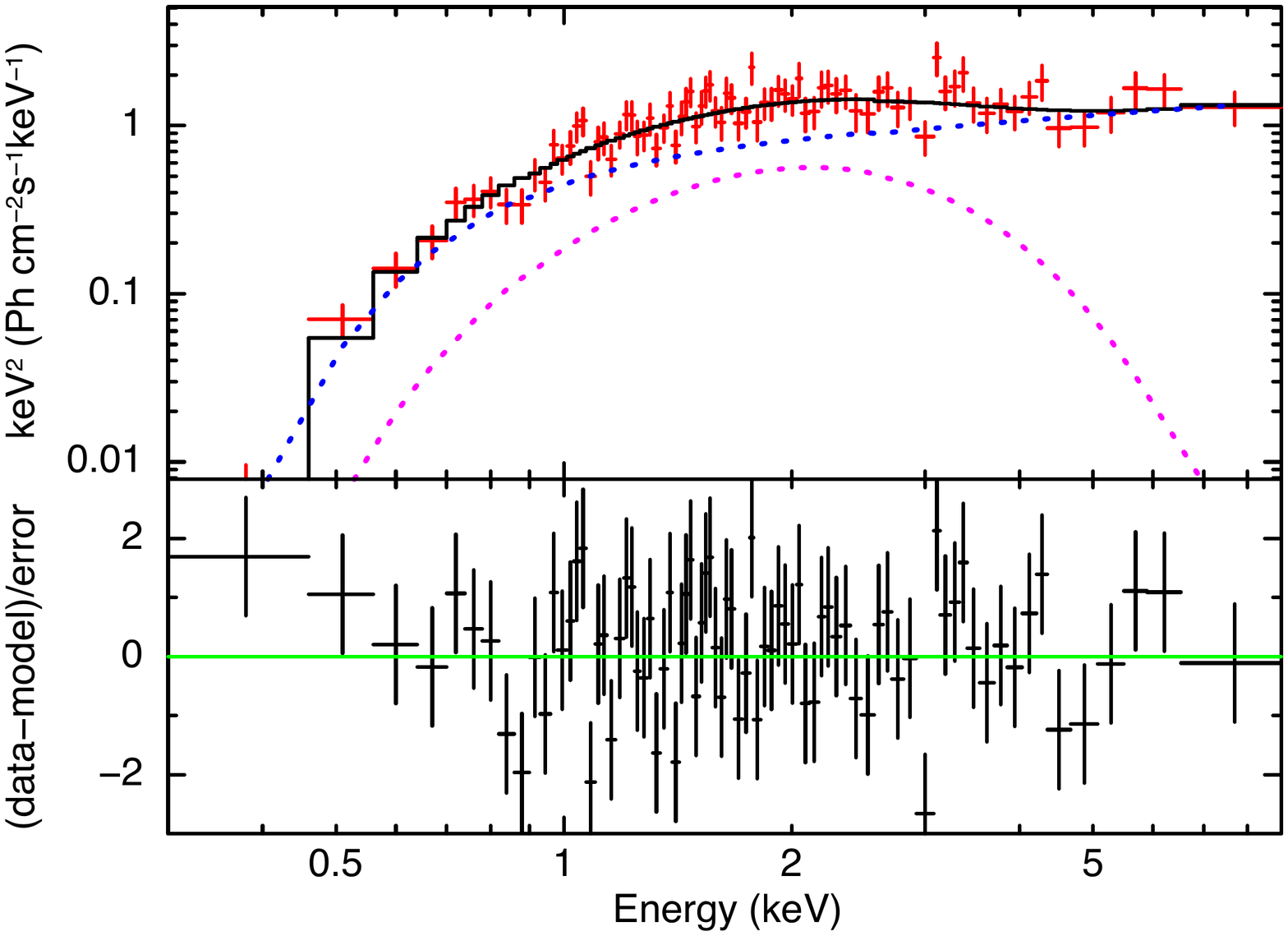}
    \end{subfigure}
    \caption{Same as in Fig. \ref{050724xspec} but for GRB~080325 in the time interval 210 - 225 s.}
	\label{080325xspec}
\end{figure*}

\begin{figure*}
    \begin{subfigure}[b]{0.49\textwidth}
    \includegraphics[width=\columnwidth,  height = 5cm]{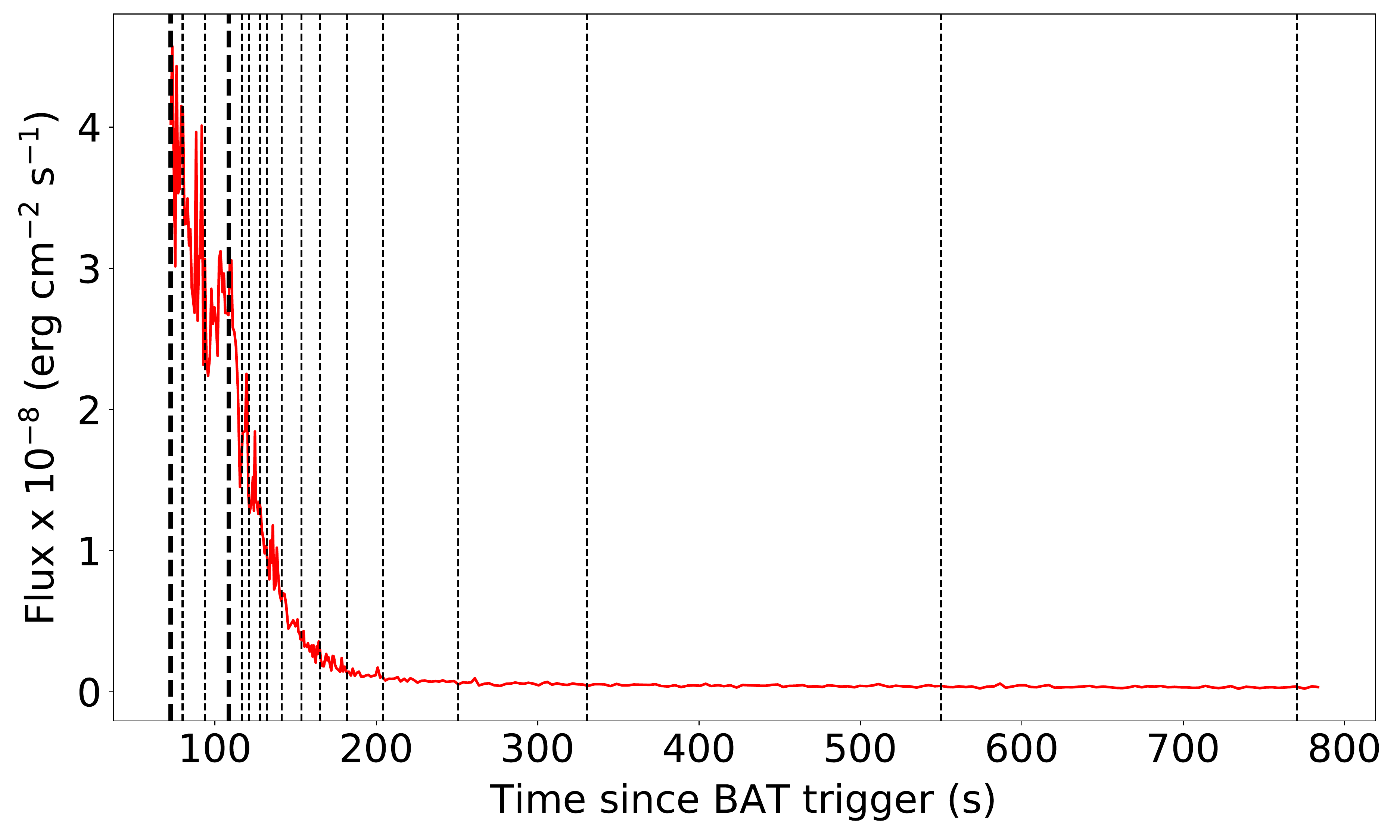}
    \end{subfigure}
    \begin{subfigure}[b]{0.49\textwidth}
    \includegraphics[width=\columnwidth,  height = 5cm]{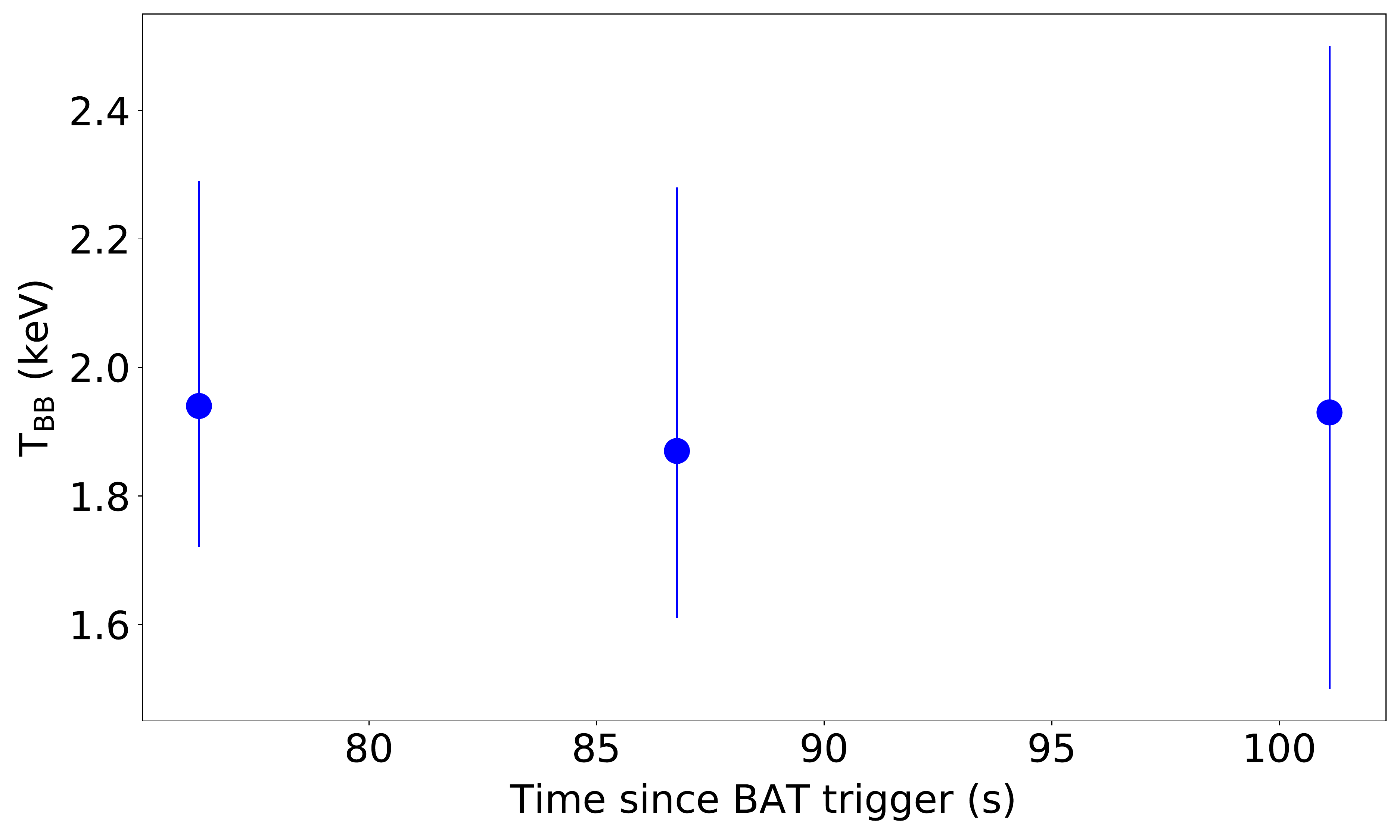}
    \end{subfigure}
    \begin{subfigure}[b]{0.49\textwidth}
     \includegraphics[width=\columnwidth, height = 5cm]{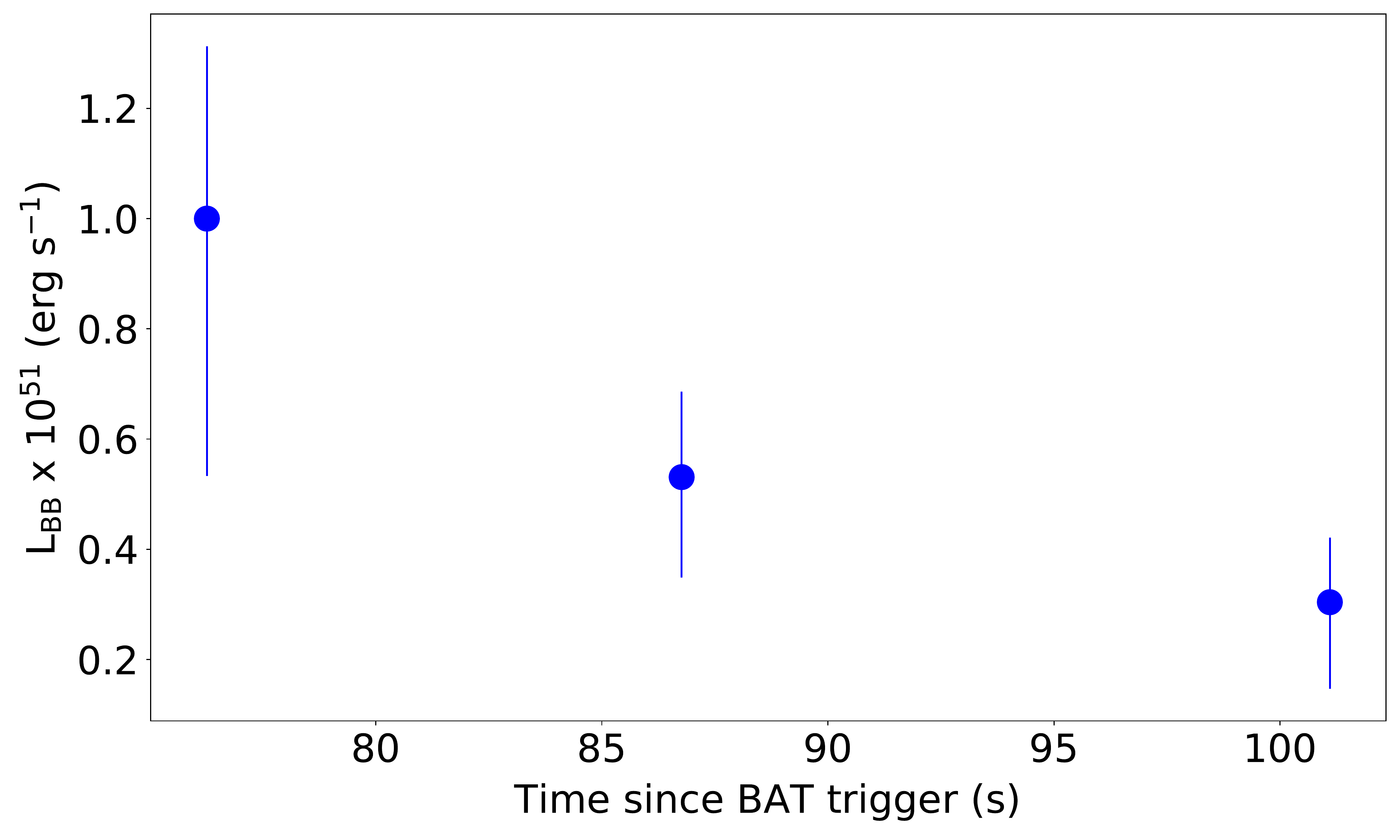}
    \end{subfigure}
    \begin{subfigure}[b]{0.49\textwidth}
     \includegraphics[width=\columnwidth, height = 5cm]{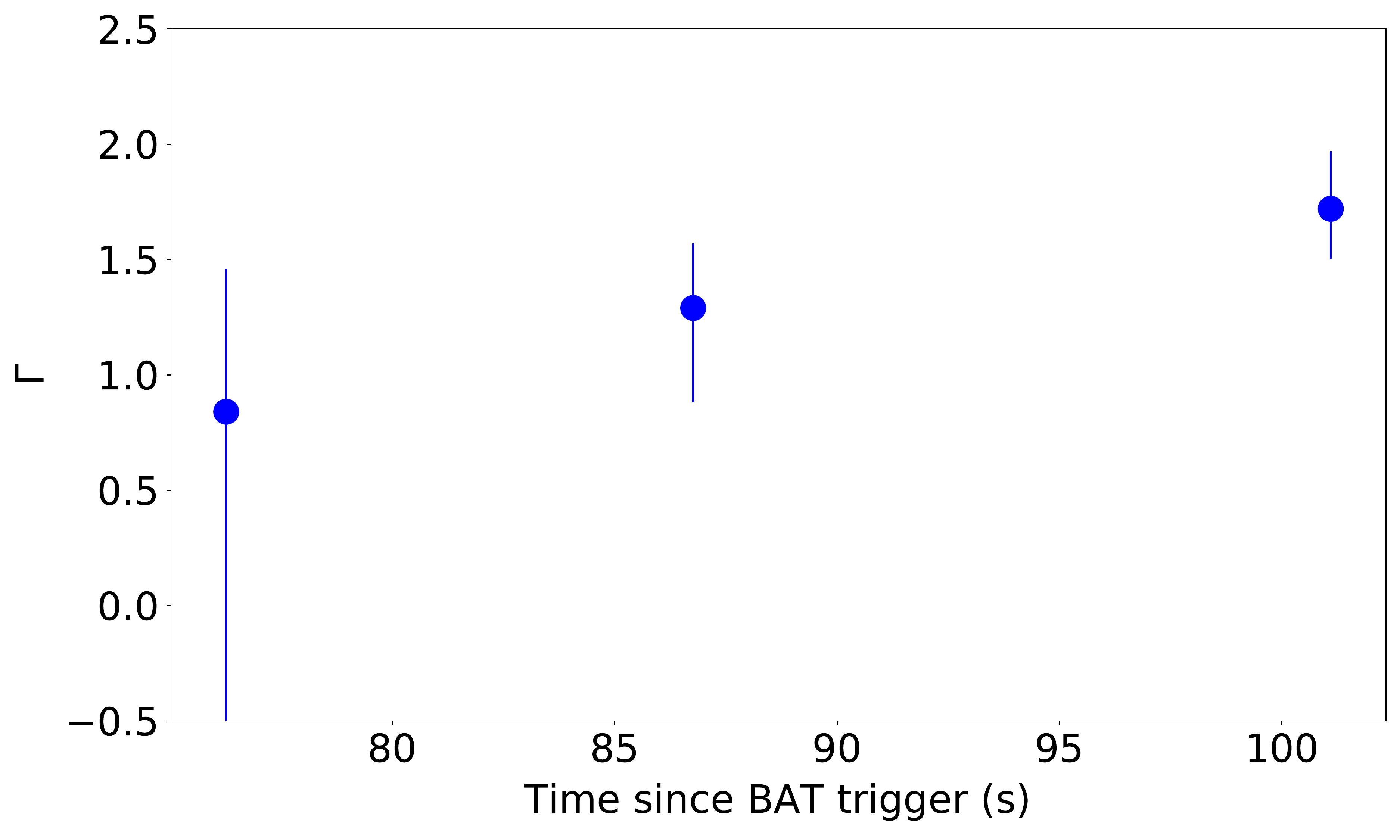}
    \end{subfigure}
    \caption{Same as in Fig. \ref{050724res} but for GRB~081221.}
	\label{081221res}
\end{figure*}

\begin{figure*}
    \begin{subfigure}[b]{0.49\textwidth}
    \includegraphics[width=\columnwidth, height = 7.6cm]{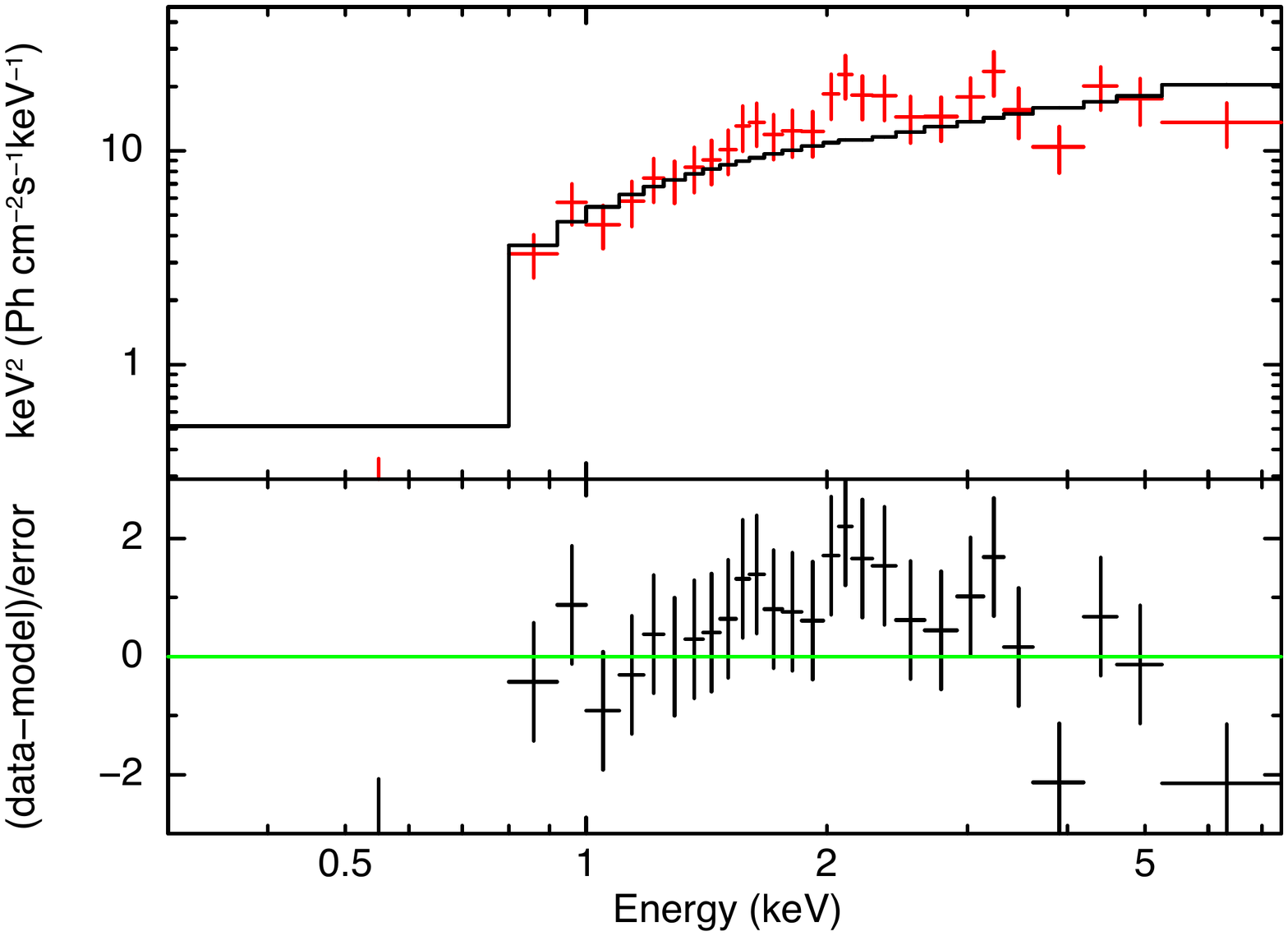}
    \end{subfigure}
    \begin{subfigure}[b]{0.49\textwidth}
     \includegraphics[width=\columnwidth, height = 7.6cm]{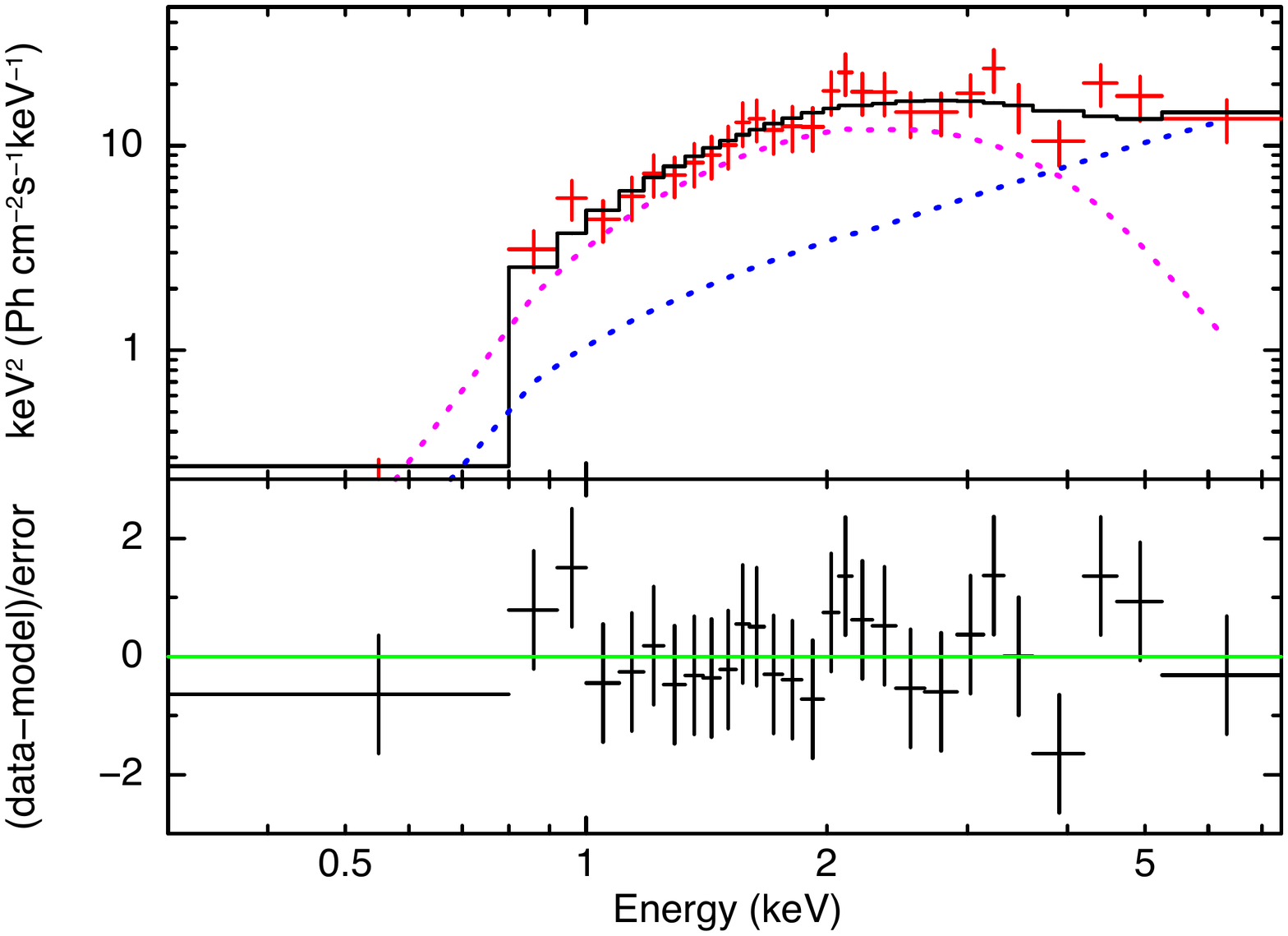}
    \end{subfigure}
    \caption{Same as in Fig.\ref{050724xspec} but for GRB 081221 in the time interval 73 - 80 s.}
	\label{081221xspec}
\end{figure*}

\begin{figure*}
    \begin{subfigure}[b]{0.49\textwidth}
    \includegraphics[width=\columnwidth,  height = 5cm]{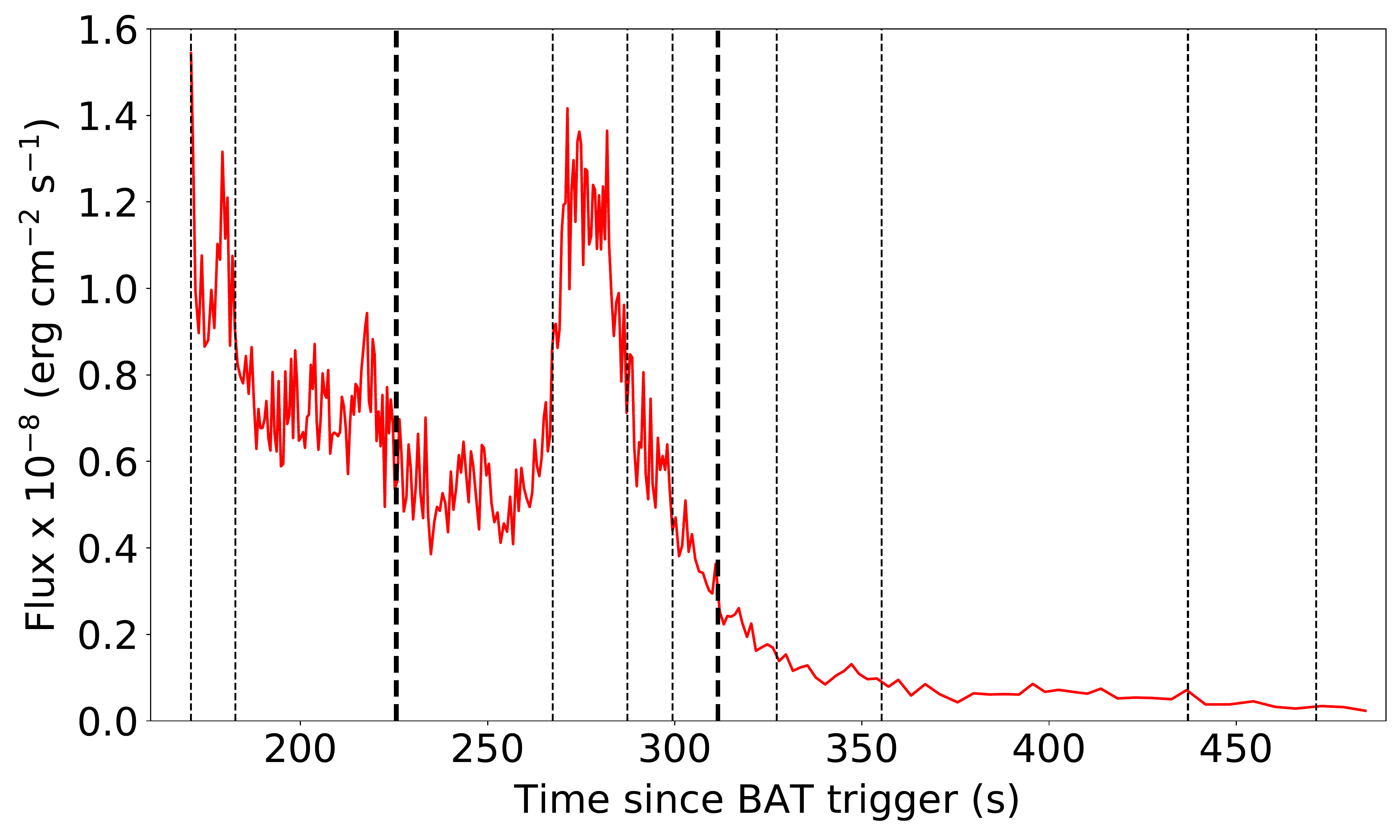}
    \end{subfigure}
    \begin{subfigure}[b]{0.49\textwidth}
    \includegraphics[width=\columnwidth,  height = 5cm]{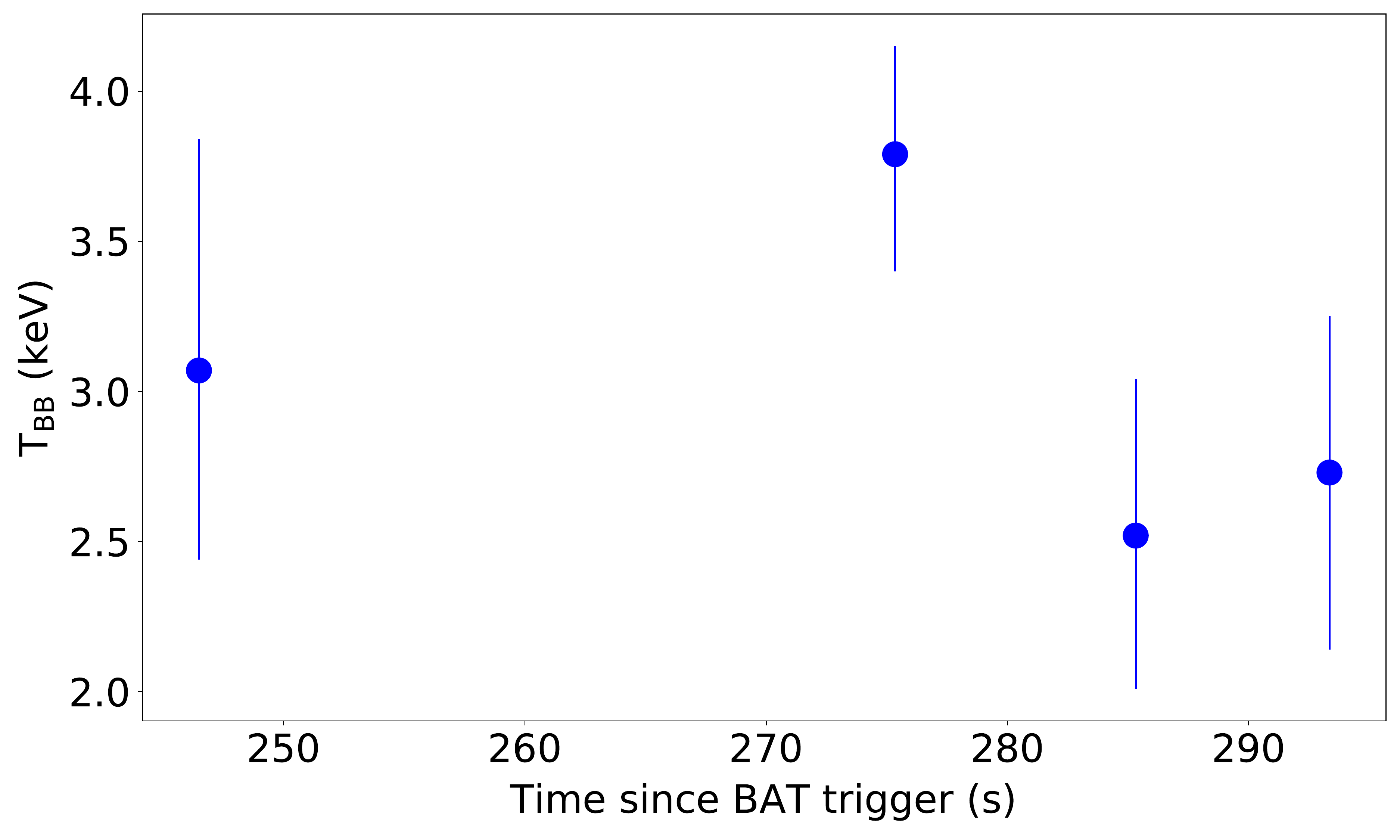}
    \end{subfigure}
    \begin{subfigure}[b]{0.49\textwidth}
     \includegraphics[width=\columnwidth, height = 5cm]{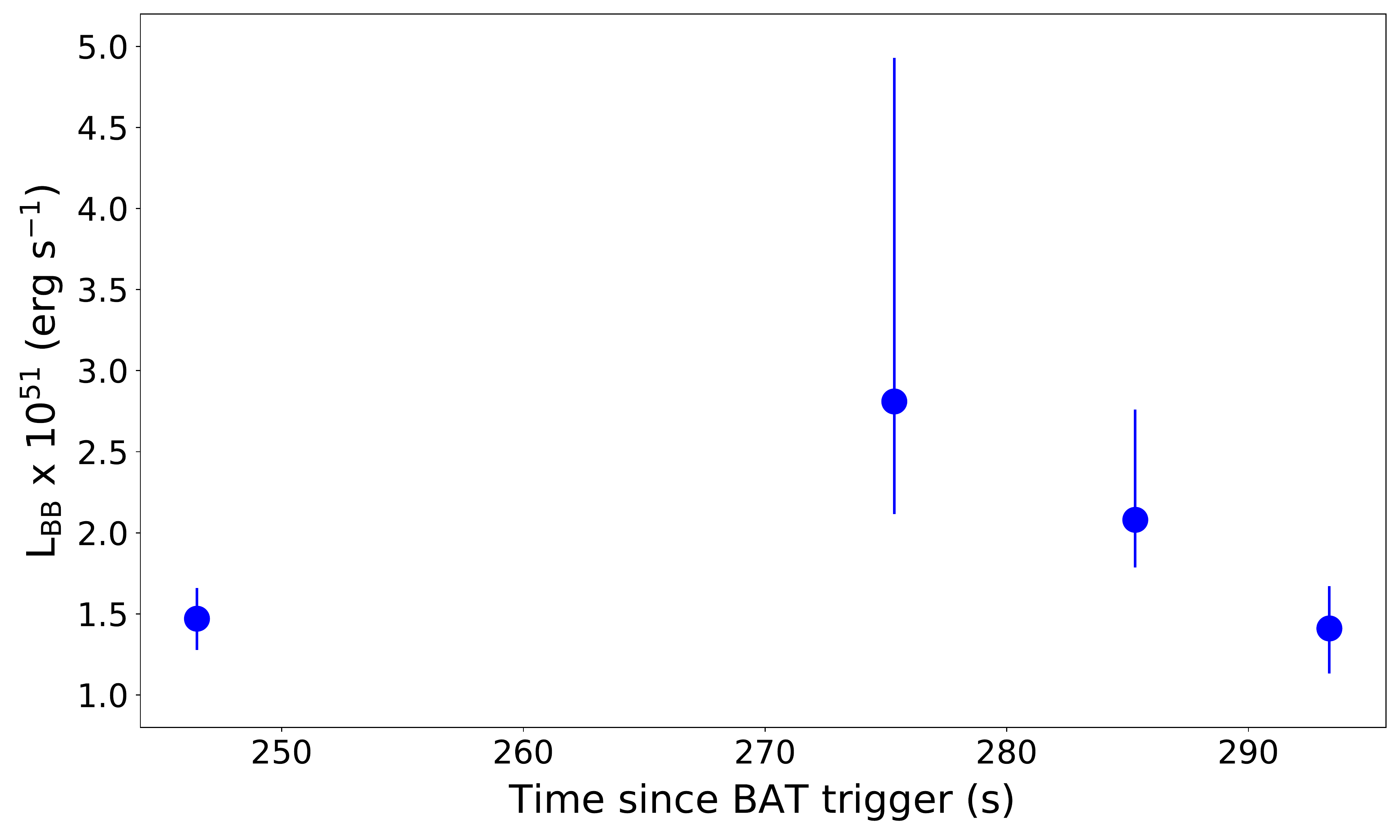}
    \end{subfigure}
    \begin{subfigure}[b]{0.49\textwidth}
     \includegraphics[width=\columnwidth, height = 5cm]{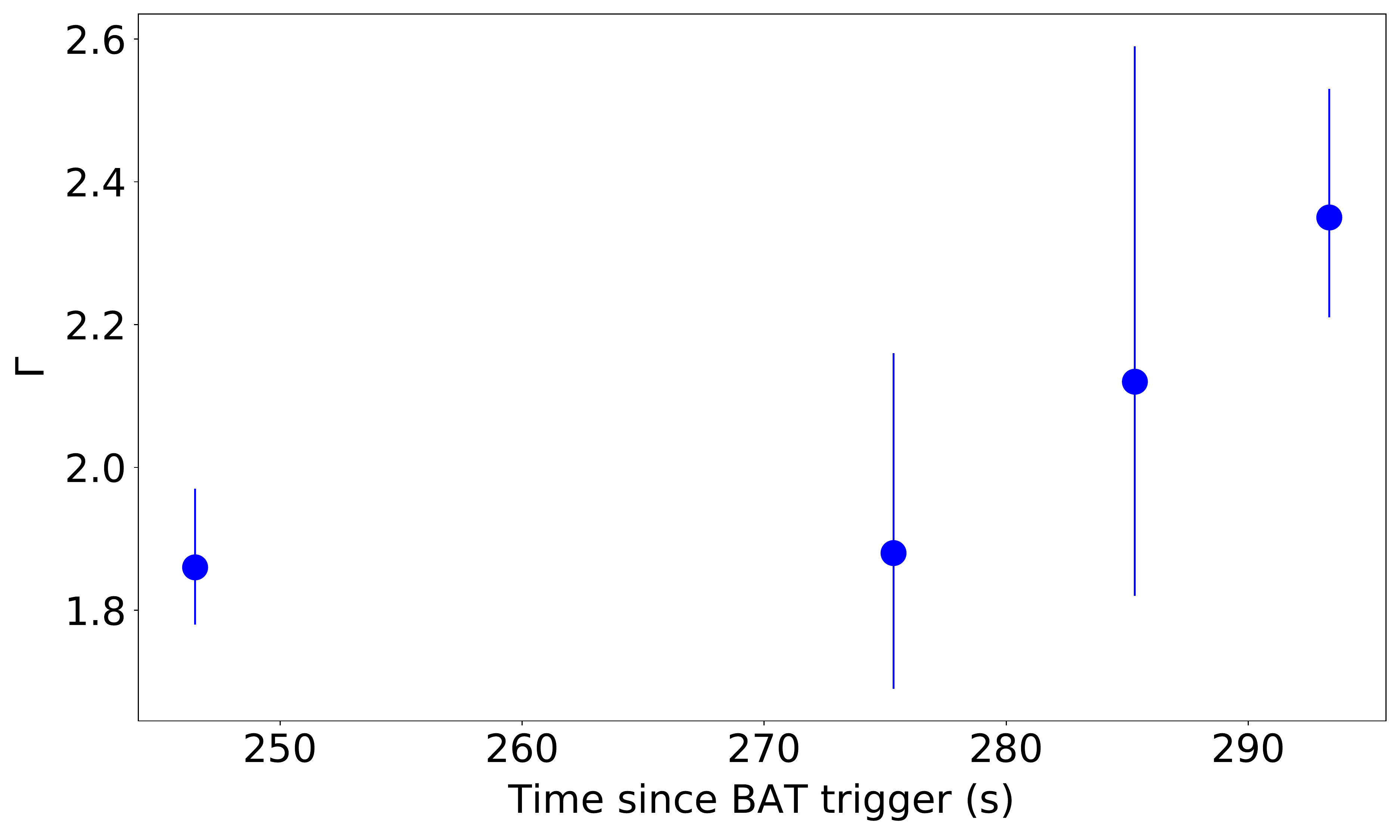}
    \end{subfigure}
    \caption{Same as in Fig. \ref{050724res} but for GRB~090516.}
	\label{090516res}
\end{figure*}

\begin{figure*}
    \begin{subfigure}[b]{0.49\textwidth}
    \includegraphics[width=\columnwidth, height = 
    7.6cm]{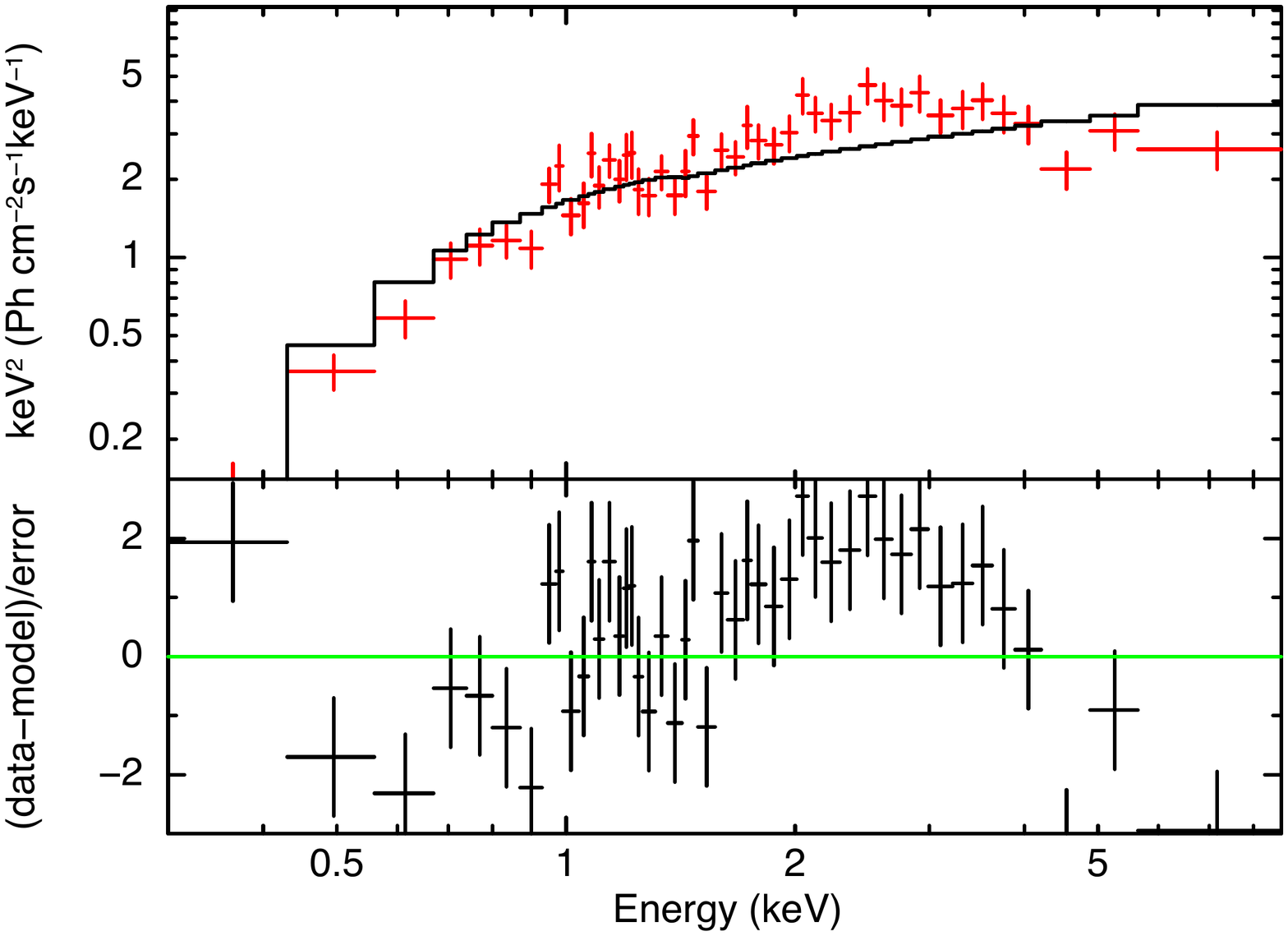}
    \end{subfigure}
    \begin{subfigure}[b]{0.49\textwidth}
     \includegraphics[width=\columnwidth, height = 7.6cm]{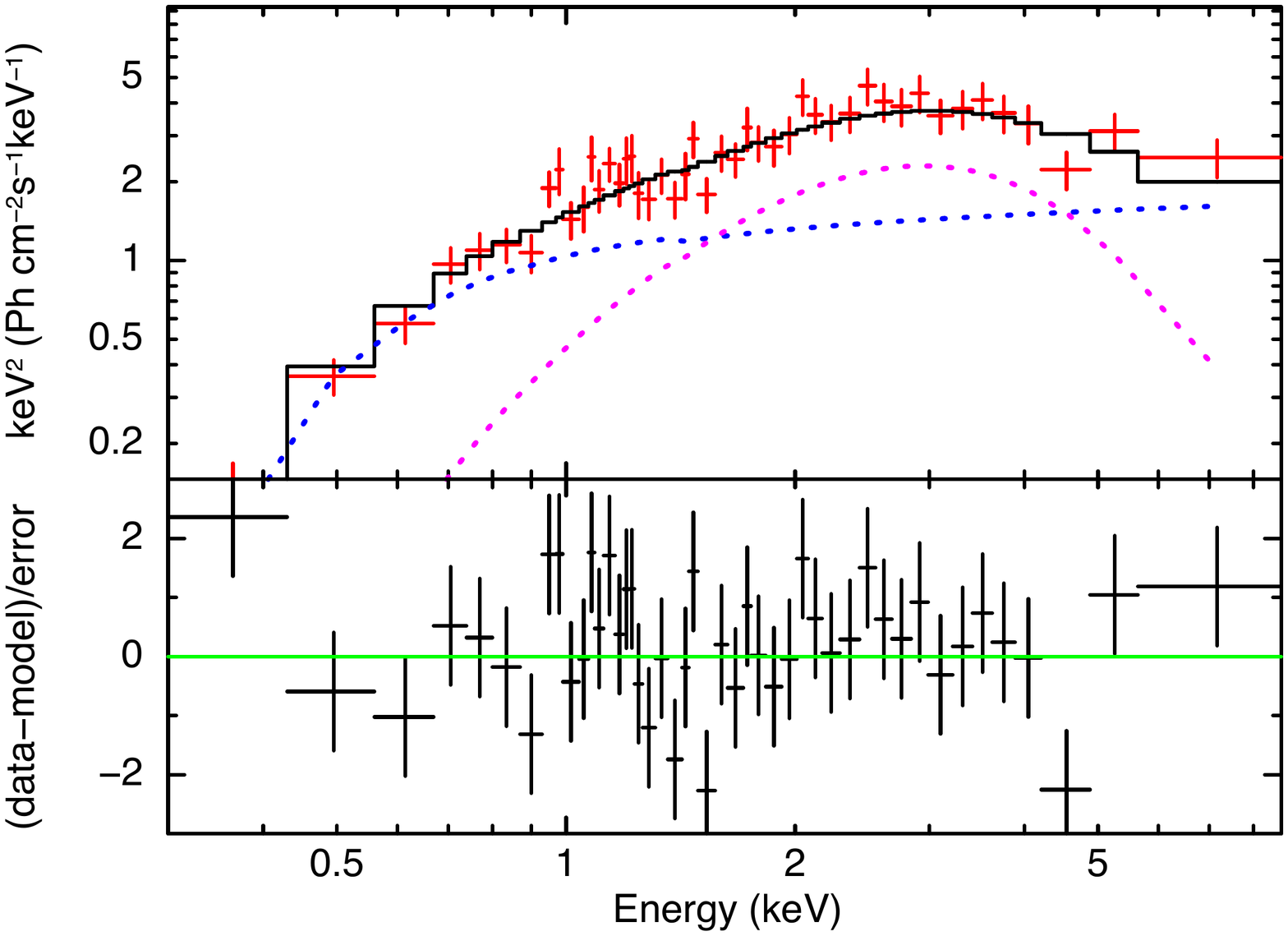}
    \end{subfigure}
    \caption{Same as in Fig. \ref{050724xspec} but for GRB 090516 in the time interval 267 - 283 s. The data have been rebinned for visual clarity. }
	\label{090516xspec}
\end{figure*}

\begin{figure*}
    \begin{subfigure}[b]{0.49\textwidth}
    \includegraphics[width=\columnwidth,  height = 5cm]{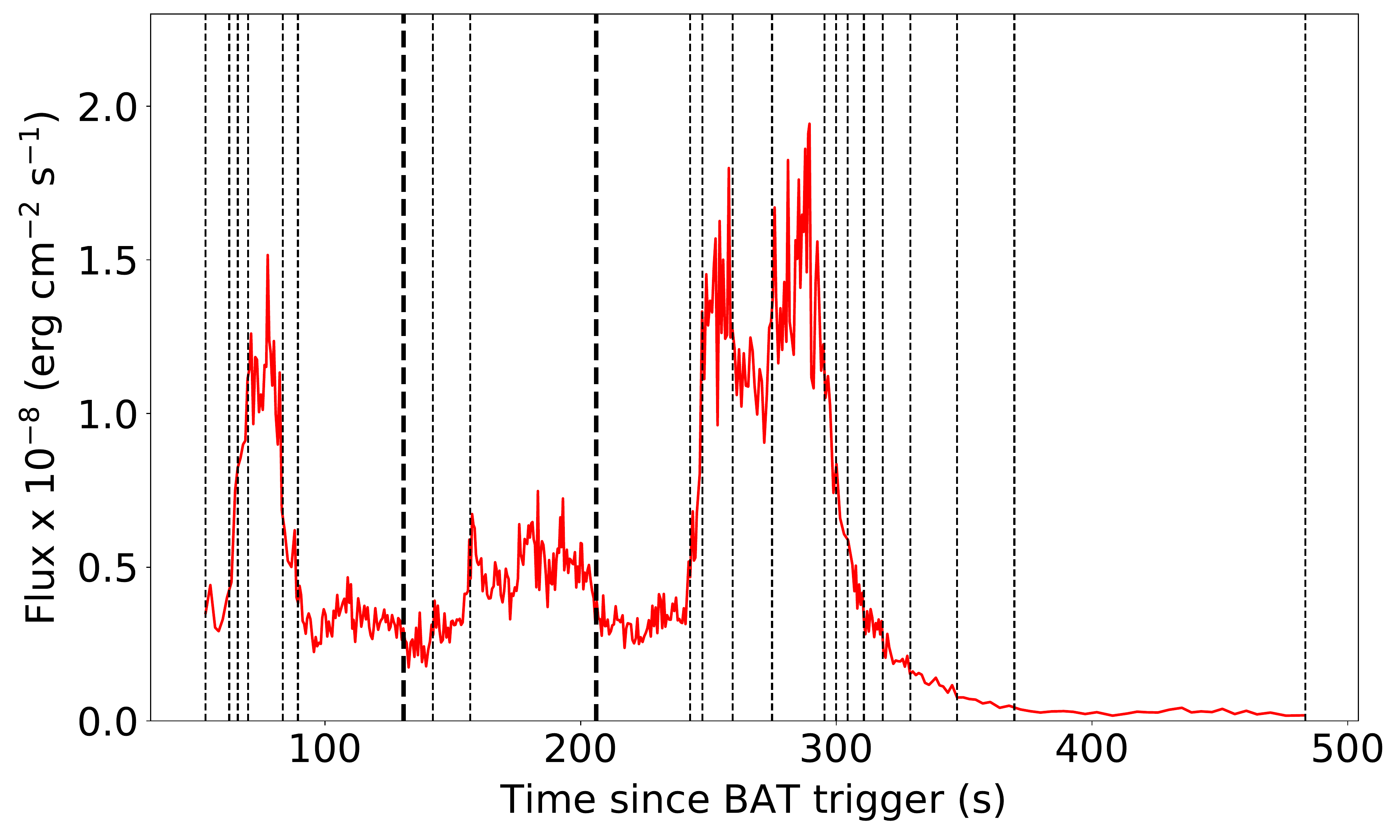}
    \end{subfigure}
    \begin{subfigure}[b]{0.49\textwidth}
    \includegraphics[width=\columnwidth,  height = 5cm]{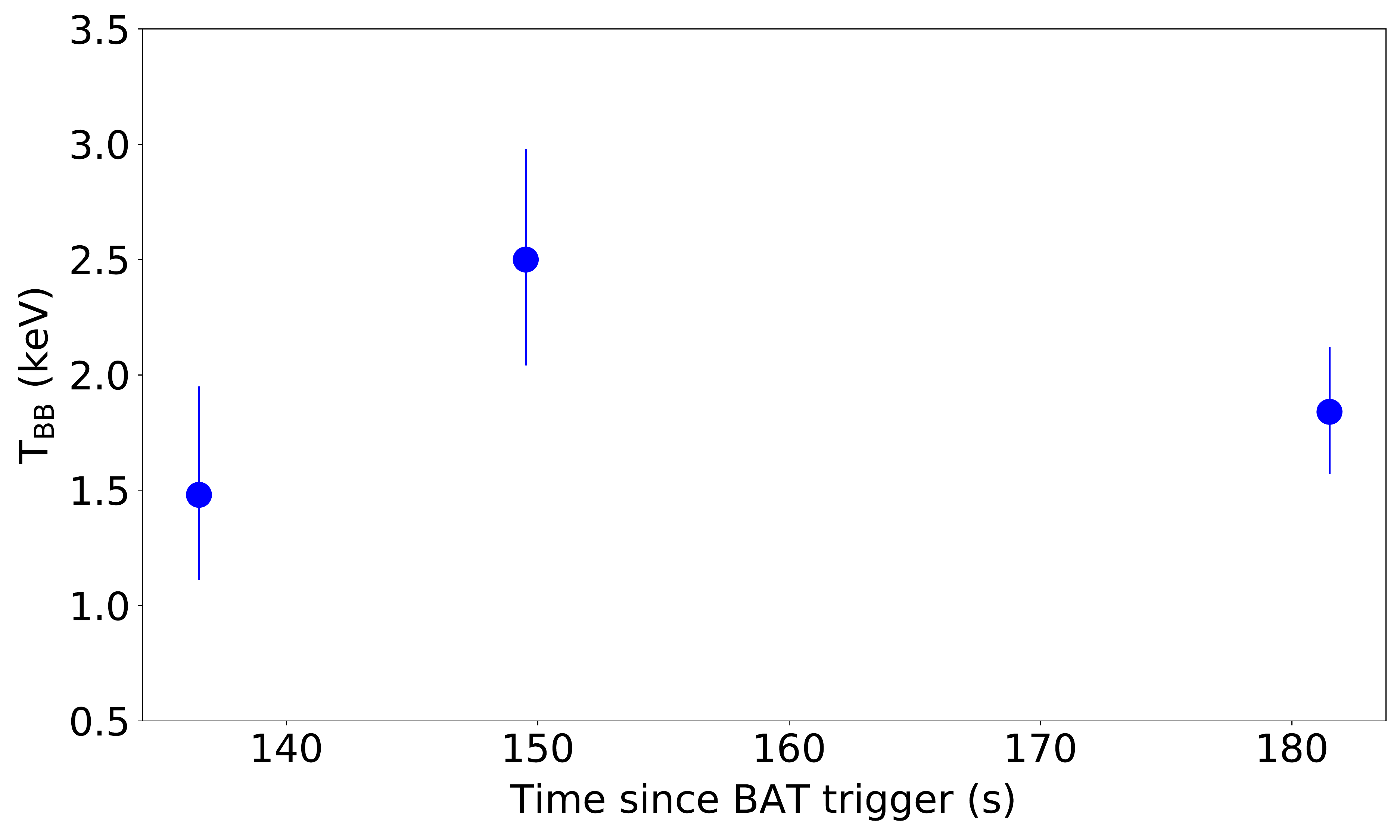}
    \end{subfigure}
    \begin{subfigure}[b]{0.49\textwidth}
     \includegraphics[width=\columnwidth, height = 5cm]{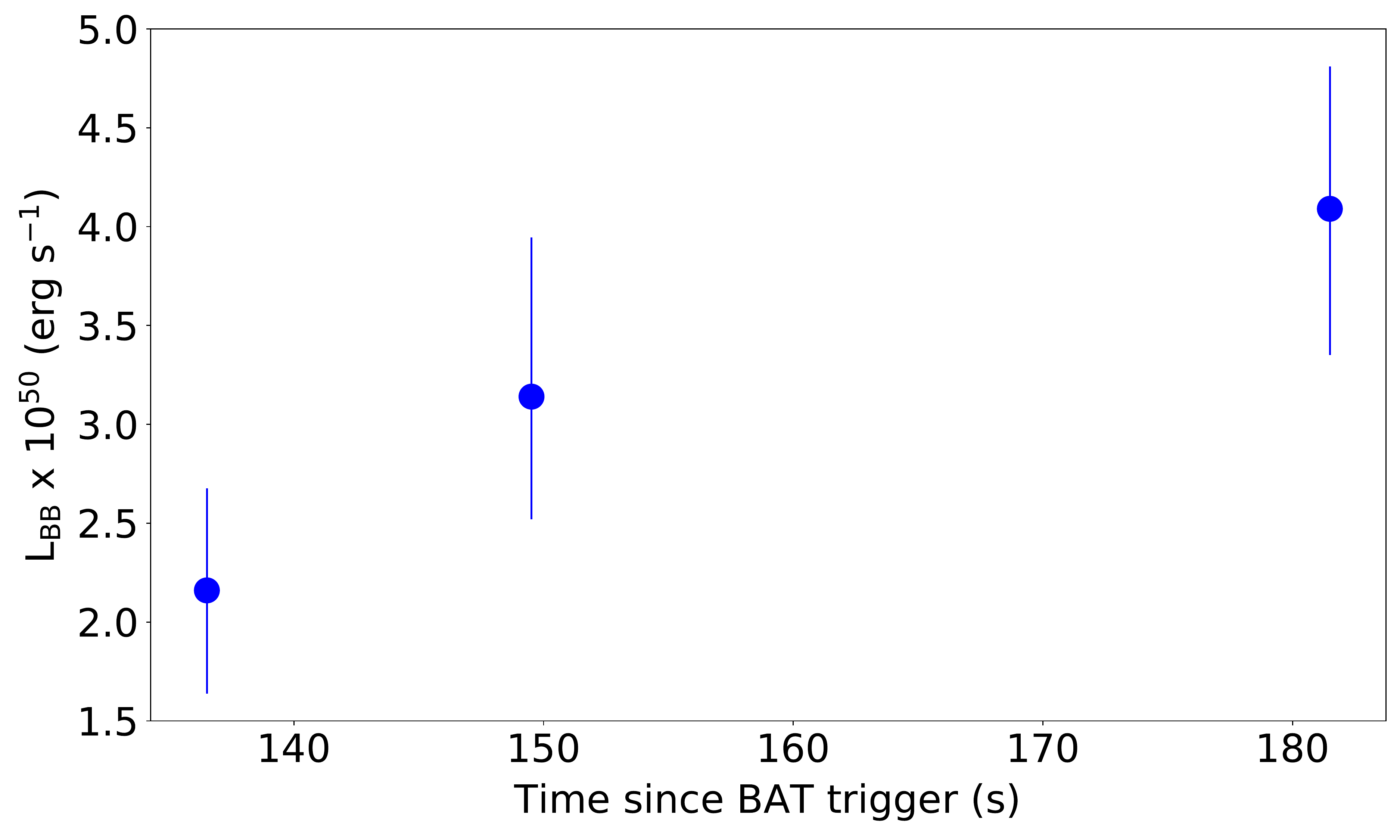}
    \end{subfigure}
    \begin{subfigure}[b]{0.49\textwidth}
     \includegraphics[width=\columnwidth, height = 5cm]{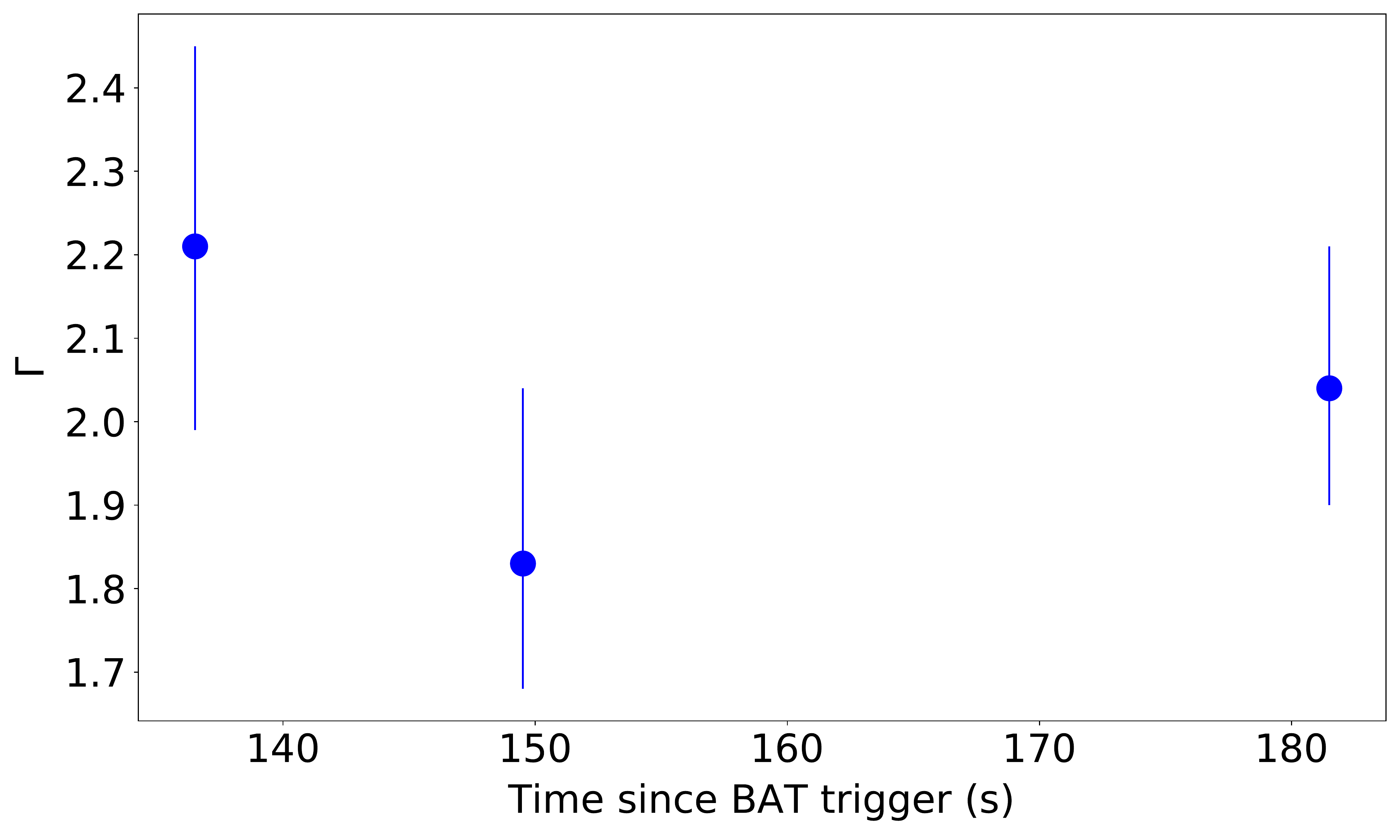}
    \end{subfigure}
    \caption{Same as in Fig. \ref{050724res} but for GRB~090715B.}
	\label{090517res}
\end{figure*}

\begin{figure*}
    \begin{subfigure}[b]{0.49\textwidth}
    \includegraphics[width=\columnwidth, height = 7.6cm]{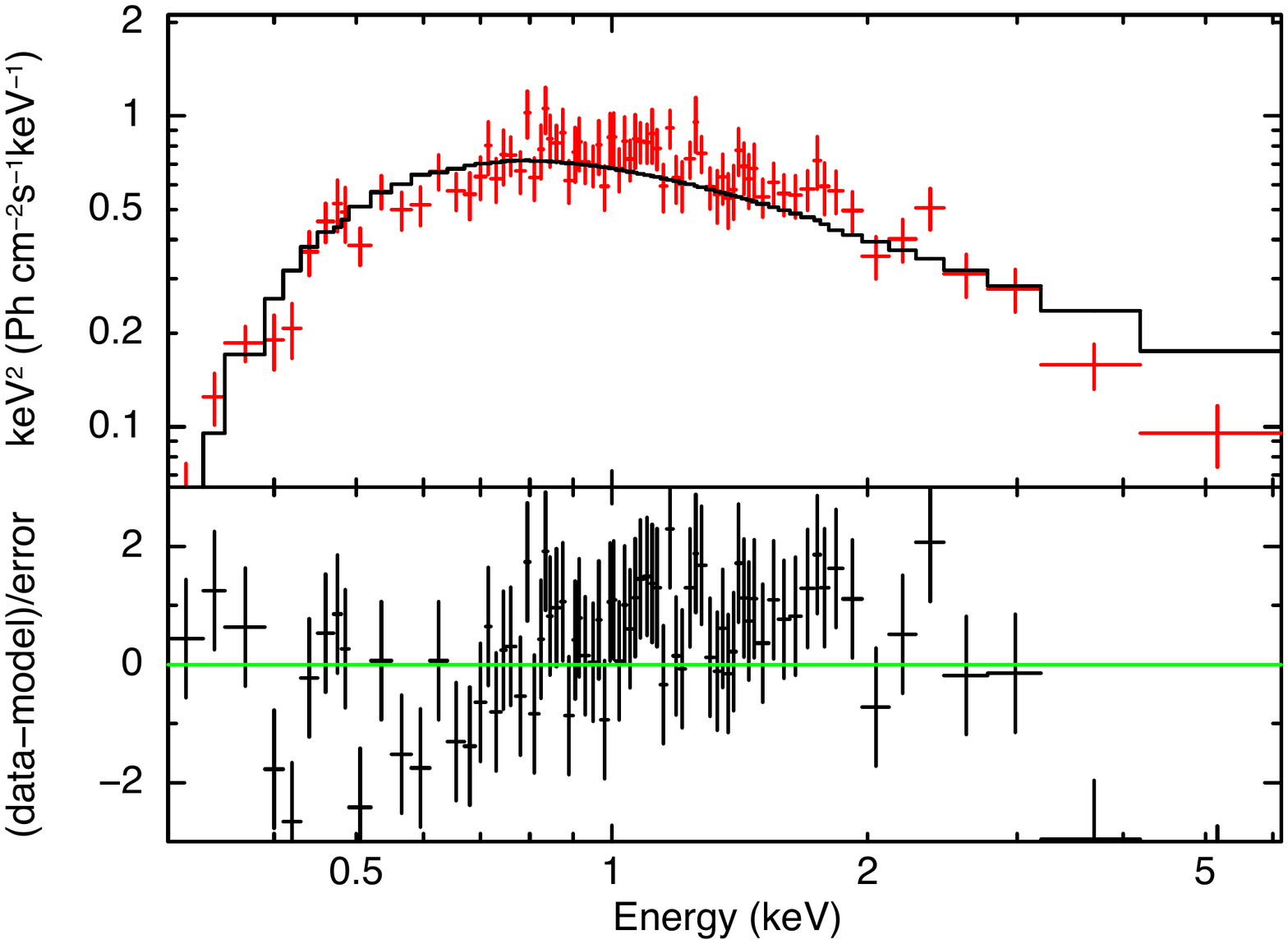}
    \end{subfigure}
    \begin{subfigure}[b]{0.49\textwidth}
     \includegraphics[width=\columnwidth, height = 7.6cm]{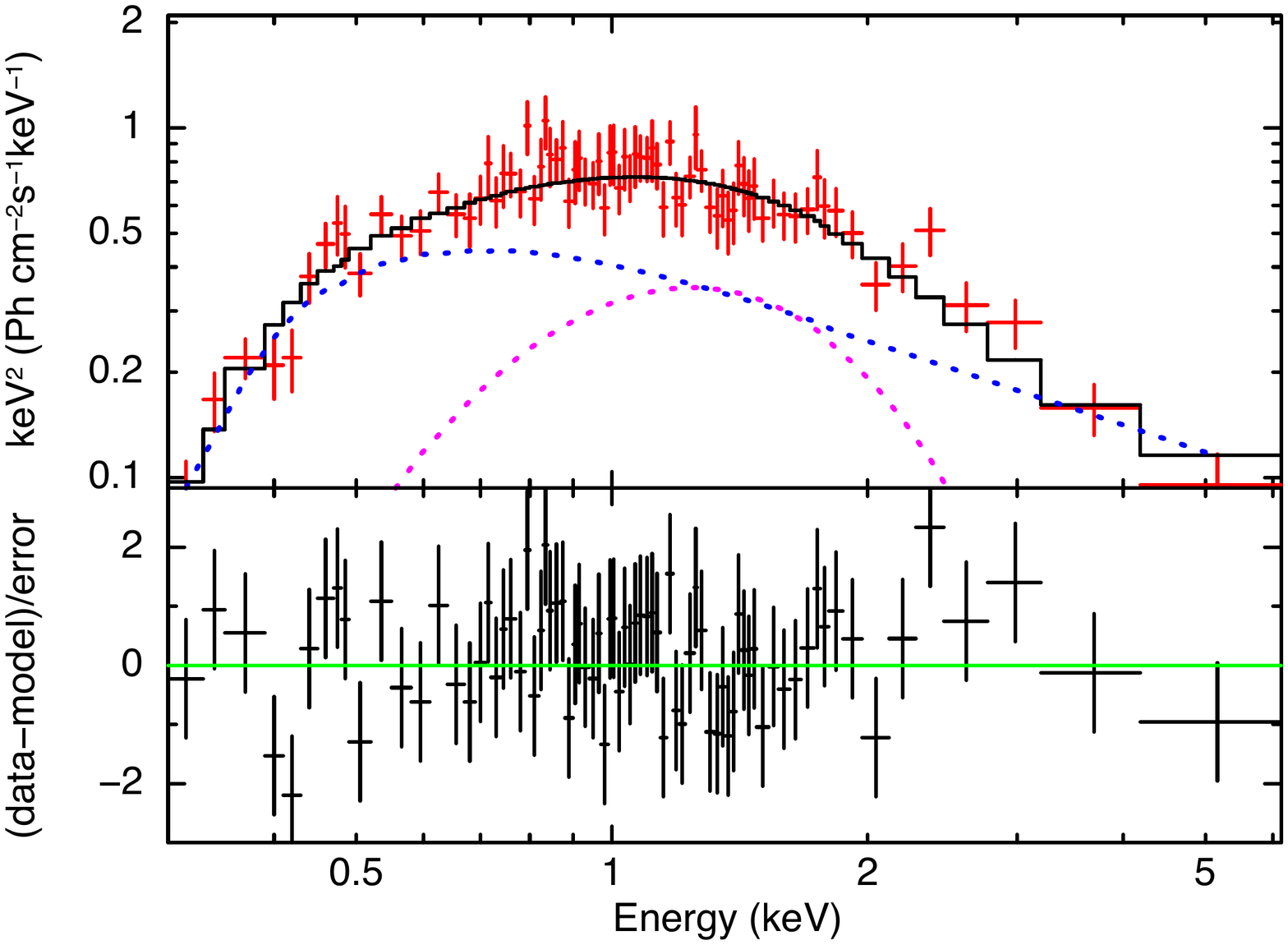}
    \end{subfigure}
    \caption{Same as in Fig. \ref{050724xspec} but for GRB 090715B in the time interval 206 - 242 s. The data have been rebinned for visual clarity.}
	\label{090715xspec}
\end{figure*}

\begin{figure*}
    \begin{subfigure}[b]{0.49\textwidth}
    \includegraphics[width=\columnwidth,  height = 5cm]{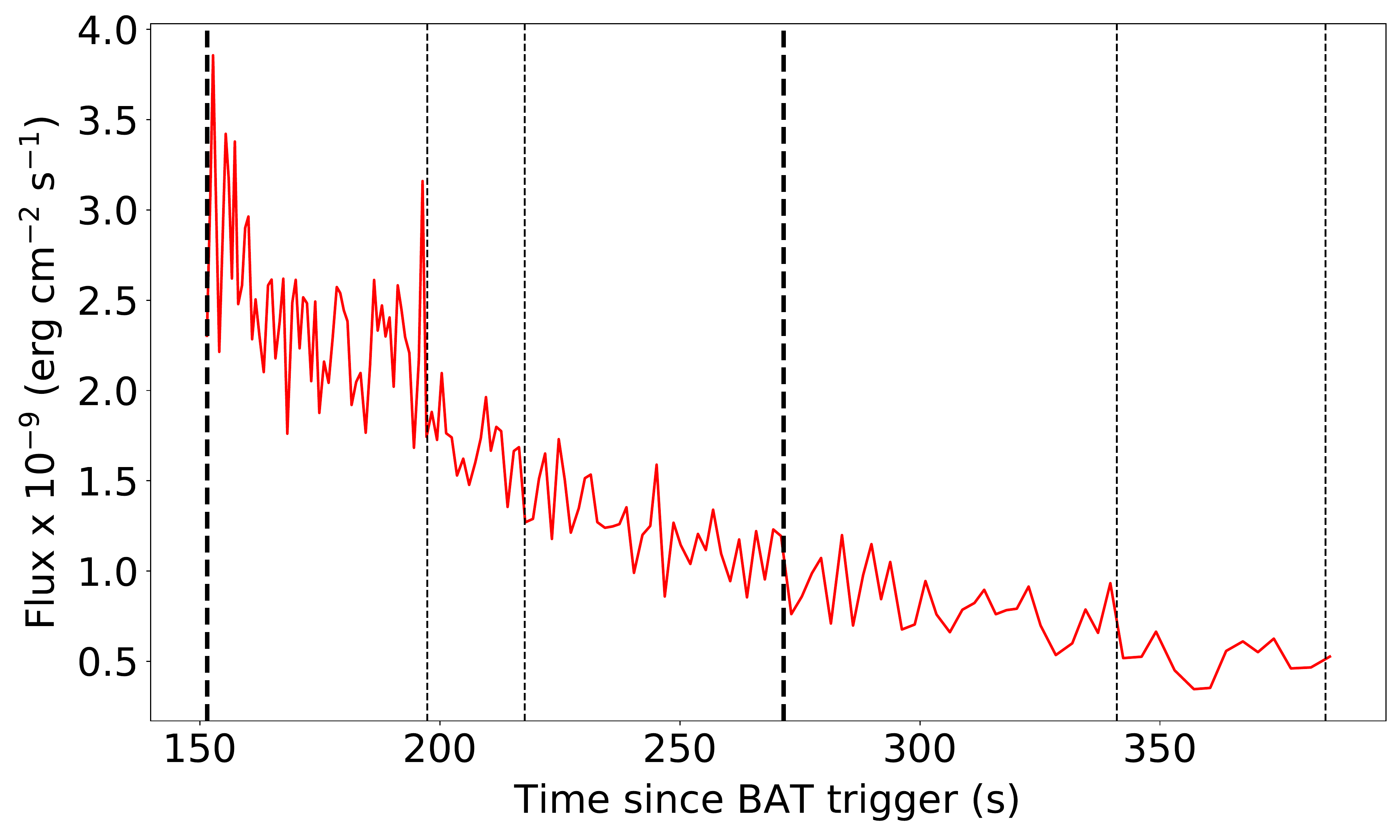}
    \end{subfigure}
    \begin{subfigure}[b]{0.49\textwidth}
    \includegraphics[width=1.1\columnwidth,  height = 5.5
cm]{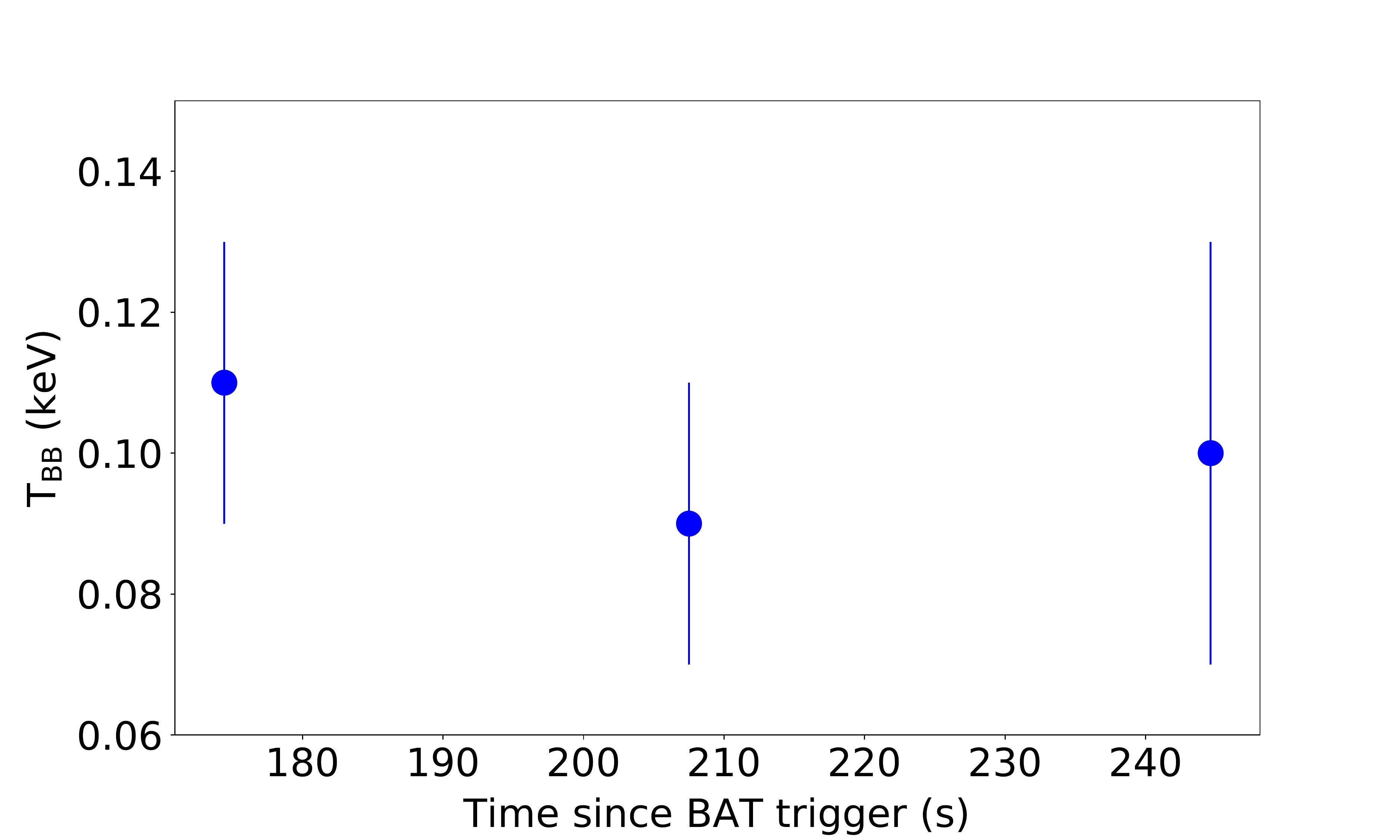}
    \end{subfigure}
    \begin{subfigure}[b]{0.49\textwidth}
     \includegraphics[width=\columnwidth, height = 5cm]{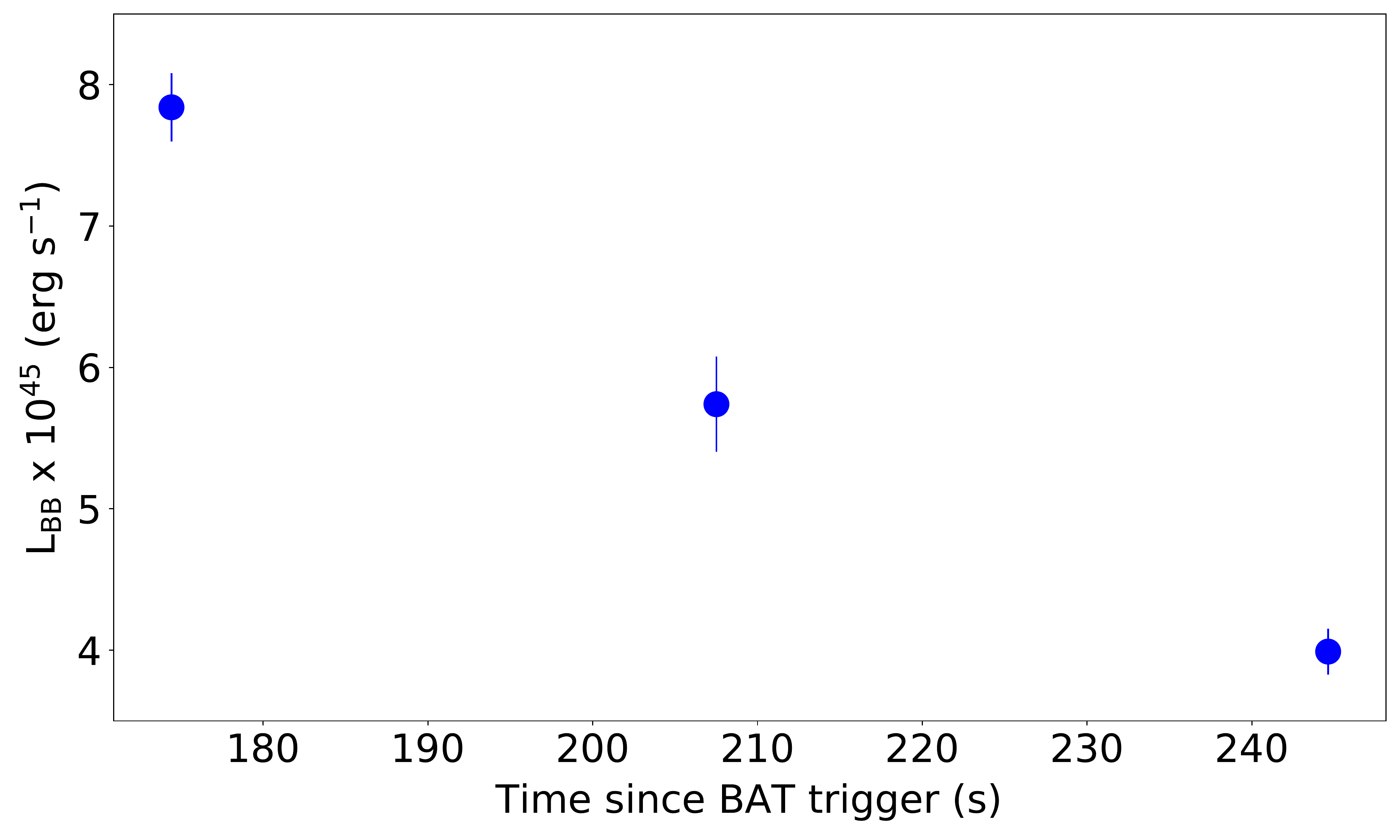}
    \end{subfigure}
    \begin{subfigure}[b]{0.49\textwidth}
     \includegraphics[width=\columnwidth, height = 5cm]{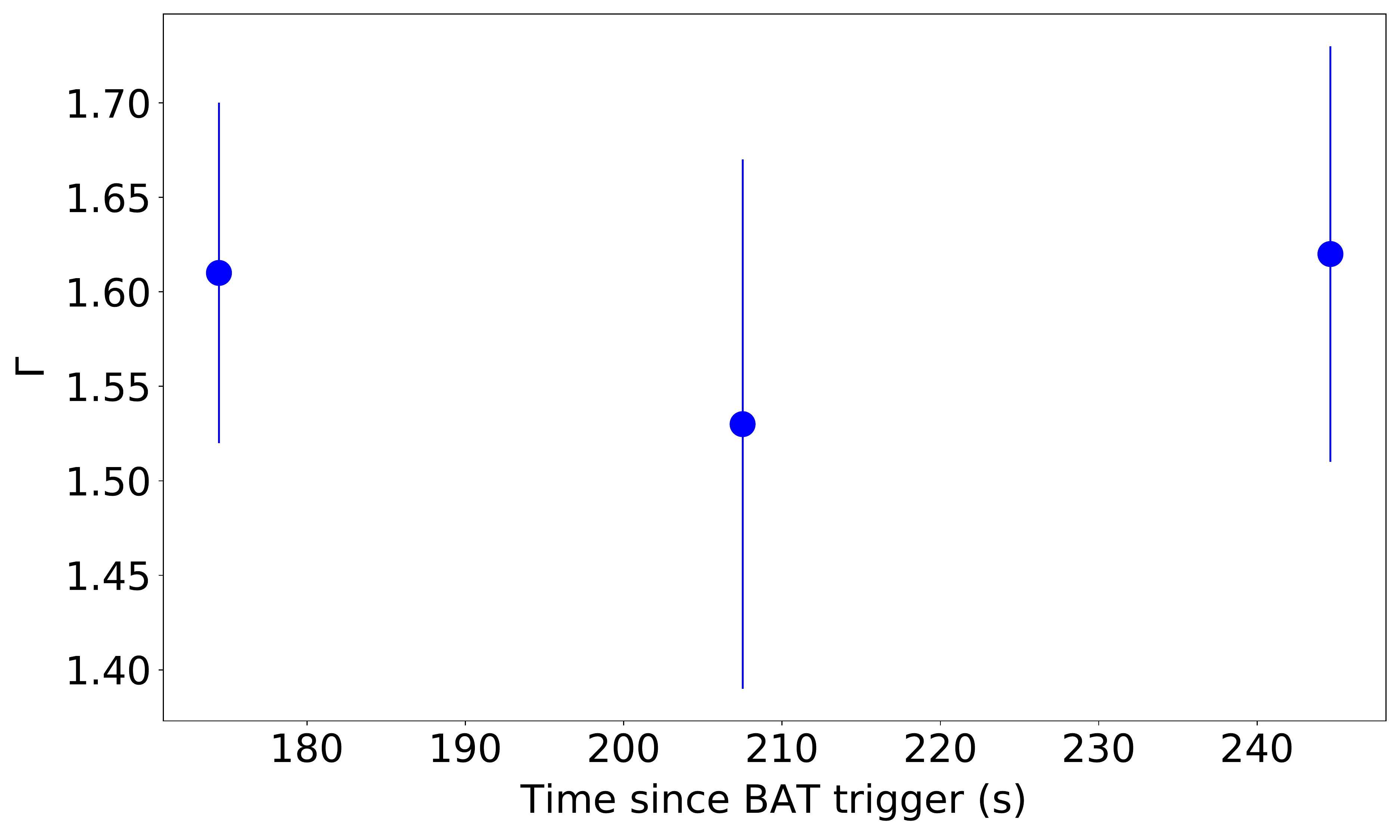}
    \end{subfigure}
    \caption{ Same as in Fig. \ref{050724res} but for GRB~171205A.}
	\label{171205res}
\end{figure*}

\begin{figure*}
    \begin{subfigure}[b]{0.49\textwidth}
    \includegraphics[width=\columnwidth, height = 7.6cm]{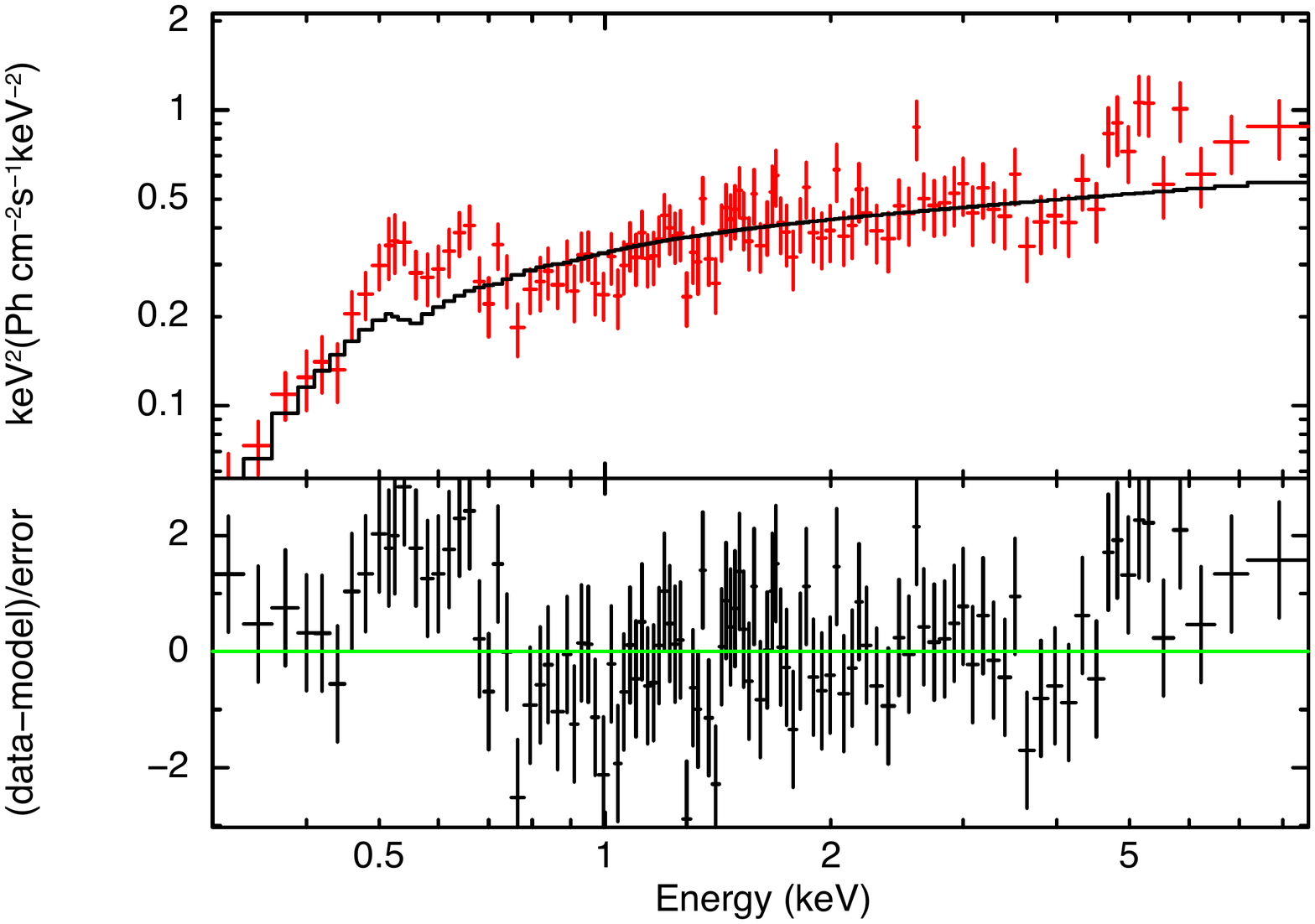}
    \end{subfigure}
    \begin{subfigure}[b]{0.49\textwidth}
     \includegraphics[width=\columnwidth, height = 7.6cm]{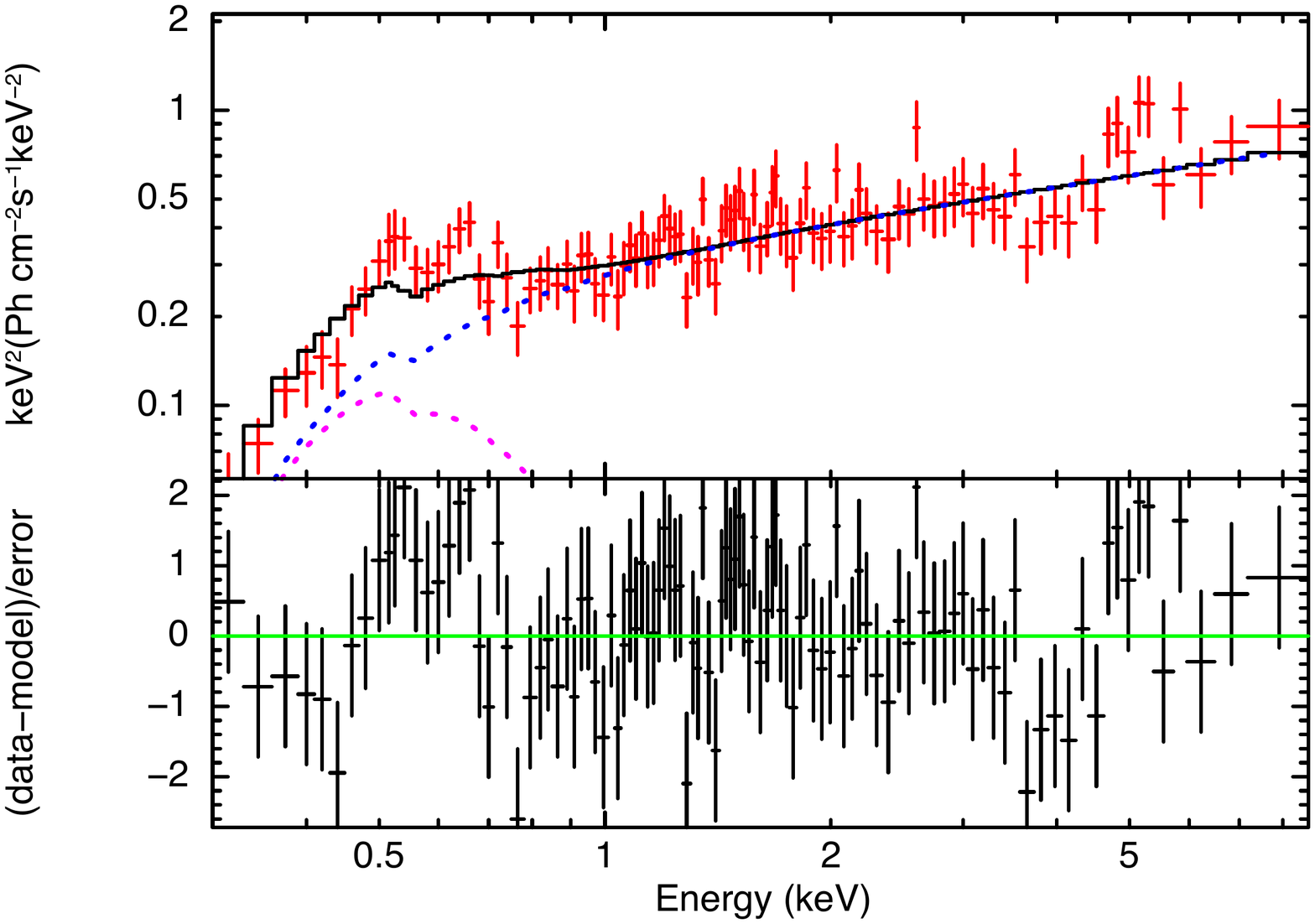}
    \end{subfigure}
    \caption{Same as in Fig. \ref{050724xspec} but for GRB 171205A in the time interval 151 - 197 s.}
	\label{171205xspec}
\end{figure*}

\begin{figure*}
    \begin{subfigure}[b]{0.49\textwidth}
    \includegraphics[width=\columnwidth,  height = 5cm]{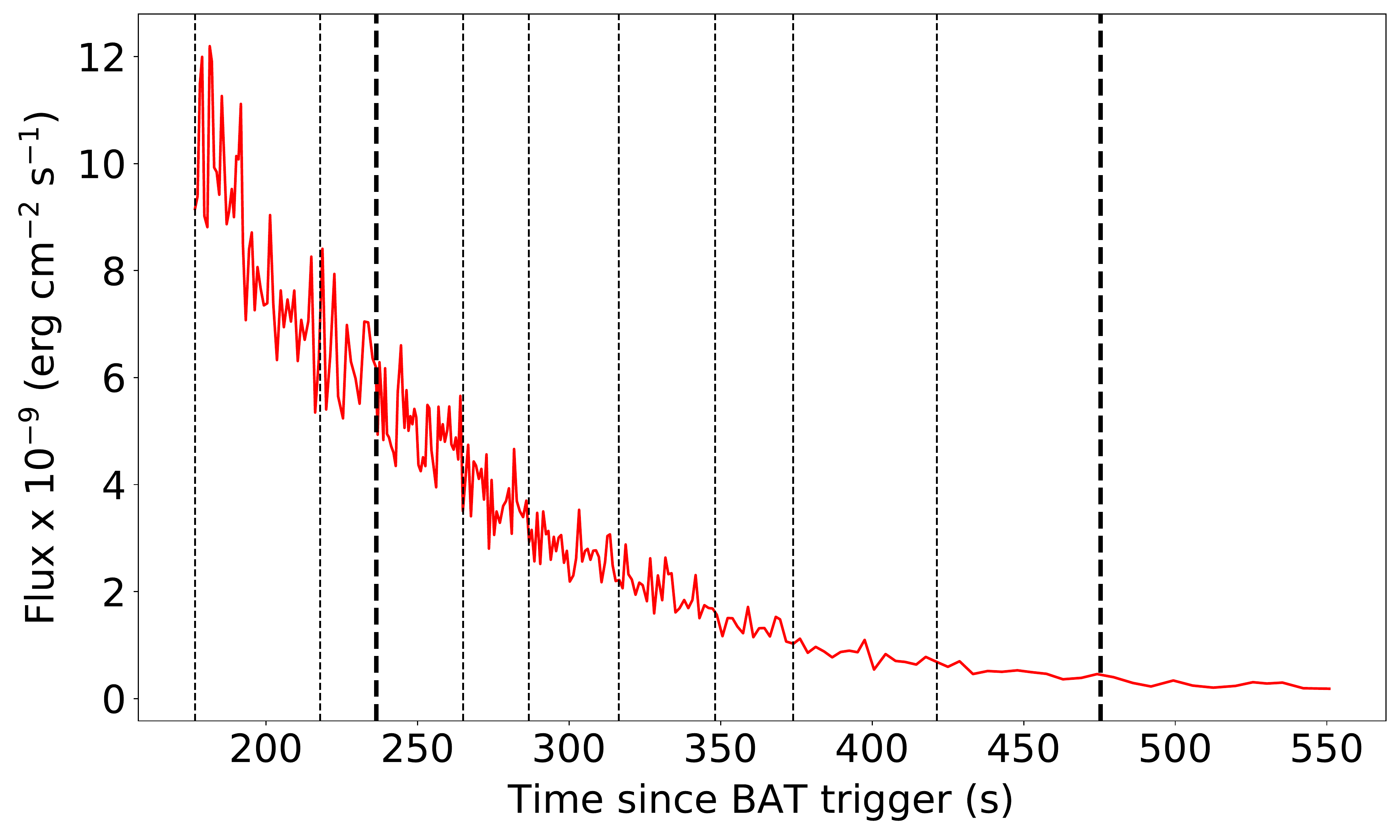}
    \end{subfigure}
    \begin{subfigure}[b]{0.5\textwidth}
    \includegraphics[width=1.1\columnwidth,  height = 5.5
cm]{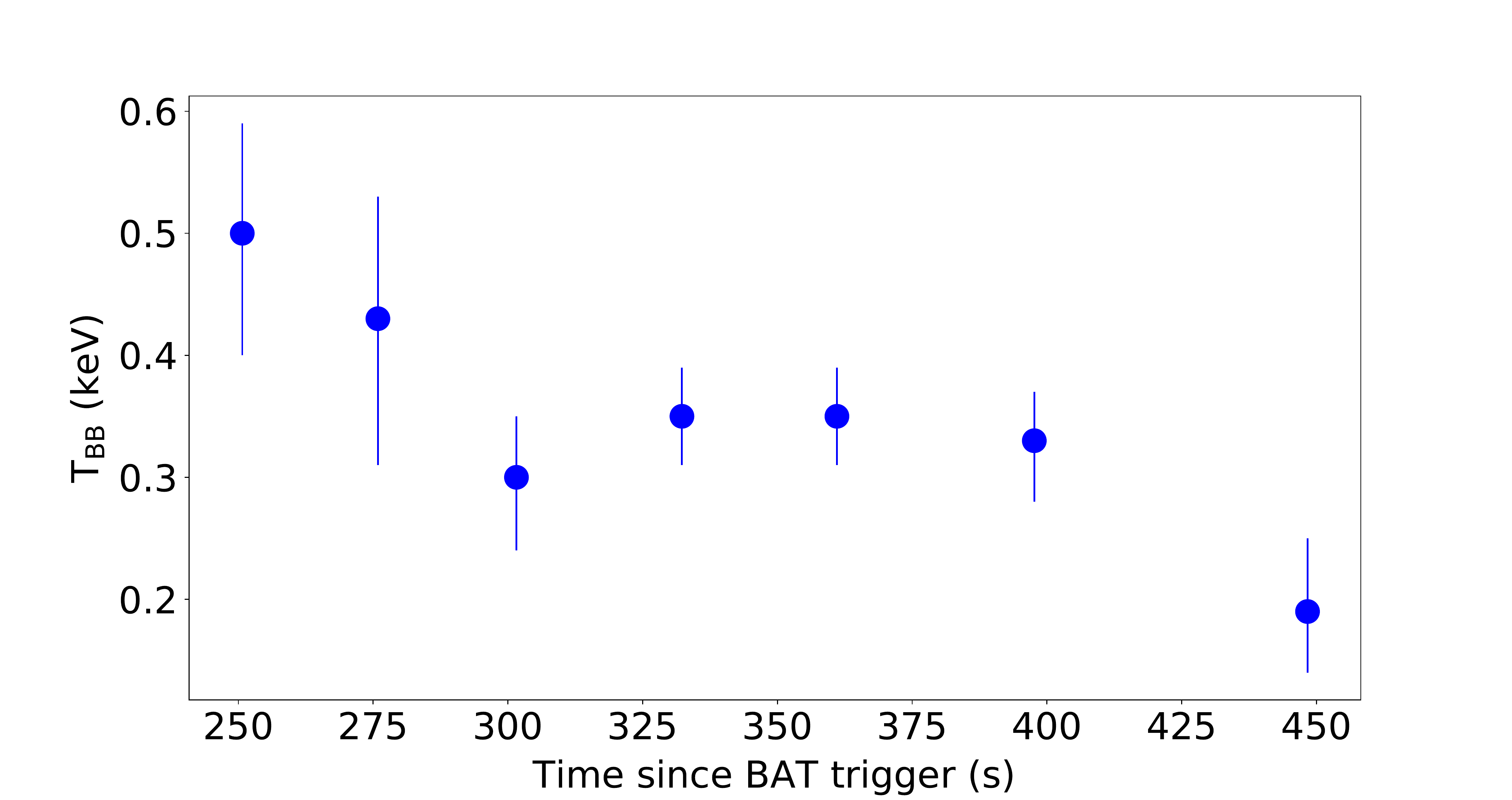}
    \end{subfigure}
    \begin{subfigure}[b]{0.49\textwidth}
     \includegraphics[width=\columnwidth, height = 5cm]{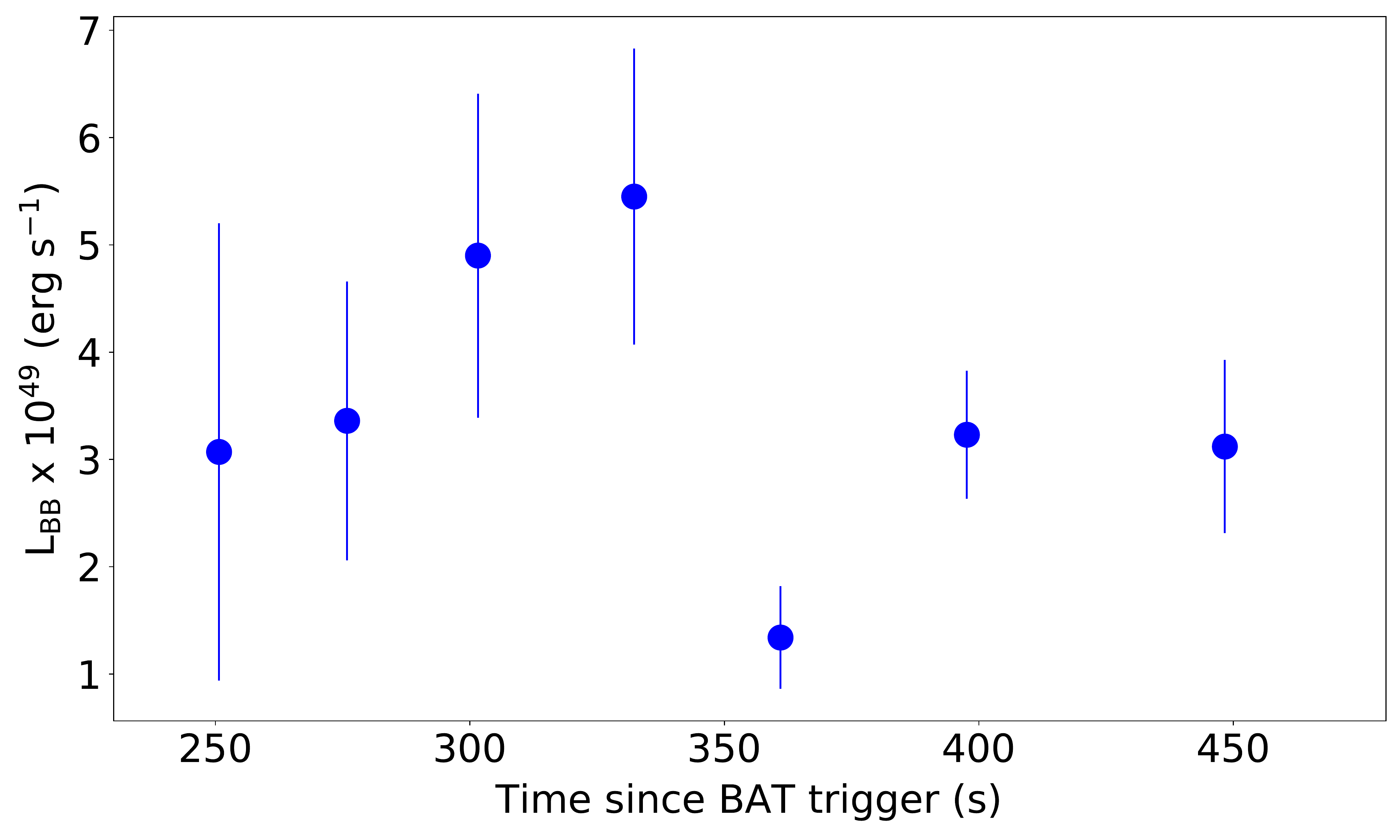}
    \end{subfigure}
    \begin{subfigure}[b]{0.49\textwidth}
     \includegraphics[width=\columnwidth, height = 5cm]{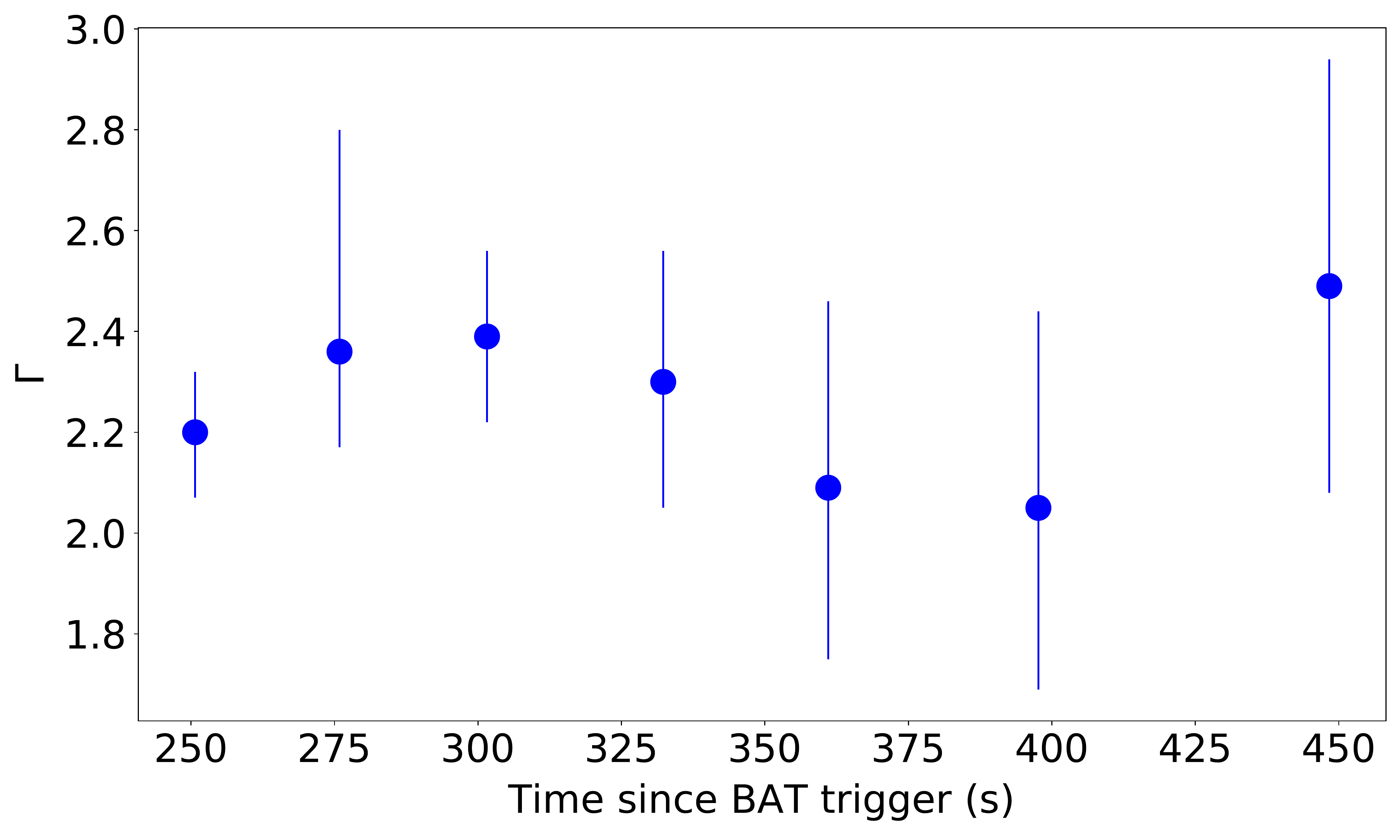}
    \end{subfigure}
    \caption{ Same as in Fig. \ref{050724res} but for GRB~171222A.}
	\label{171222res}
\end{figure*}

\begin{figure*}
    \begin{subfigure}[b]{0.49\textwidth}
    \includegraphics[width=\columnwidth, height = 7.6cm]{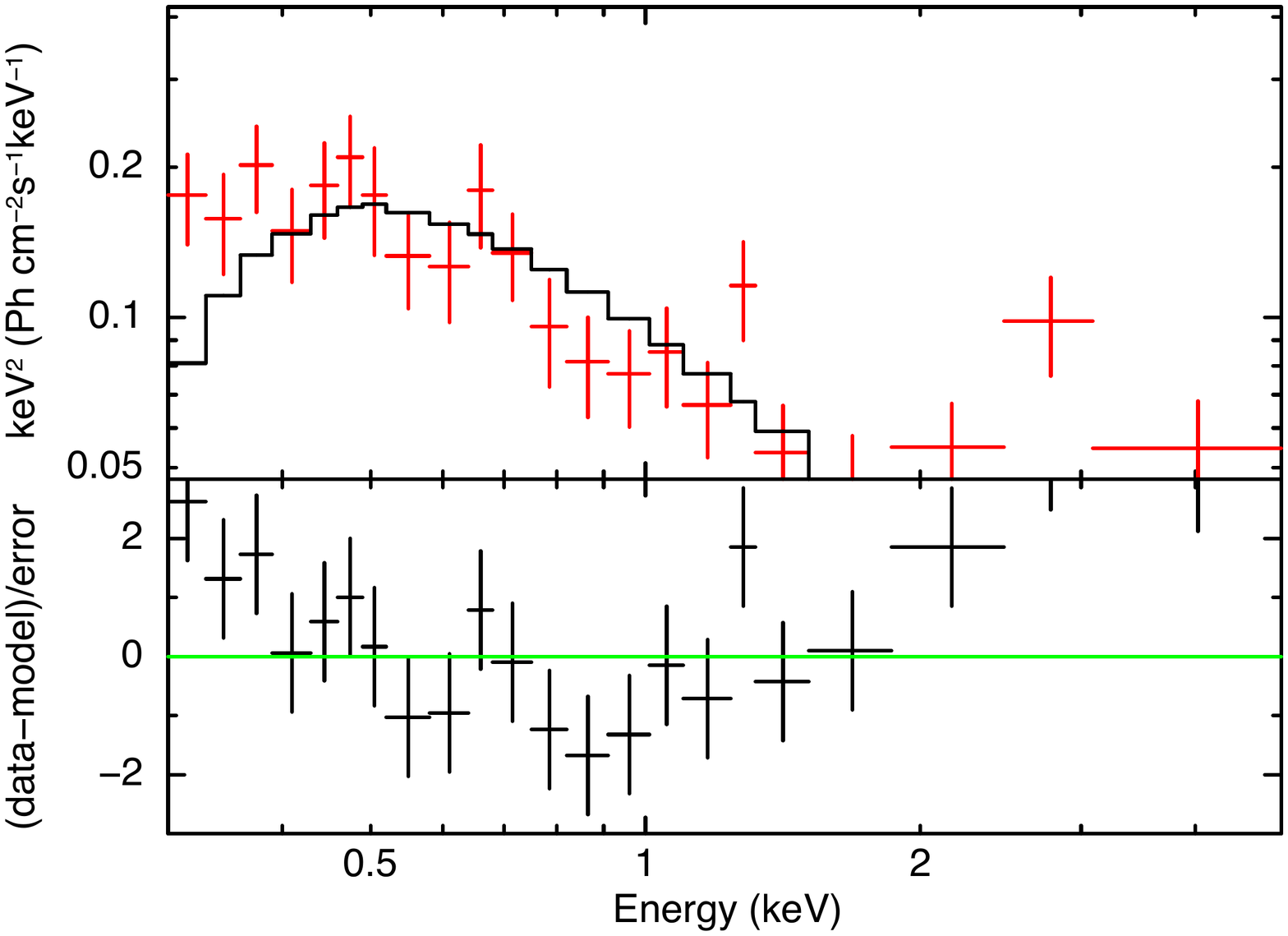}
    \end{subfigure}
    \begin{subfigure}[b]{0.49\textwidth}
     \includegraphics[width=\columnwidth, height = 7.6cm]{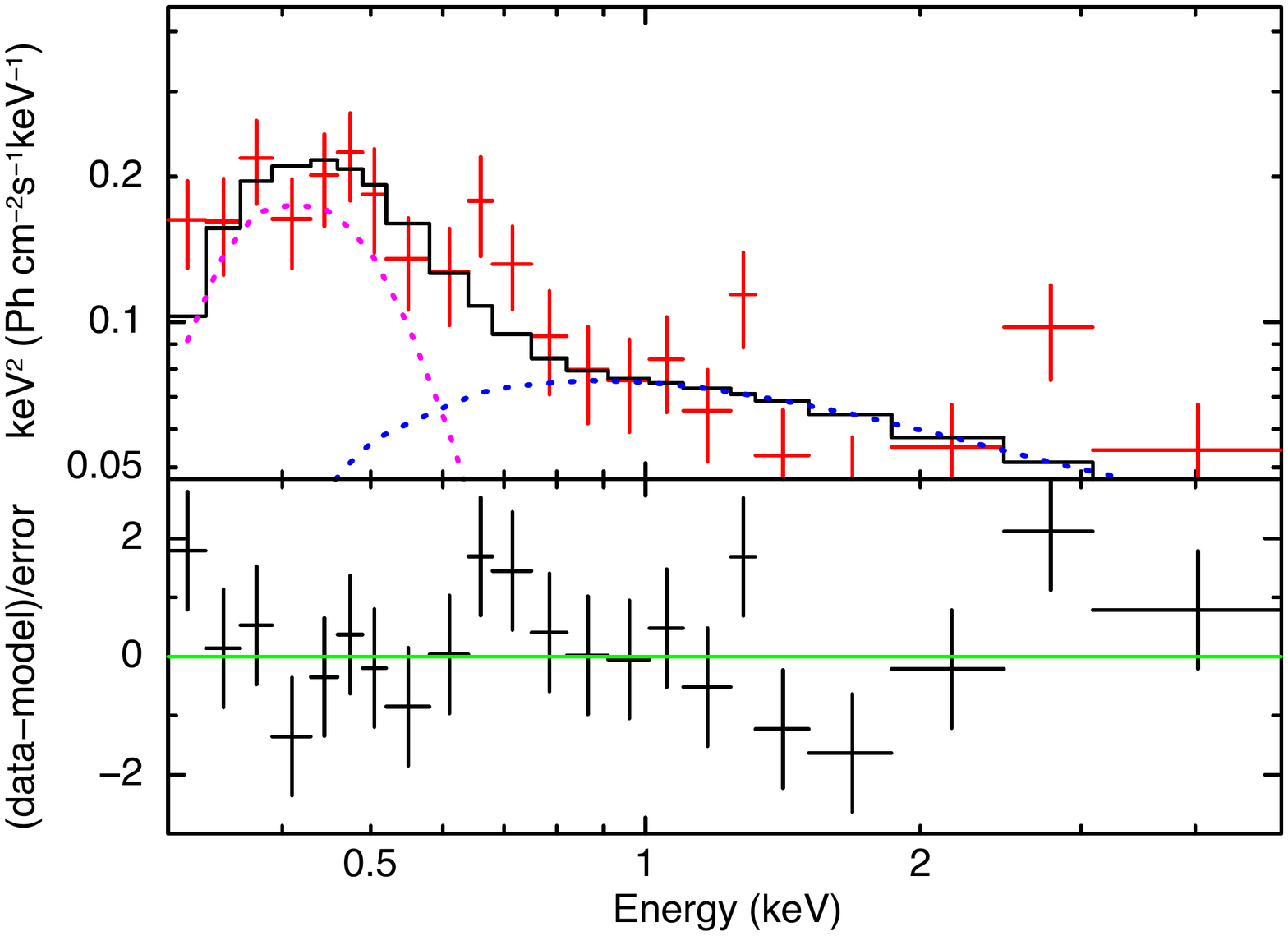}
    \end{subfigure}
    \caption{Same as in Fig. \ref{050724xspec} but for GRB 171222A in the time interval 421 - 475 s.}
	\label{171222xspec}
\end{figure*}

\begin{figure*}
    \begin{subfigure}[b]{0.49\textwidth}
     \includegraphics[width=\columnwidth, height = 5cm]{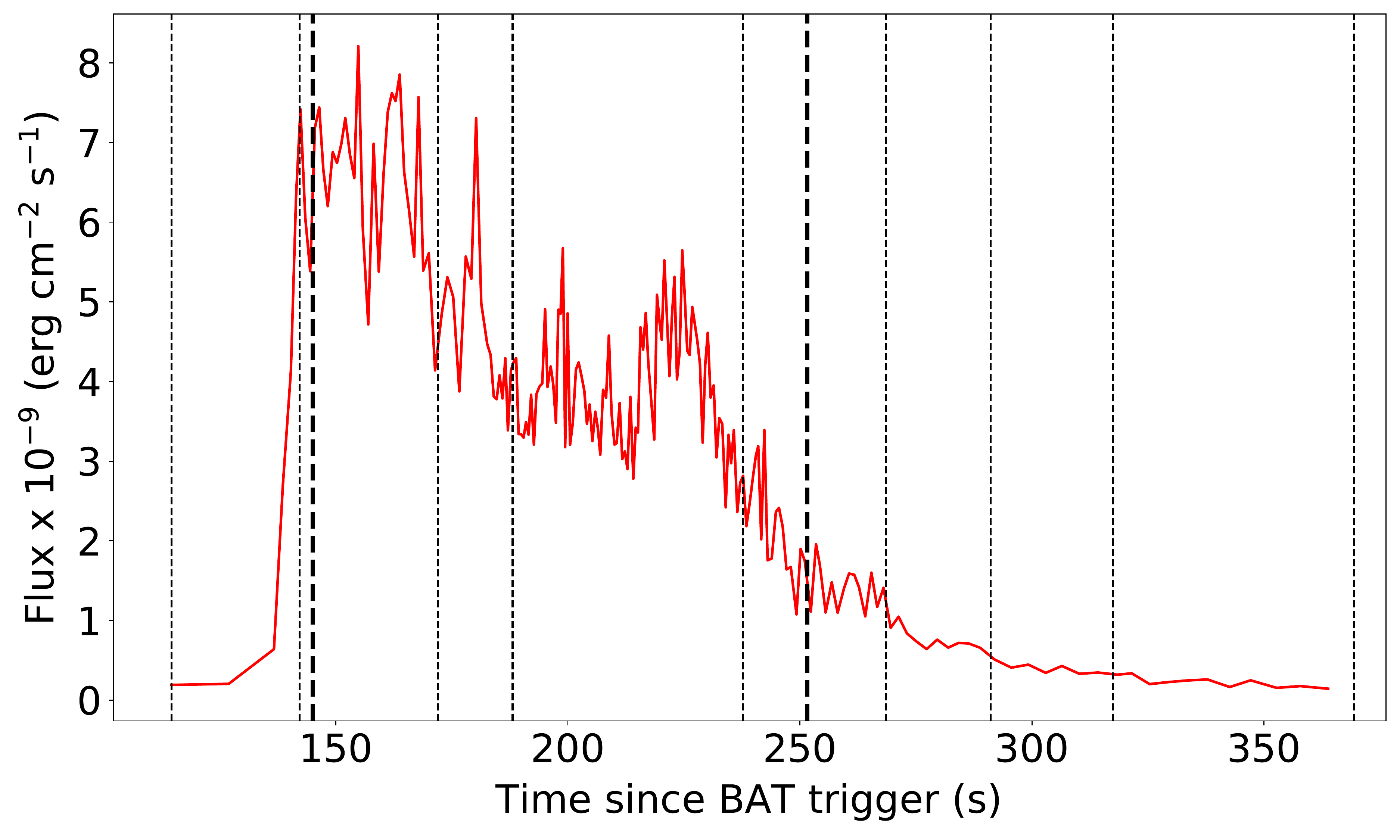}
    \end{subfigure}
    \begin{subfigure}[b]{0.49\textwidth}
    \includegraphics[width=\columnwidth,  height = 5cm]{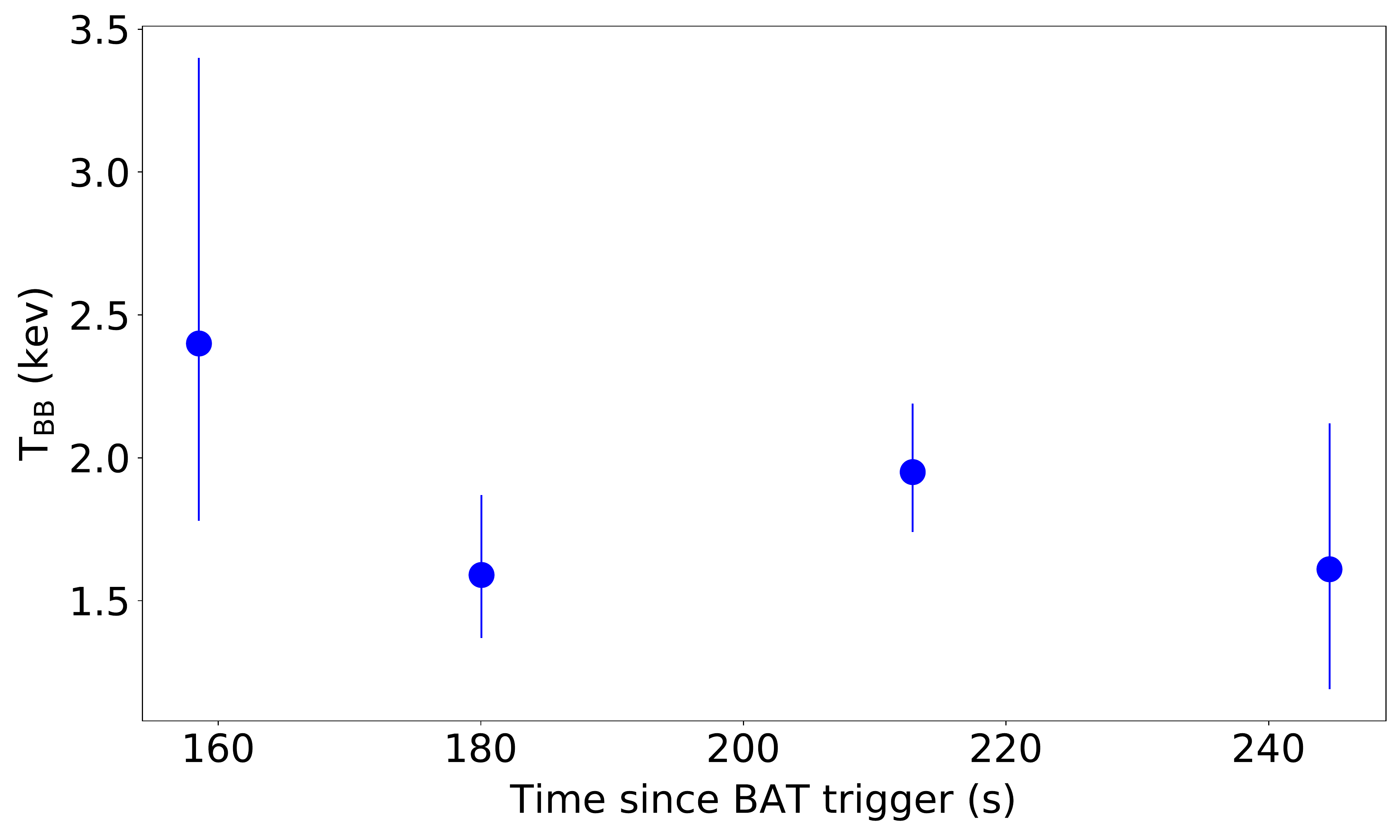}
    \end{subfigure}
    \begin{subfigure}[b]{0.49\textwidth}
     \includegraphics[width=\columnwidth, height = 5cm]{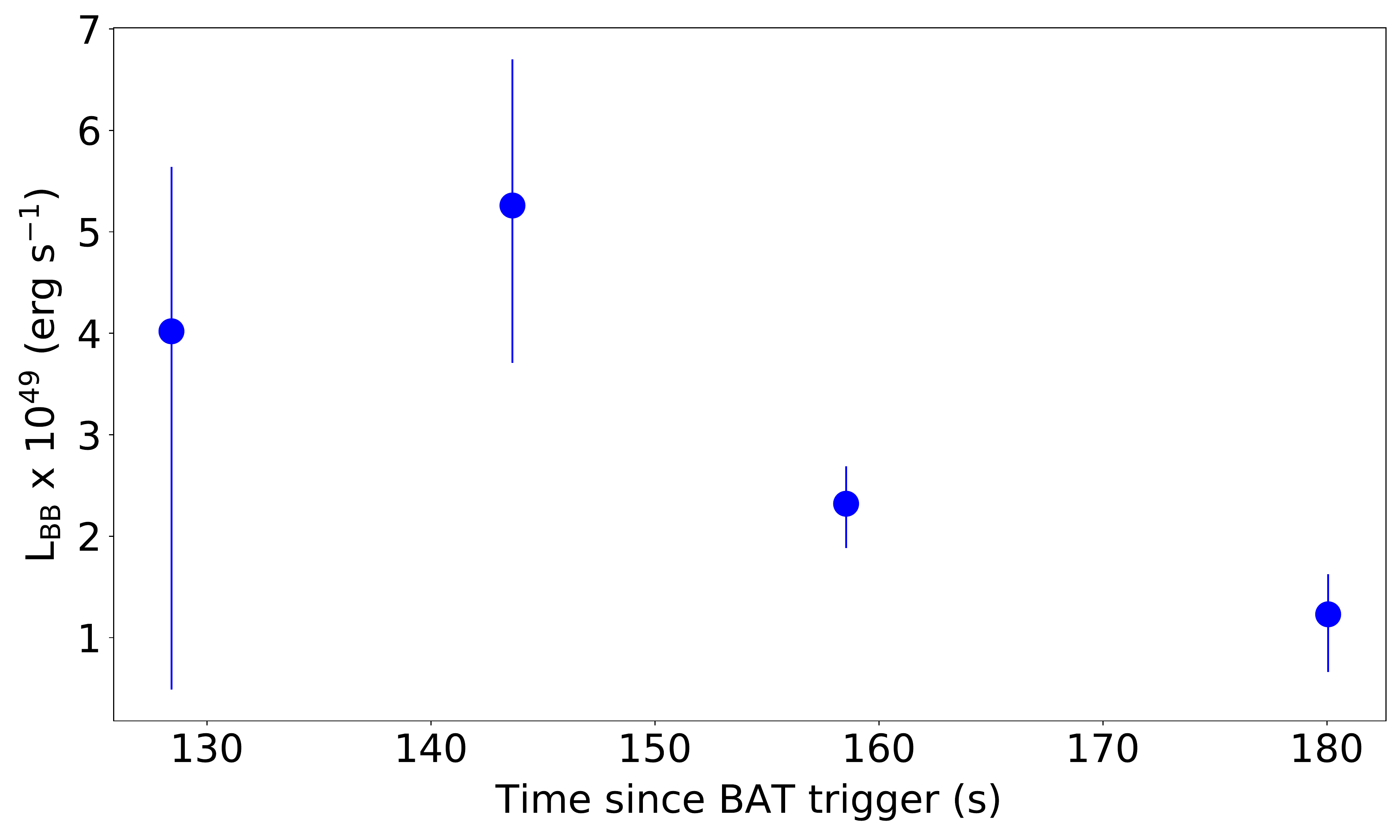}
    \end{subfigure}
    \begin{subfigure}[b]{0.49\textwidth}
     \includegraphics[width=\columnwidth, height = 5cm]{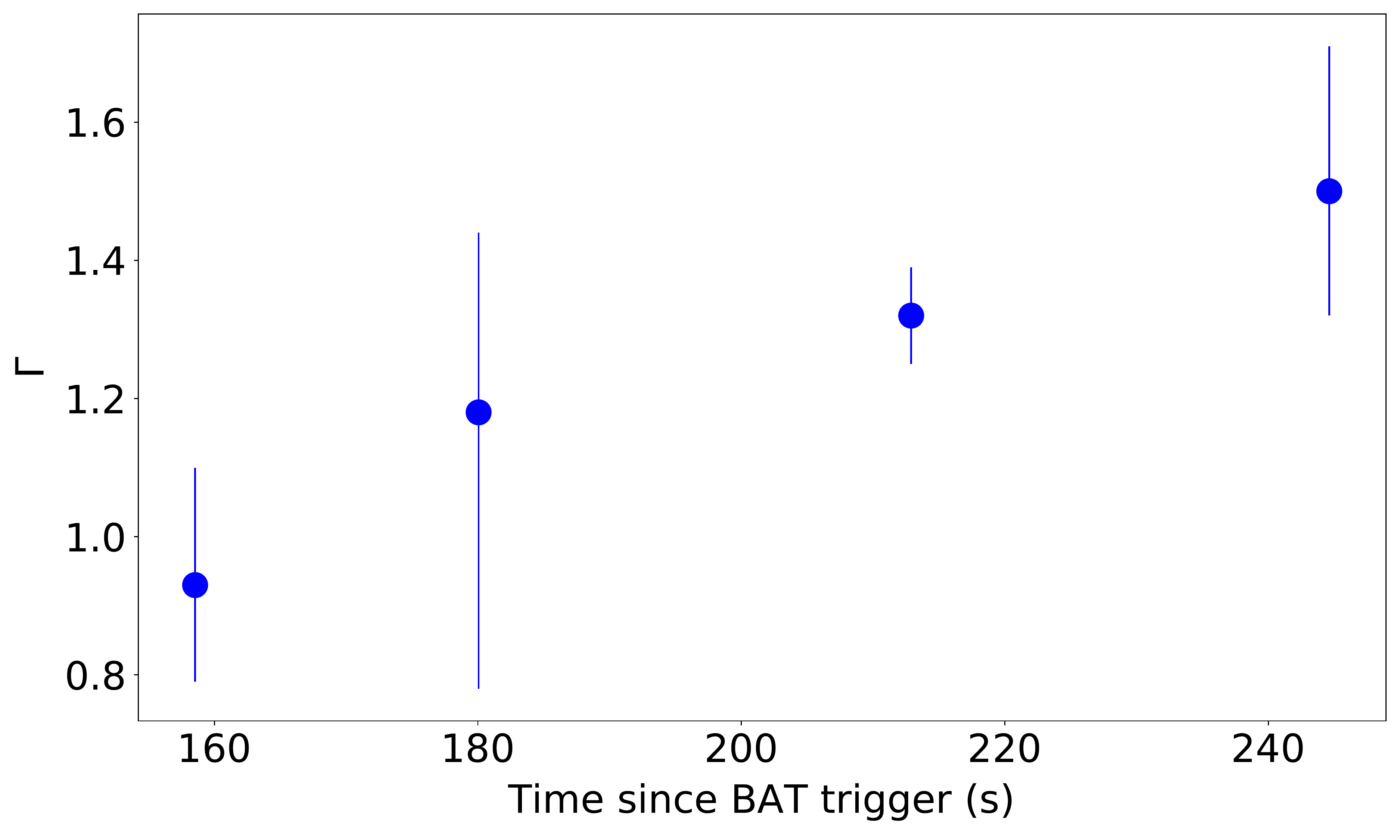}
    \end{subfigure}
    \caption{Same as in Fig. \ref{050724res} but for GRB~180329B.}
	\label{180329res}
\end{figure*}

\begin{figure*}
    \begin{subfigure}[b]{0.49\textwidth}
    \includegraphics[width=\columnwidth, height = 7.6cm]{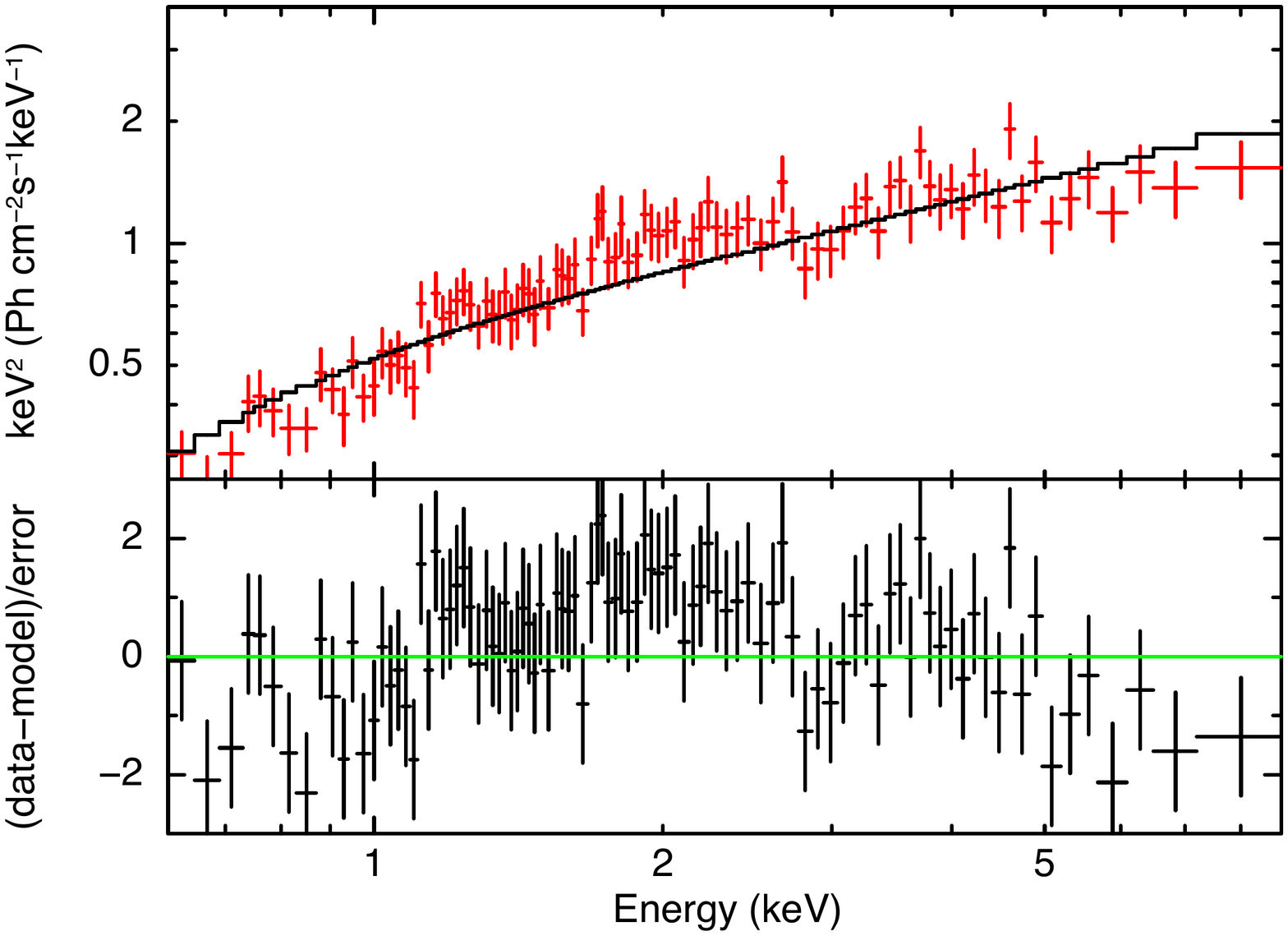}
    \end{subfigure}
    \begin{subfigure}[b]{0.49\textwidth}
     \includegraphics[width=\columnwidth, height = 7.6cm]{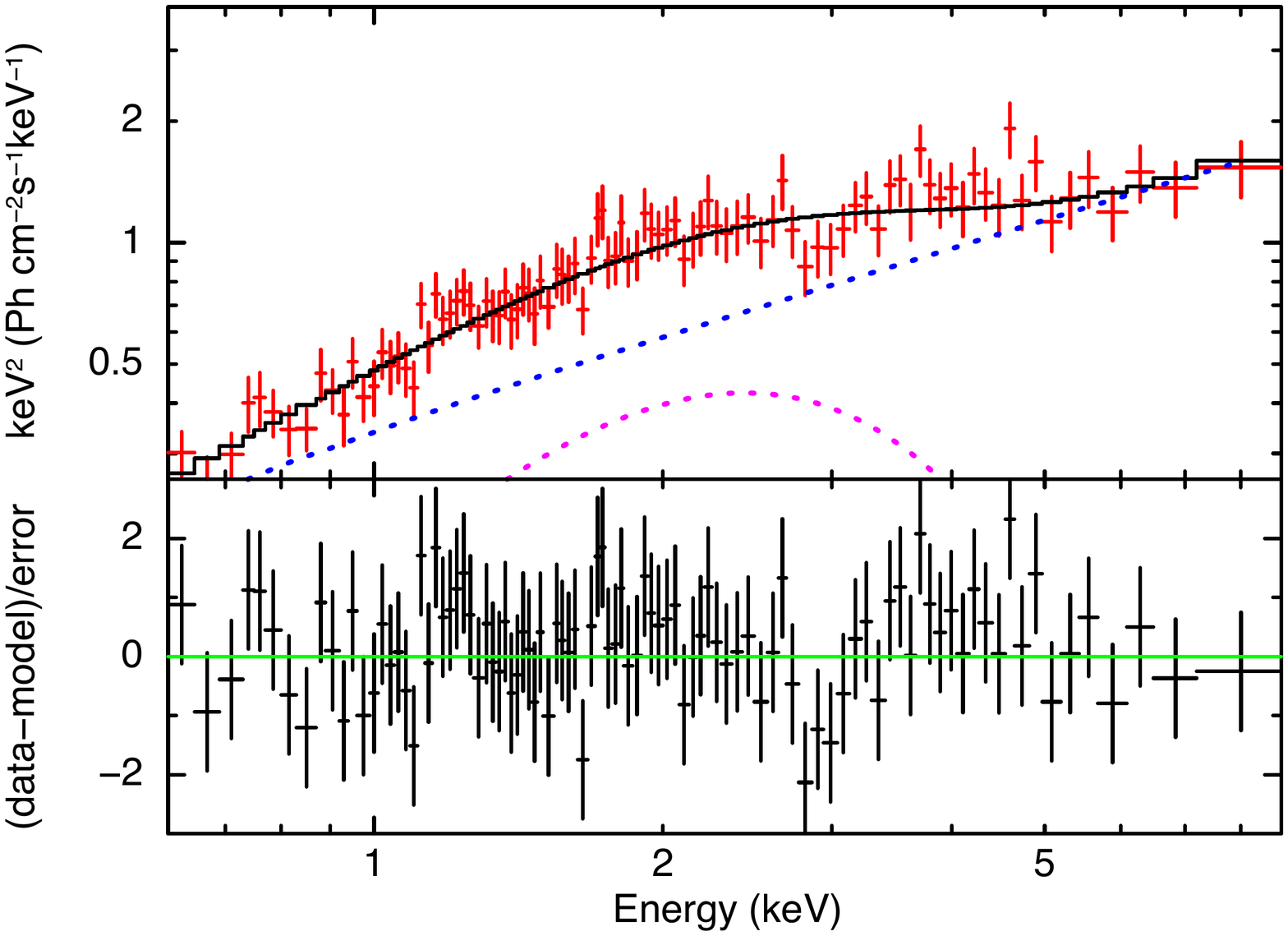}
    \end{subfigure}
    \caption{Same as in Fig. \ref{050724xspec} but for GRB 180329B in the time interval 188 - 237 s. The data have been rebinned for visual clarity.}
	\label{180329xspec}
\end{figure*}

\begin{figure*}
    \begin{subfigure}[b]{0.49\textwidth}
     \includegraphics[width=\columnwidth, height = 5cm]{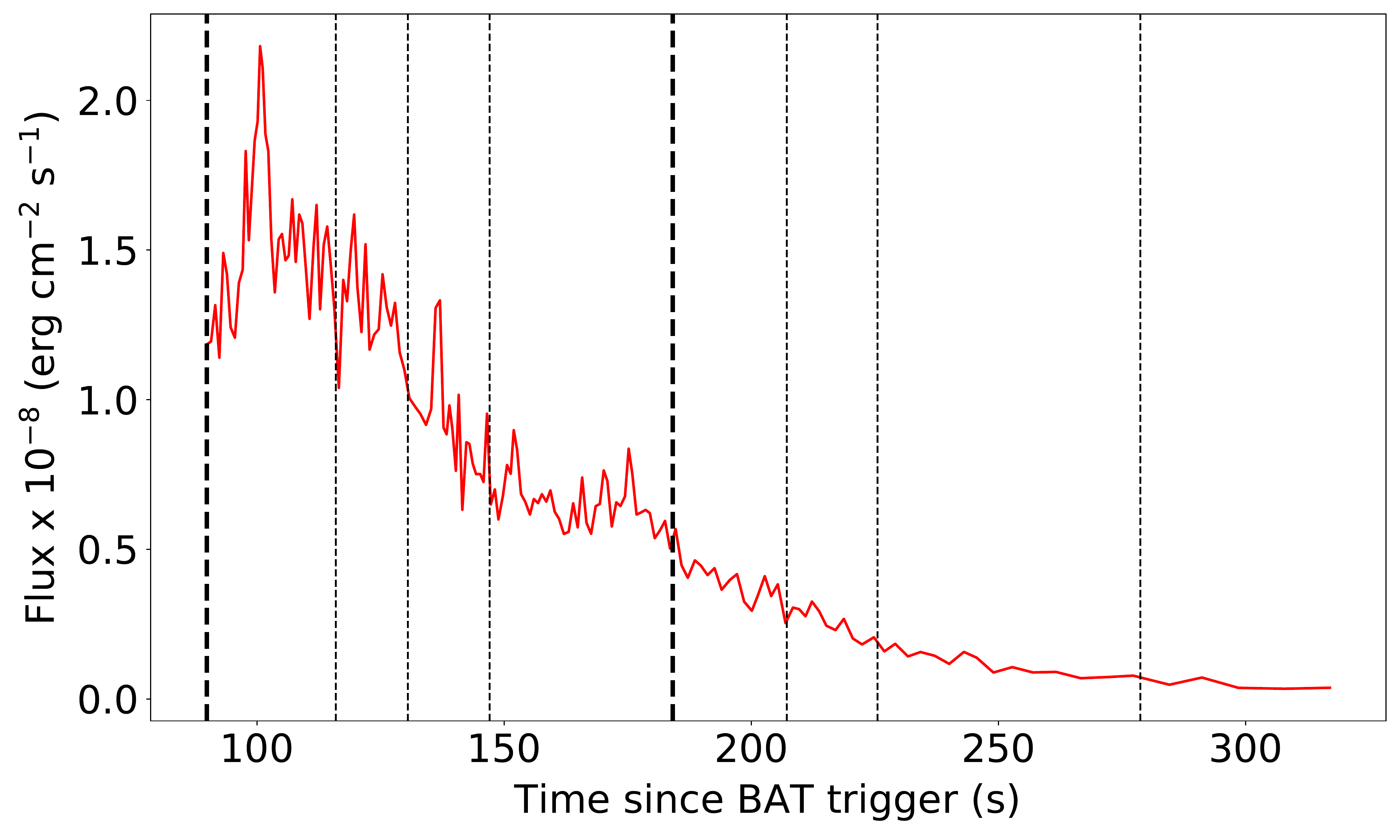}
    \end{subfigure}
    \begin{subfigure}[b]{0.49\textwidth}
    \includegraphics[width=\columnwidth,  height = 5cm]{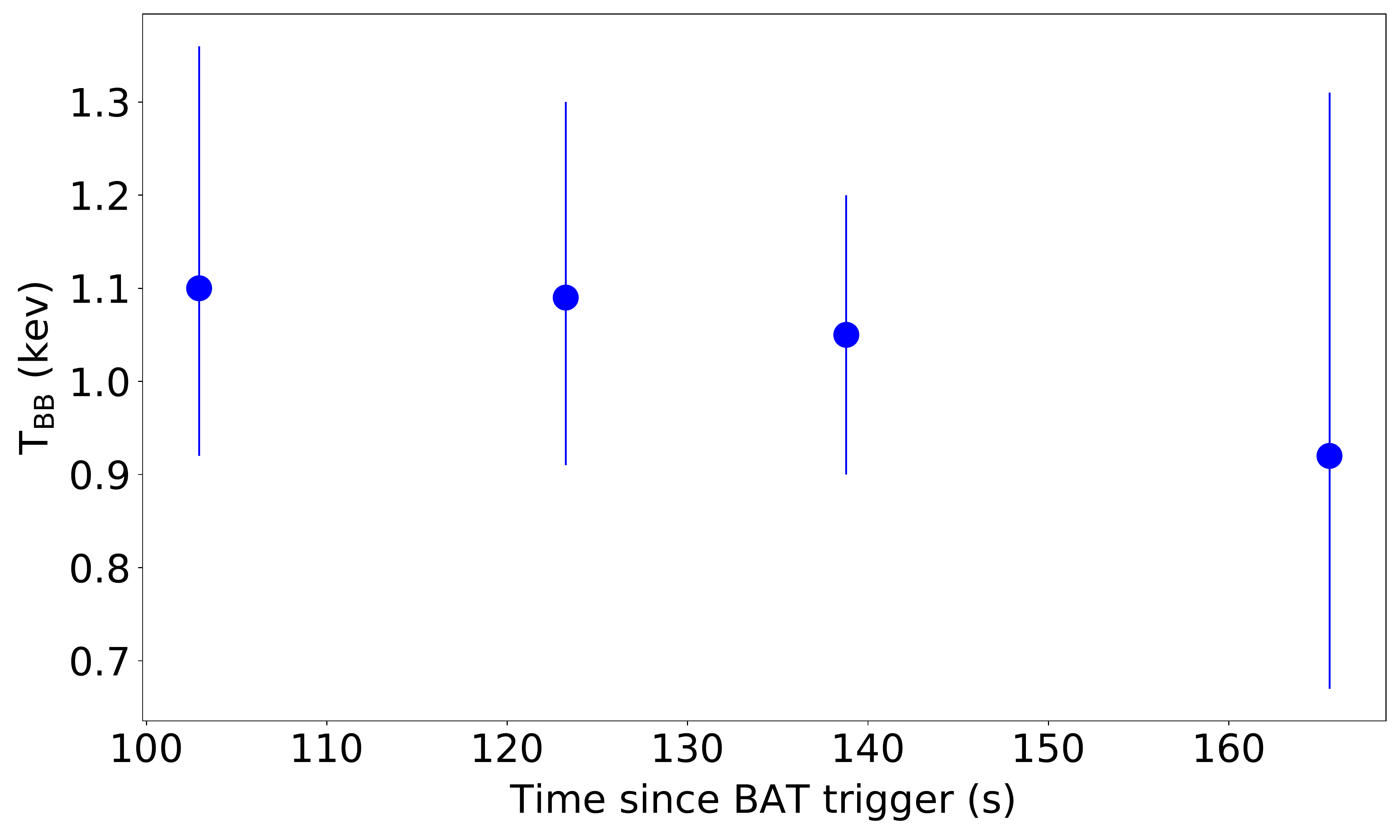}
    \end{subfigure}
    \begin{subfigure}[b]{0.49\textwidth}
     \includegraphics[width=\columnwidth, height = 5cm]{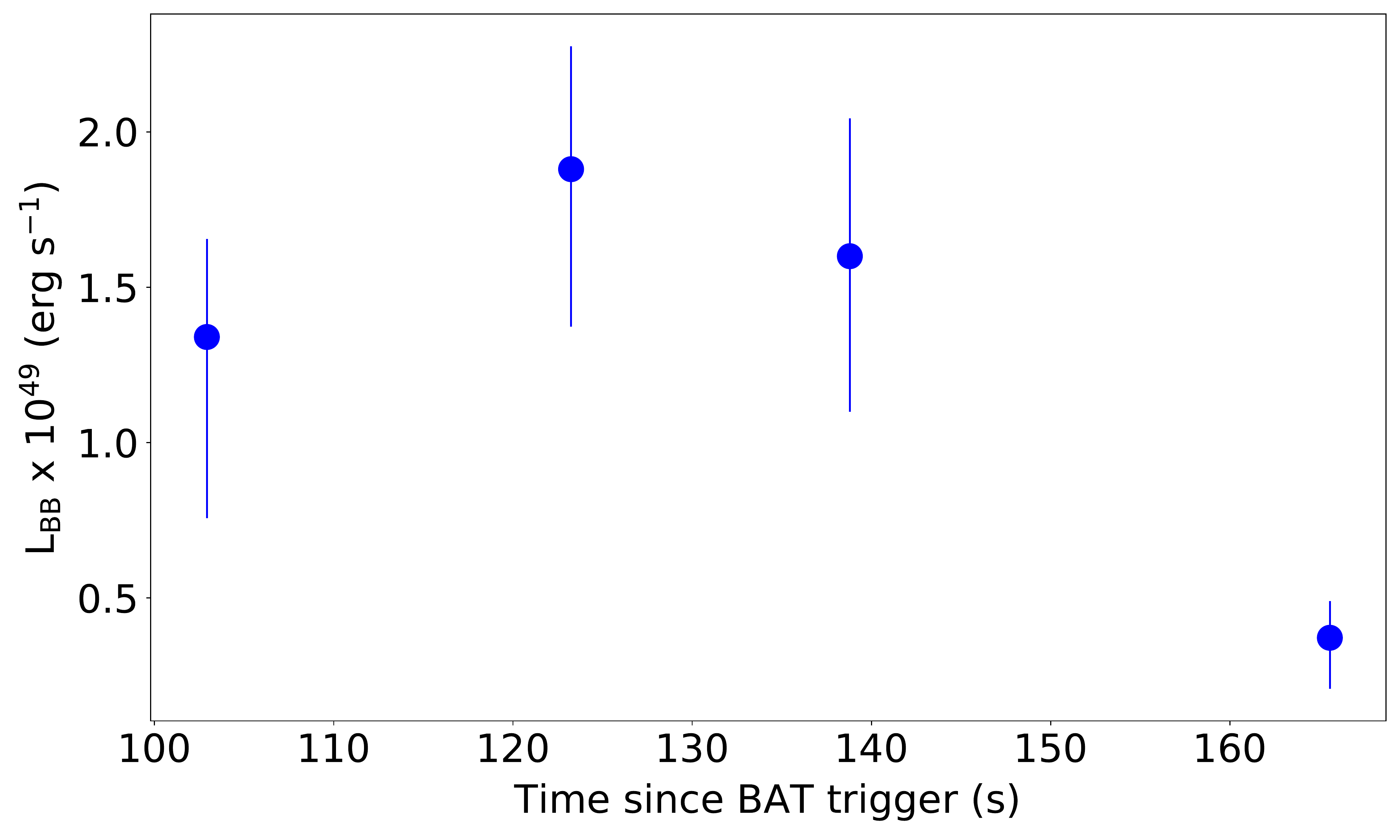}
    \end{subfigure}
    \begin{subfigure}[b]{0.49\textwidth}
     \includegraphics[width=\columnwidth, height = 5cm]{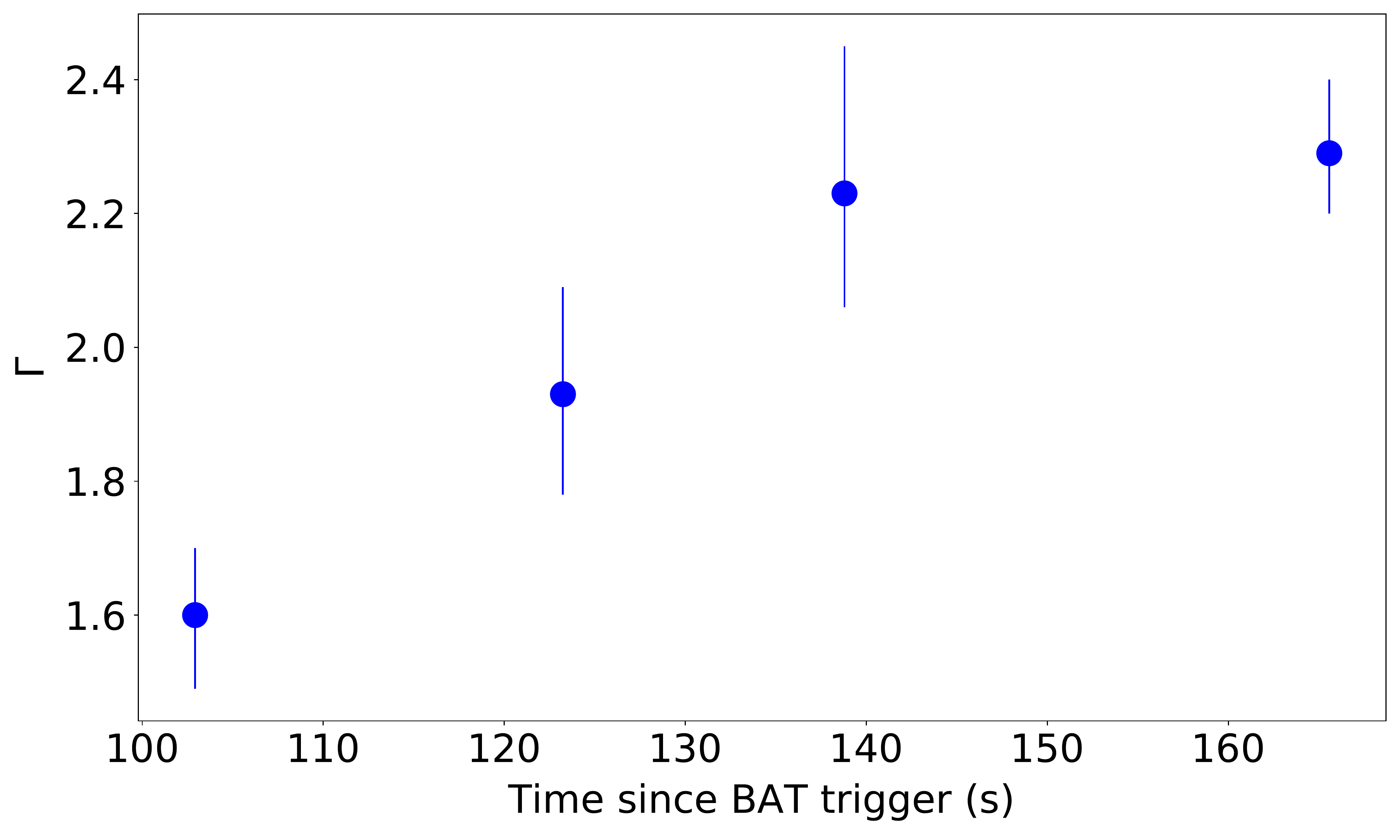}
    \end{subfigure}
    \caption{Same as in Fig. \ref{050724res} but for GRB 180620B.}
	\label{180620res}
\end{figure*}

\begin{figure*}
    \begin{subfigure}[b]{0.49\textwidth}
    \includegraphics[width=\columnwidth, height = 7.6cm]{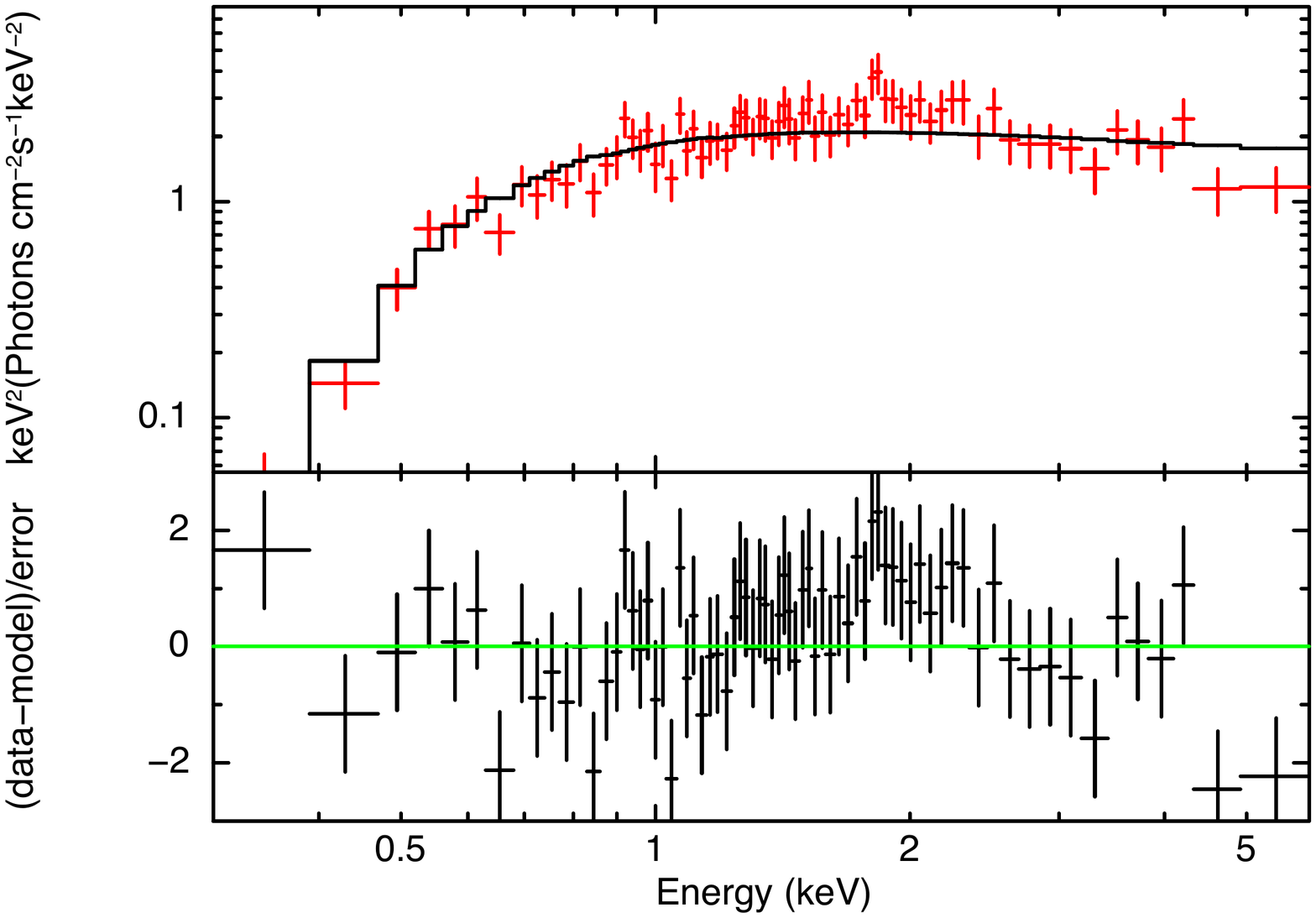}
    \end{subfigure}
    \begin{subfigure}[b]{0.49\textwidth}
     \includegraphics[width=\columnwidth, height = 7.6cm]{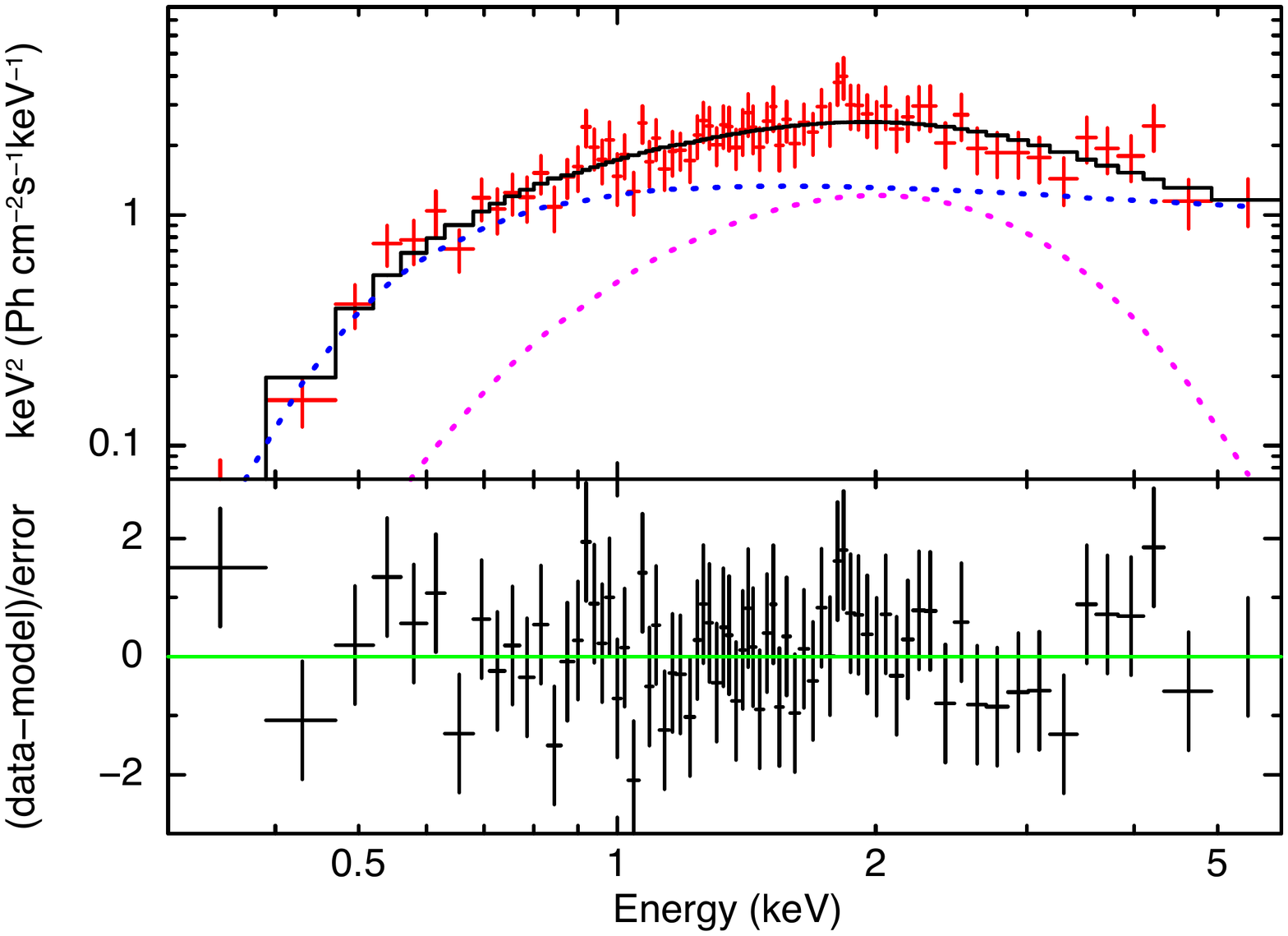}
    \end{subfigure}
    \caption{Same as in Fig. \ref{050724xspec} but for GRB 180620B in the time interval 130 - 147 s. The data have been rebinned for visual clarity.}
	\label{180620xspec}
\end{figure*}

Here we present the best-fit parameters and spectral fits for all GRBs with detected blackbody components. Light curves together with the evolution of the rest-frame temperatures, blackbody luminosities and photon indices are shown in Figs. \ref{050724res}, \ref{071031res}, \ref{080325res}, \ref{081221res}, \ref{090516res}, \ref{090517res}, \ref{171205res}, \ref{171222res}, \ref{180329res}, \ref{180620res}. Spectral fits with and without a blackbody for the time interval where the blackbody is most significant are shown in Figs. \ref{050724xspec}, \ref{071031xspec}, \ref{080325xspec}, \ref{081221xspec}, \ref{090516xspec}, \ref{090715xspec}, \ref{171205xspec}, \ref{171222xspec}, \ref{180329xspec}, \ref{180620xspec}. Further notes on the individual GRBs are provided below. 

\paragraph*{GRB~050724:} The WT observations start $\sim 80~\rm{s}$ after the BAT trigger and the light curve decays smoothly over the course of the observation. The blackbody is significant at $> 3 \sigma$ from the start of the observation until $ \sim 200$~s. The temperature of the blackbody is approximately constant around $0.9 \pm 0.14 ~ \rm{keV}$, while the photon index varies between 0.7 and 2.2 with relatively large uncertainties and no clear trend in time. 

\paragraph*{GRB~071031:} The WT observations start $\sim 100\ \rm{s}$ after the BAT trigger. The light curve is characterized by a large flare at the beginning and two smaller flares with peaks at $\sim 200 ~\rm{s}$ and $\sim 250\ \rm{s}$. The simulations show that the blackbody is significant at $> 3 \sigma$ from 120~s until 215~s. The temperature of the blackbody evolves  from $2.74 \pm 0.43 ~ \rm{keV}$ to $1.77 \pm 0.32 ~ \rm{keV}$. The photon index softens from $1.49 \pm 0.12$ to $2.05 \pm 0.09$. 

\paragraph*{GRB~080325:} The WT observations start $\sim 150 ~ \rm{s}$ after the BAT trigger. There are two flares superposed on the decaying light curve, one at the beginning and one peaking at $\sim 220 ~ \rm{s}$. The simulations reveal a blackbody at $> 3 \sigma$ significance from the start of the observations until $\sim 250 ~ \rm{s}$. The temperature of the blackbody is poorly constrained and consistent with being constant around $1.5 ~ \rm{keV}$, while the photon index softens from $1.05 \pm 0.21$ to $1.88 \pm 0.16$. 

\paragraph*{GRB~081221:} The WT observations start $\sim 75 ~ \rm{s}$ after the BAT trigger. The light curve decays smoothly and the blackbody is significant at $> 3 \sigma$ from the beginning until $\sim 110 ~ \rm{s}$. The temperature of the blackbody is poorly constrained and is consistent with being constant at $\sim 1.9 ~ \rm{keV}$, while the photon index softens from $0.8 \pm 1.9$ to $ 1.72 \pm 0.23$. 

\paragraph*{GRB~090516:} The WT observations start $\sim 170 ~ \rm{s}$ after the BAT trigger. The light curve is characterized by a large flare peaking at $\sim 270 ~ \rm{s}$. The simulations show a blackbody at $ > 3 \sigma$ significance from $225 ~ \rm{s}$ until $300 ~ \rm{s}$, i.e. covering the time of the flare. The temperature of the blackbody is one of the highest in the sample, starting at $3.07 \pm 0.7 ~ \rm{keV}$ and showing some evidence of tracking the flare.  The photon index softens from $1.86 \pm 0.1$ to $2.35 \pm 0.16$. 

\paragraph*{GRB~090715B:} The WT observations start $\sim 50 ~ \rm{s}$ after the BAT trigger. The light curve is erratic and characterized by two prominent flares; one at the beginning and one peaking at $\sim 250 ~ \rm{s}$. The blackbody is detected  at $> 3 \sigma$ from $130~  \rm{s}$ to $ 206 ~ \rm{s}$, which corresponds to the time interval between the two big flares. The blackbody temperature is in the range 1.48 -- 2.5~keV and the photon index is in range 1.83--2.21.

\paragraph*{GRB~171205A:} The WT observations start $\sim 150 ~ \rm{s}$ after the BAT trigger and the light curve decays smoothly  throughout the observation. The blackbody is detected with $> 3 \sigma$ significance from 150~s until $\sim 270 \rm{s}$. The temperature is consistent with being constant at $\sim 0.1 \ \rm{keV}$ while the photon index has a value around 1.6 and is also consistent with being constant.

\paragraph*{GRB~171222A:} The WT observations start $\sim 170 ~ \rm{s}$ after the BAT trigger and the light curve decays smoothly  throughout the observation. The blackbody is detected with $> 3 \sigma$ significance from 250~s until the end of the observation. The temperature decays from $0.5 \pm 0.1 \ \rm{keV}$ to $0.19 \pm 0.05 \ \rm{keV}$ while the photon index has a value around 2.2 and shows no evolutionary trend within the uncertainties. 

\paragraph*{GRB~180329B:} The WT observations start $\sim 110 ~ \rm{s}$ after the BAT trigger. There are small flares superposed on the decaying light curve until $\sim 250 ~ \rm{s}$ and the simulations show a blackbody with $> 3 \sigma$ significance from $145 ~ \rm{s}$ until $250 ~ \rm{s}$. The temperature of the blackbody is initially poorly constrained at $2.4 \pm 0.81 ~ \rm{keV}$ and reaches a value of $1.61 \pm 0.47 ~ \rm{keV}$ at the end. The photon index softens from $0.93 \pm 0.15$ to $ 1.5 \pm 0.20$. 

\paragraph*{GRB~180620B:} The WT observations start $\sim 90 ~ \rm{s}$ after the BAT trigger, showing a smoothly decaying light curve with a small flare at the beginning. The blackbody is detected from the beginning until $\sim 184 ~ \rm{s}$. The temperature of the blackbody has large uncertainties and is consistent with staying constant around $\sim 1 ~ \rm{keV}$. The photon index softens from $ 1.6 \pm 0.11$ to $ 2.29 \pm 0.10$.  

\section{Light curves of GRBs with significant blackbody components reported in V18}
\label{lcv18sec}

\begin{figure*}
    \begin{subfigure}[b]{0.49\textwidth}
    \includegraphics[width=\columnwidth,  height = 4.5cm]{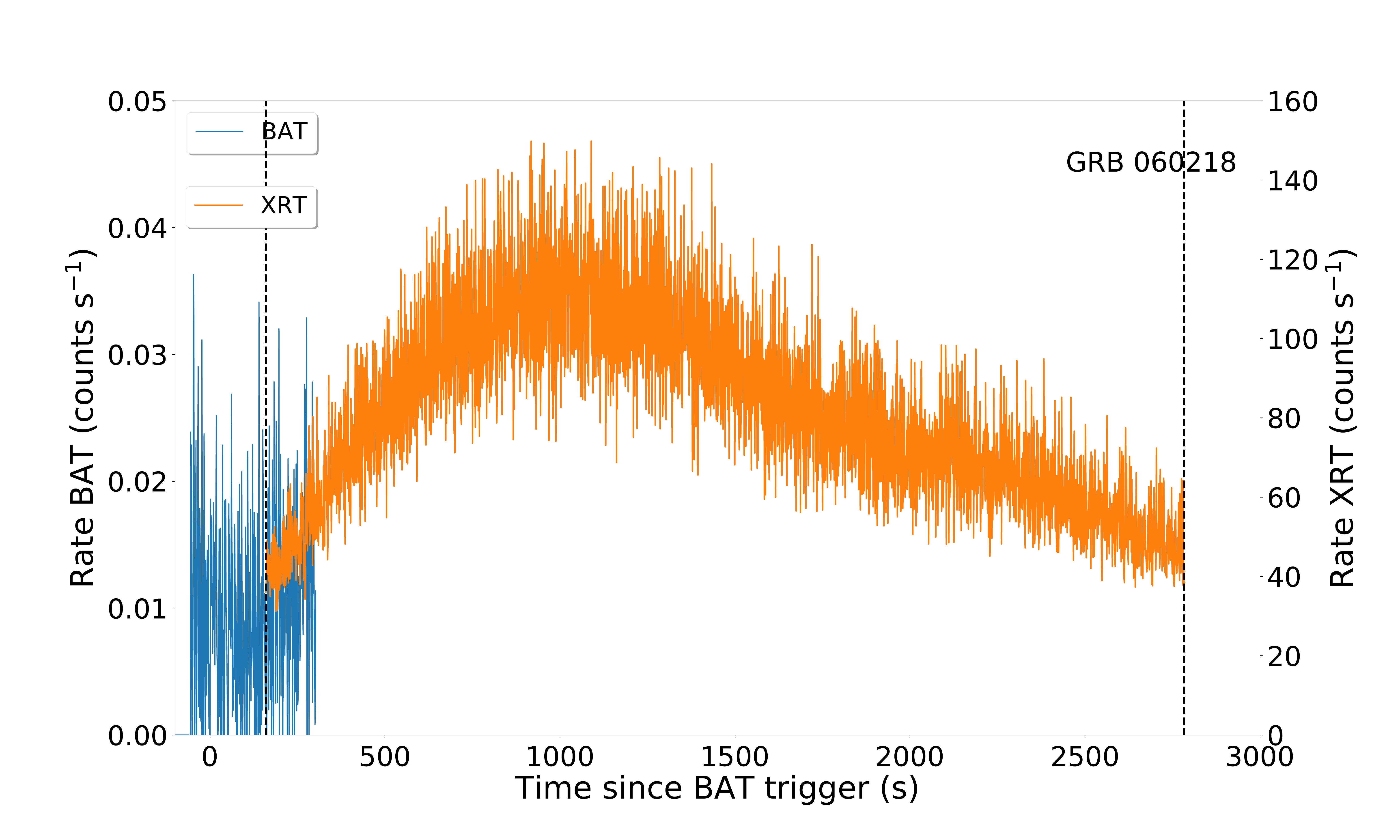}
    \end{subfigure}
    \begin{subfigure}[b]{0.49\textwidth}
     \includegraphics[width=\columnwidth, height = 4.5cm]{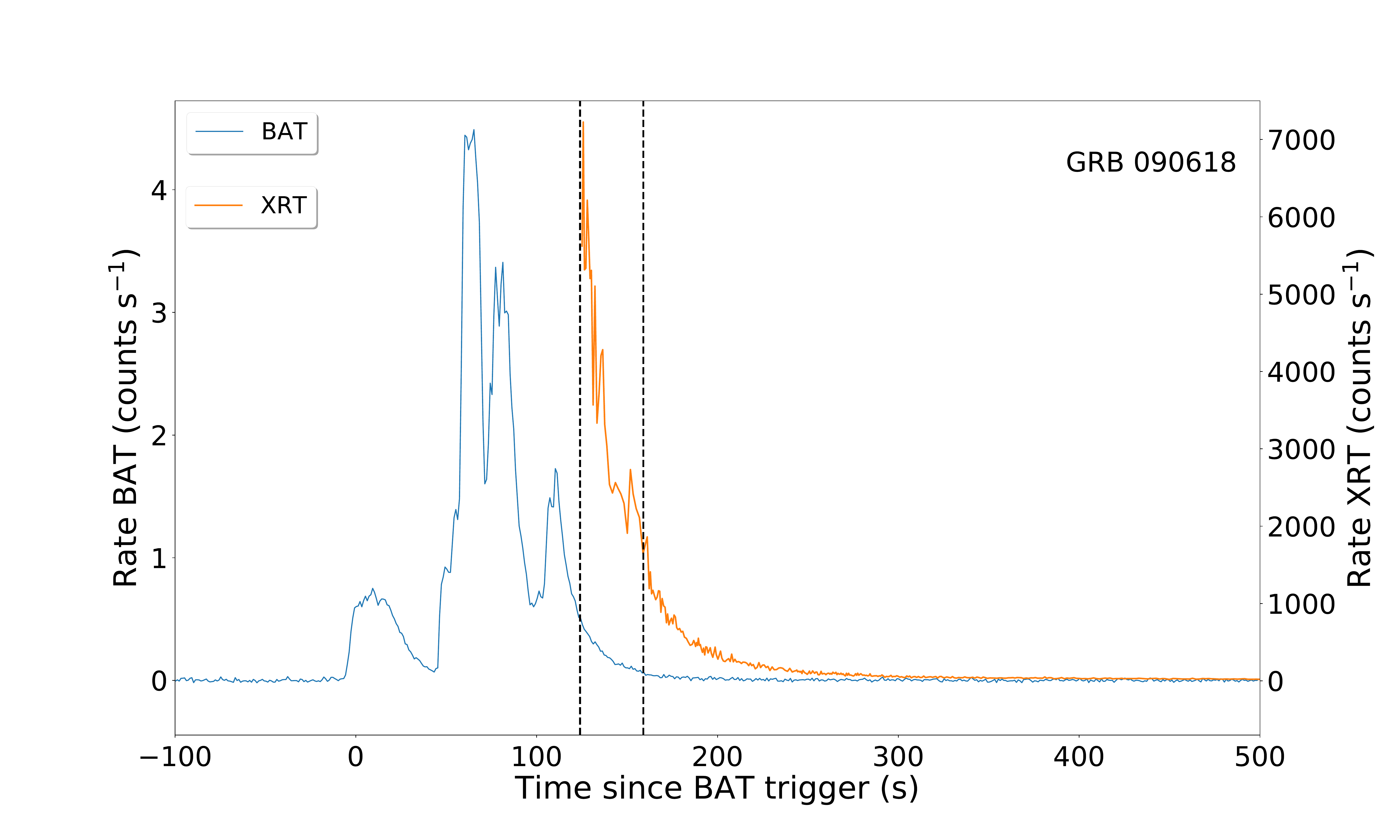}
    \end{subfigure}
        \begin{subfigure}[b]{0.49\textwidth}
     \includegraphics[width=\columnwidth, height = 4.5cm]{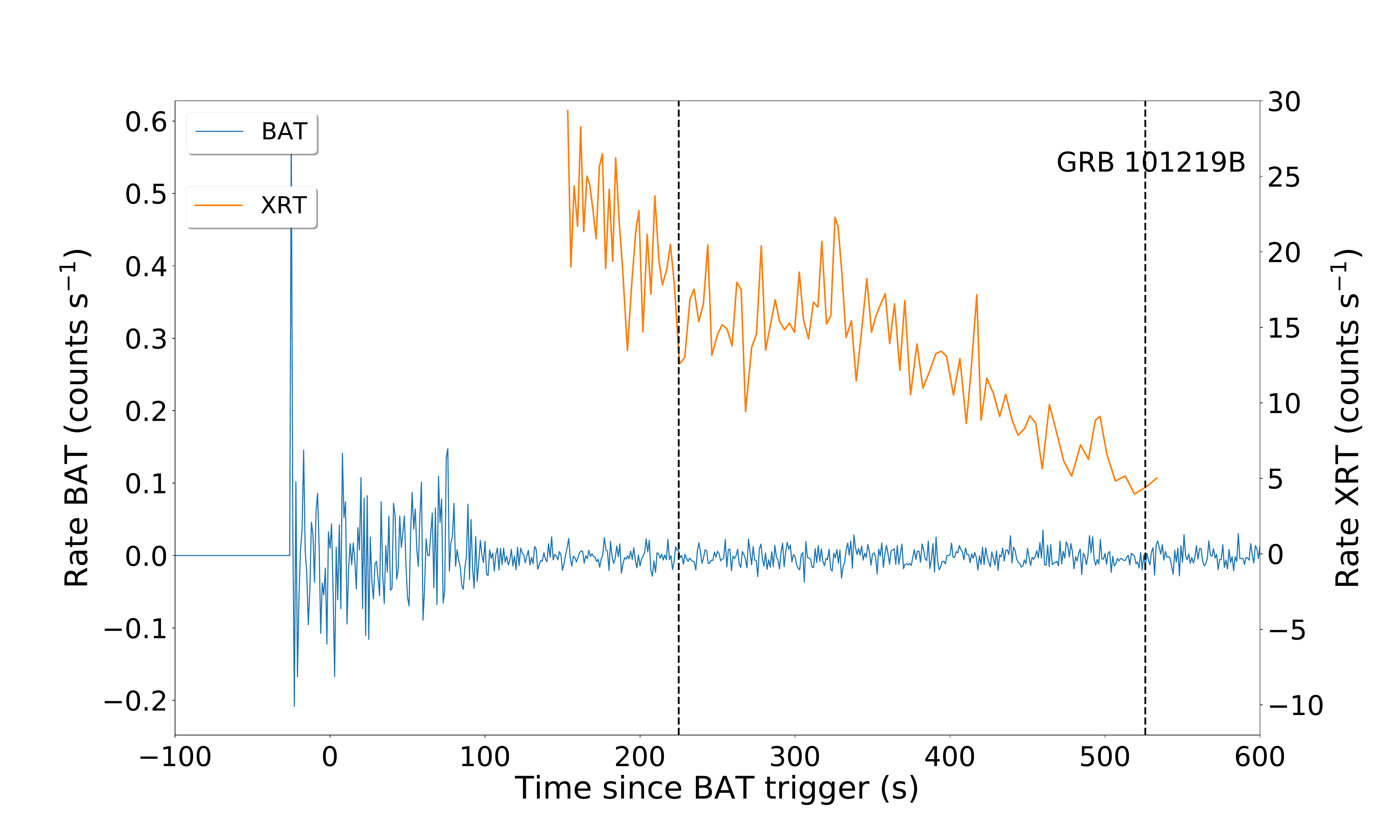}
    \end{subfigure}
        \begin{subfigure}[b]{0.49\textwidth}
     \includegraphics[width=\columnwidth, height = 4.5cm]{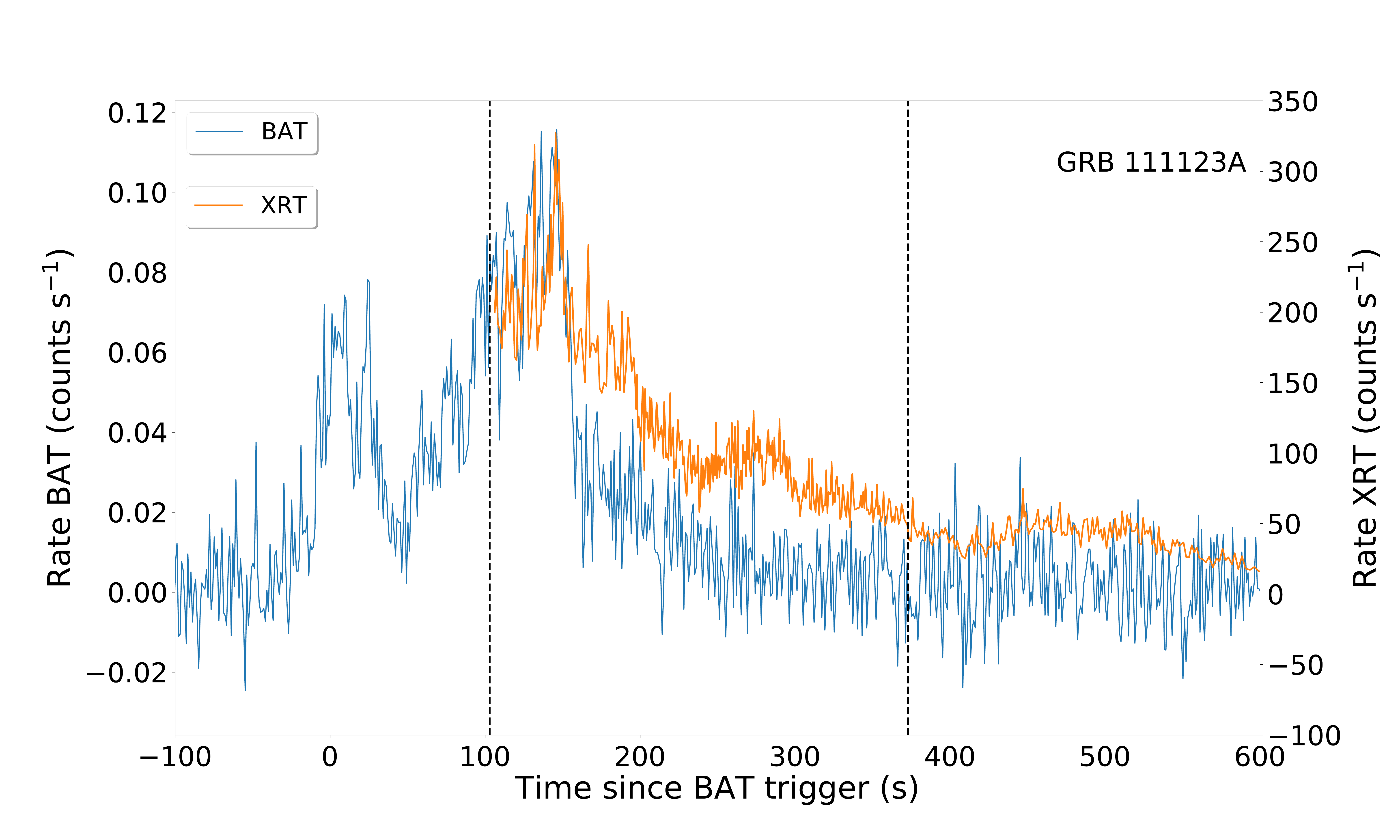}
    \end{subfigure}
        \begin{subfigure}[b]{0.49\textwidth}
     \includegraphics[width=\columnwidth, height = 4.5cm]{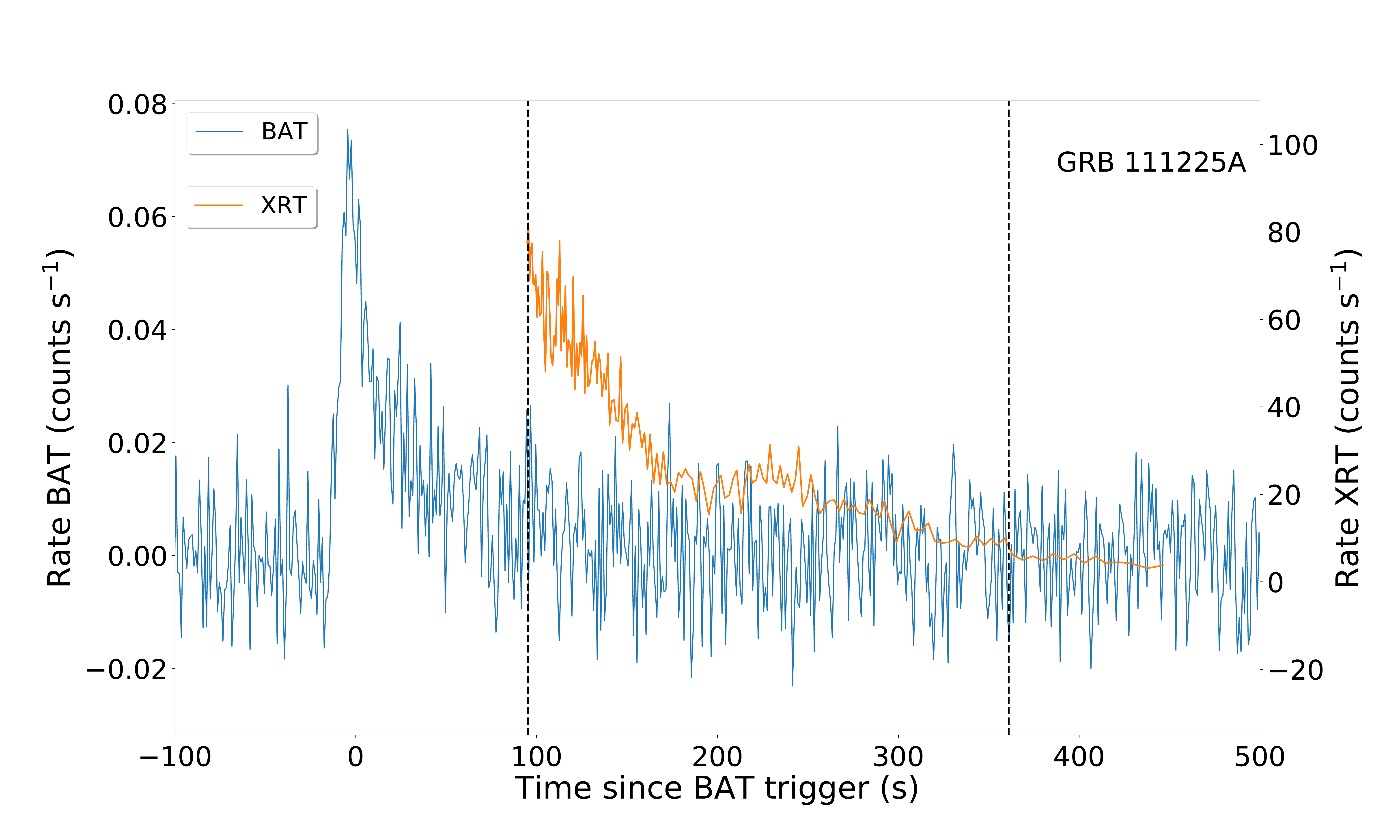}
    \end{subfigure}
        \begin{subfigure}[b]{0.49\textwidth}
     \includegraphics[width=\columnwidth, height = 4.5cm]{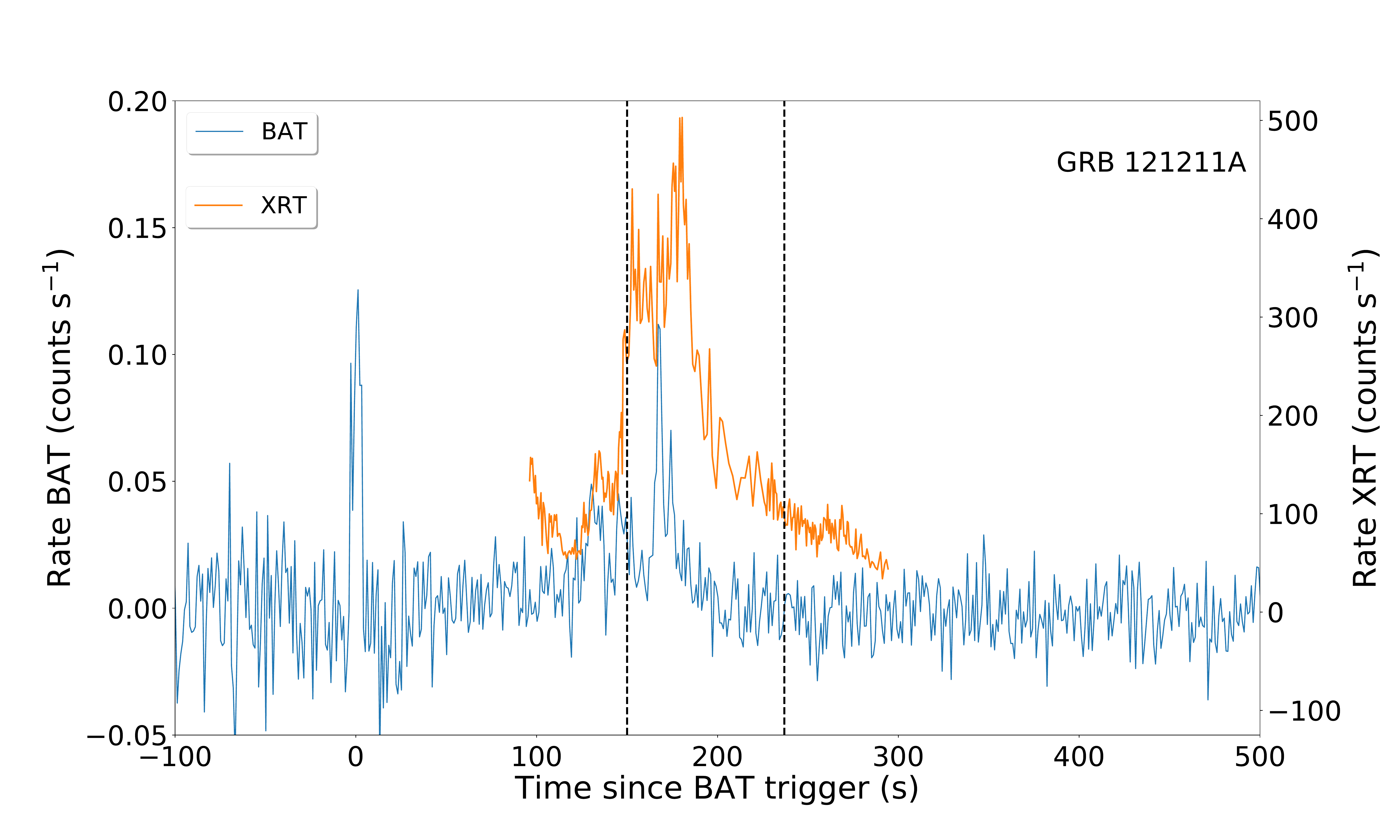}
    \end{subfigure}
        \begin{subfigure}[b]{0.49\textwidth}
     \includegraphics[width=\columnwidth, height = 4.5cm]{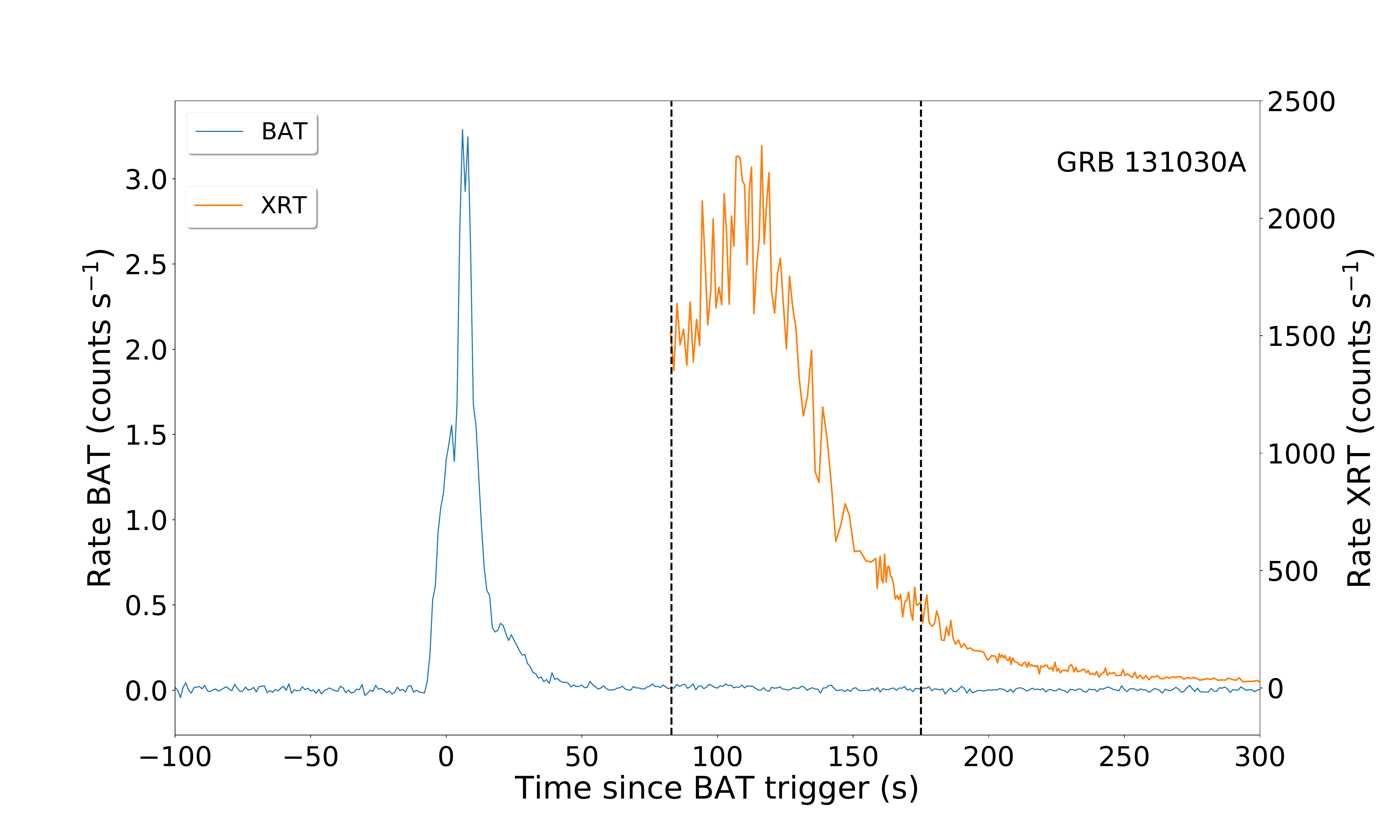}
    \end{subfigure}
        \begin{subfigure}[b]{0.49\textwidth}
     \includegraphics[width=\columnwidth, height = 4.5cm]{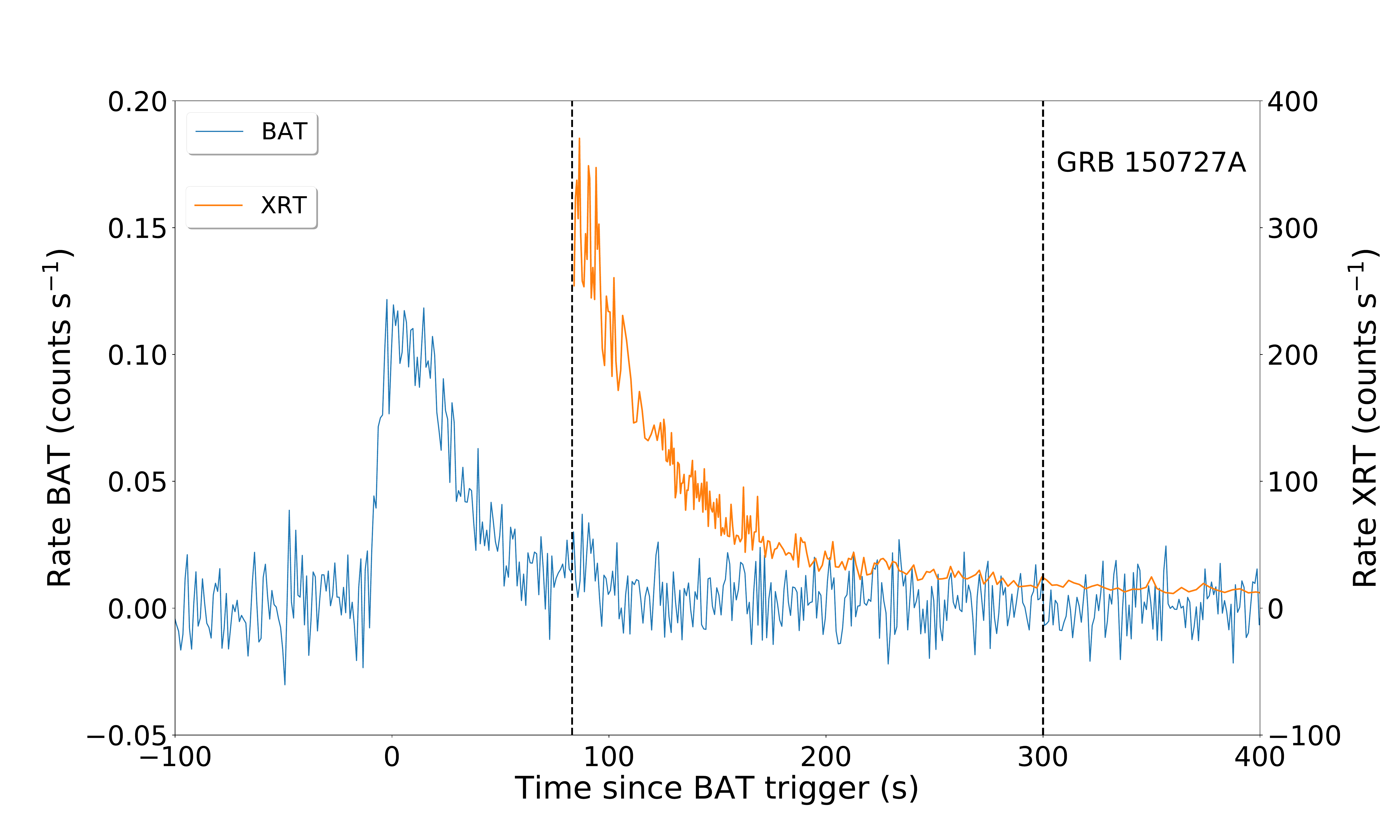}
    \end{subfigure}
        \begin{subfigure}[b]{0.49\textwidth}
     \includegraphics[width=\columnwidth, height = 4.5cm]{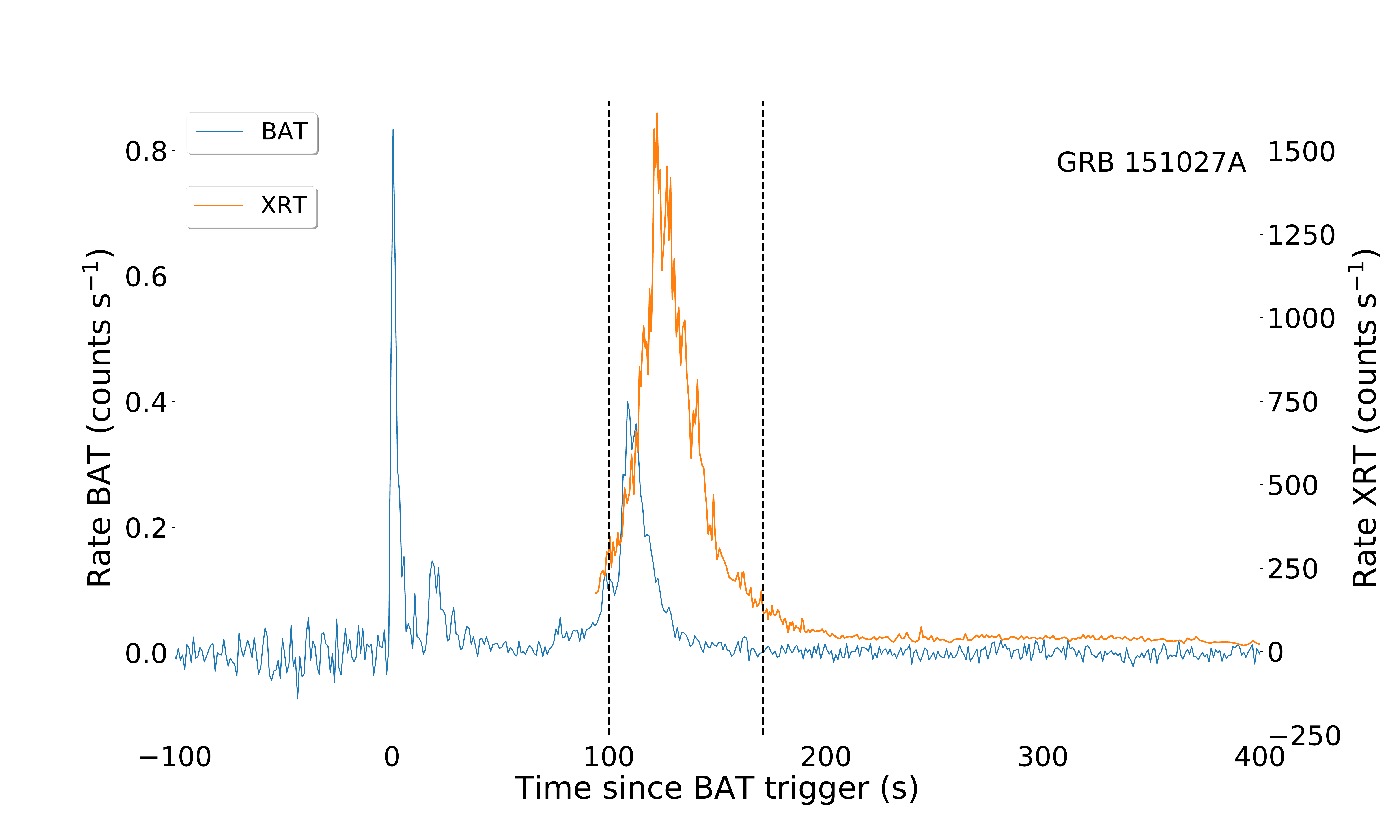}
    \end{subfigure}
    \caption{BAT and XRT light curves for GRBs with significant blackbody components analysed in V18. The black dashed lines mark the time interval where the blackbody is significant for each GRB.}
	\label{lcv18}
\end{figure*}

In Fig.~\ref{lcv18} we present the BAT+XRT light curves of the nine GRBs with blackbody components that were discussed in V18. 

%%%%%%%%%%%%%%%%%%%%%%%%%%%%%%%%%%%%%%%%%%%%%%%%%%

% Don't change these lines
\bsp	% typesetting comment
\label{lastpage}
\end{document}